\documentclass[reprint, amsmath,amssymb,aps]{revtex4-2}

\usepackage{graphicx}
\usepackage{dcolumn}
\usepackage{bm}
\usepackage{xcolor} 

\newcommand{\lso}{Li$_2$S}
\newcommand{\lst}{Li$_2$S$_2$}
\newcommand{\lsf}{Li$_2$S$_4$}
\newcommand{\lss}{Li$_2$S$_6$}
\newcommand{\lse}{Li$_2$S$_8$}
\newcommand{\TMNC}{TM--N$_4$--C}
\newcommand{\FeNC}{Fe--N$_4$--C}
\newcommand{\ZnNC}{Zn--N$_4$--C}

\begin{document}

\title{Accelerating the theoretical study of Li-polysulphide adsorption on single-atom catalysts via machine learning approaches}

\author{Eleftherios I. Andritsos}
\email{lefteri.andritsos@kcl.ac.uk}
\affiliation{Department of Physics, King’s College London, Strand, London, WC2R 2LS, UK}

\author{Kevin Rossi}
\email{kevin.rossi@epfl.ch}
\affiliation{Laboratory of Nanochemistry for Energy, Institute of Chemistry, Ecole Polytechnique F\'{e}d\'{e}rale de Lausanne, Lausanne, 1015, CH}

\begin{abstract}
Unlocking the design of Li-S batteries where no shuttle effects appears, and thus their energy storage capacity does not diminish over time, would enable the manufacturing of energy storage devices more performant than the current Li-ion commercial ones.
Computational screening of Li-polysulphide (LiPS) adsorption on single-atom catalyst (SAC) substrates is of great aid to the design of lithium--sulphur batteries which are robust against the LiPS shuttling from the cathode to the anode and the electrolyte.
To aid this process, we develop a machine learning protocol to accelerate the systematic mapping of dominant local minima found with DFT calculations, and, in turn, fast screen LiPS adsorption properties on SACs.
We first validate the approach by probing the potential energy surface for Li-polysulphides adsorbed on graphene decorated with a Fe SAC bound to four nitrogen atoms. 
We identify minima whose binding energy is better or on par with the one previously reported in the literature. 
We then move to analyse the adsorption trends on \ZnNC SAC and observe similar adsorption strength and behaviour with the \FeNC \ SAC, highlighting the good predictive power of our protocol.

\end{abstract}

\maketitle

\section{Introduction}

Lithium--sulphur (Li--S) batteries are an emerging technology in the field of energy storage. 
Their high-density energy storage capacity, of $\sim$2500 W h kg$^{-1}$ theoretical specific energy \cite{Ji2010rev}, is indeed much higher than that of Li-ion or other commercial batteries.
However, in practice, the Li--S battery rate capability and life cycle degrades severely over time because of the so called ``shuttle'' effect.
The latter is caused by lithium polysulphide (LiPS) molecules remaining in the electrolyte during the charge--discharge process, leading to loss of active material and reduced battery performance.

Controlling the LiPS rate of migration to the anode and the reaction kinetics in the cathode is important for tackling the issue of the ``shuttle'' effect. 
Several strategies have been adopted to reduce the diffusivity of LiPS to the electrolyte, mainly focusing on the development of cathode host materials that can strongly bind the LiPS while also improving the reaction kinetics \cite{Tao2016,Ren2019,Zhou2020rev}.
However, regardless of the vast amount of research in the field over the recent years, prevention of battery degradation still remains a challenge. \par

To overcome the issue, recent studies focus on single-atom catalyst (SAC) materials.
The latter provide strong polysulphide adsorption and improve the reaction kinetics, thanks to their exceptional electrocatalytic properties.
Recent studies indeed suggest that SAC with atomically dispersed transition metals (TMs) on graphene lattices are particularly promising candidates.
In particular, calculations on the effect of SACs with TM--N$_4$--C formation (TM = Co, Fe, Mn, Ru, V, W, Zn, and C as a graphene lattice) on the LiPS adsorption, the Gibbs free energy change and the reaction activation barrier \cite{Zhang2019Fe, Wang2019, Zeng2019_FeSAC, Du2019, Zhou2020_SAC, WangC2020, Li2020, Shao2020, Andritsos2021}, predict a remarkable improvement in the cathode properties which could potentially reduce the LiPS ``shuttle'' effect. \par

Numerical methods provide ground to screen SAC adsorption properties.
The identification of the global energy minimum LiPS-SAC structure, a stepping stone to predict adsorption trends, is however rather challenging.
Several different initial guess structures and structural optimisations have to be considered in order to identify the global minimum structure.
For example, in high-order LiPS intermediates (\lsf, \lss \ and \lse) the final binding energies varies over 0.5 eV \cite{Jand2016} depending on the initial LiPS structure that is used as input in the relaxation.
By the same token, reported values of \lss \ binding energy on Fe-based SAC, one of the most examined SAC material for Li--S batteries, vary by about 0.6 eV, from $-0.9$ eV to $-$1.5 eV \cite{Zeng2019_FeSAC, Zhang2019Fe, Zhou2020_SAC, Andritsos2021}.
The large discrepancy in the results, raises concerns regarding the quality of the protocols adopted to identify the minimum energy structure, and adds uncertainty when comparing SAC results from different studies. \par  

\begin{figure*}[t]
    \centering
    \includegraphics[width=0.9\textwidth]{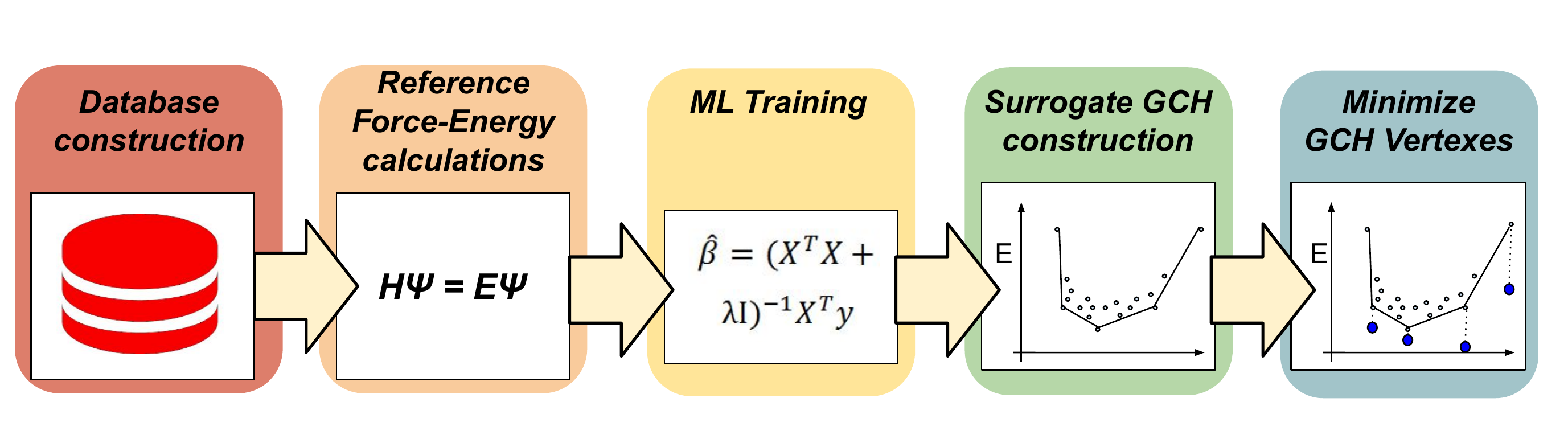}
    \caption{Schematic representation of the five steps describing the binding energy prediction workflow in this study.}
    \label{fig:wflow}
\end{figure*}

Here, we demonstrate the use of machine learning (ML) methods to accelerate the prediction of LiPS binding energies on SAC materials.
In particular, we develop a protocol which hinges on a supervised/unsupervised machine learning strategy, in combination with density functional theory (DFT) calculations, to efficiently screen the stable structures of Li$_2$S$_n$ ($n = 1, 2, 4, 6, 8)$ on \TMNC \ (TM = Fe, Zn) anchoring materials.
We adopt unsupervised and supervised machine learning to isolate a structurally diverse and energetically favourable set of structures from a database encompassing roto-translations of LiPSs on a graphene-based SAC substrate.
We then show that by relaxing these structures we identify SAC/LiPS geometries whose binding energy is on-par or lower with those reported in the literature.
We thus demonstrate how machine learning methods of different nature can accelerate the systematic identification of the energetically favourable LiPS/SAC substrate geometries. \par

The proposed approach gauges on the recent advances in ML methods applied to the modelling of catalysts and materials for energy applications \cite{Chi2020_rev,Liu2021_rev}.
While application of ML methods emerged as a new paradigm to accelerate the exploration of complex interactions between small and medium molecules with a variety of binding sites, \cite{Rossi2020, Tran2018, Gu2018} data-driven approaches in the field of Li--S batteries still remain too under-exploited \cite{AndritsosRoadmap}.
ML modelling in Li--S batteries has been used to examine the impact of materials and battery design on the battery performance \cite{Kilic2020}, and the LiPS adsorption on two-dimensional layer-structured MoSe$_2$ and WSe$_2$ by calculating the binding energy of LiPSs for arbitrary spacial configurations at random sites on the host material \cite{Zhang2021}.
A recent study also combined a machine learning framework, based on a modified Crystal Graph Convolutional Neural Network and DFT calculations to offer rational design of SACs for Li--S batteries \cite{Lian2021}.
The authors focus mainly on the low-order LiPS and S$_8$ to reveal the binding energy pattern of LiPSs on a wide range of SACs. \par

\section{Computational Methods}

The workflow we adopt in this study is visualised in Figure \ref{fig:wflow}. While we refer the interested reader to the dedicated subsections ( \ref{SOAP}--\ref{GCH} ) for detailed technical description on each technique, we here briefly summarise the key aspect in each step of the workflow:

\begin{enumerate}
    \item Database construction:
    \begin{enumerate}
        \item Firstly, we define a supercell space and relax the forces on atoms in the LiPS and in the SAC, each considered as an isolated system,  at DFT level. This ensures that the LiPS and SAC structure are more likely to be energetically favourable when modelling LiPS adsorption on the SAC.
        \item We define a 3D grid in the supercell, where translations of the LiPS will be placed.
        The grid spacing is 0.5 \AA \ along the plane parallel to the substrate, and 0.25 \AA \ along the axis perpendicular to the substrate. 
        For each translation we apply rotation of $10^{\circ}$ in the solid angle centred at the centre of mass of the polysulphide, and obtain the full set of structures.
        The final number roto-translation is reduced and differs for each LiPS, as we apply cutoff criteria so that no LiPS atom is closer than 1.5 \AA \ to atoms in the substrate.
    \end{enumerate}
    
    \item Reference Force-Energy calculations:
    \begin{enumerate}
        \item We map each structure generated in the previous step into a high dimensional feature space. 
        The latter is defined through a Smooth Overlap of Atomic Orbitals (SOAP) representation \cite{SOAP}, employing seven angular and six radial components and a radial cut-off of 6.5 \AA. 
    
        \item We then select a maximum of 10\% of the most different structures, in terms of SOAP representation, within the full set of LiPS roto-translations along the substrate, by means of a furthest point sampling (FPS) algorithm \cite{Imbalzano2018}.

        \item We calculate from single-point DFT calculations the binding energy corresponding to each of these structures.
    \end{enumerate}
    
    \item ML training: we train a Kernel Ridge regressor \cite{KRR} to predict the binding energies of selected LiPS--SAC configuration, based on its SOAP representation. 
    We utilise the regressor to map predicted binding energies in the full set of roto-translations  generated in the initial step of this protocol.
    
    \item Surrogate Generalised Convex Hull (GCH) construction: We build a GCH which utilises the first principal component analysis (PCA)\cite{Pearson1901} component projection of the SOAP features assigned to each LiPS--SAC as the free thermodynamic variable, and the predicted binding energy as the energetic coordinate.
    
    \item Minimise GCH Vertexes: We perform DFT geometry optimisation to the structures which lie at vertexes of the GCH. We make this choice under the heuristic hypothesis that vertexes of the GCH are the most likely to relax into dissimilar low-energy structures.\cite{Anelli2018}
\end{enumerate}

\subsection{Materials} \label{sec:Materials}

\begin{figure}[t]
    \centering
    \begin{flushleft} \hspace{0.7cm} \textbf{A} \end{flushleft} 
    {\includegraphics[width=0.8\linewidth]{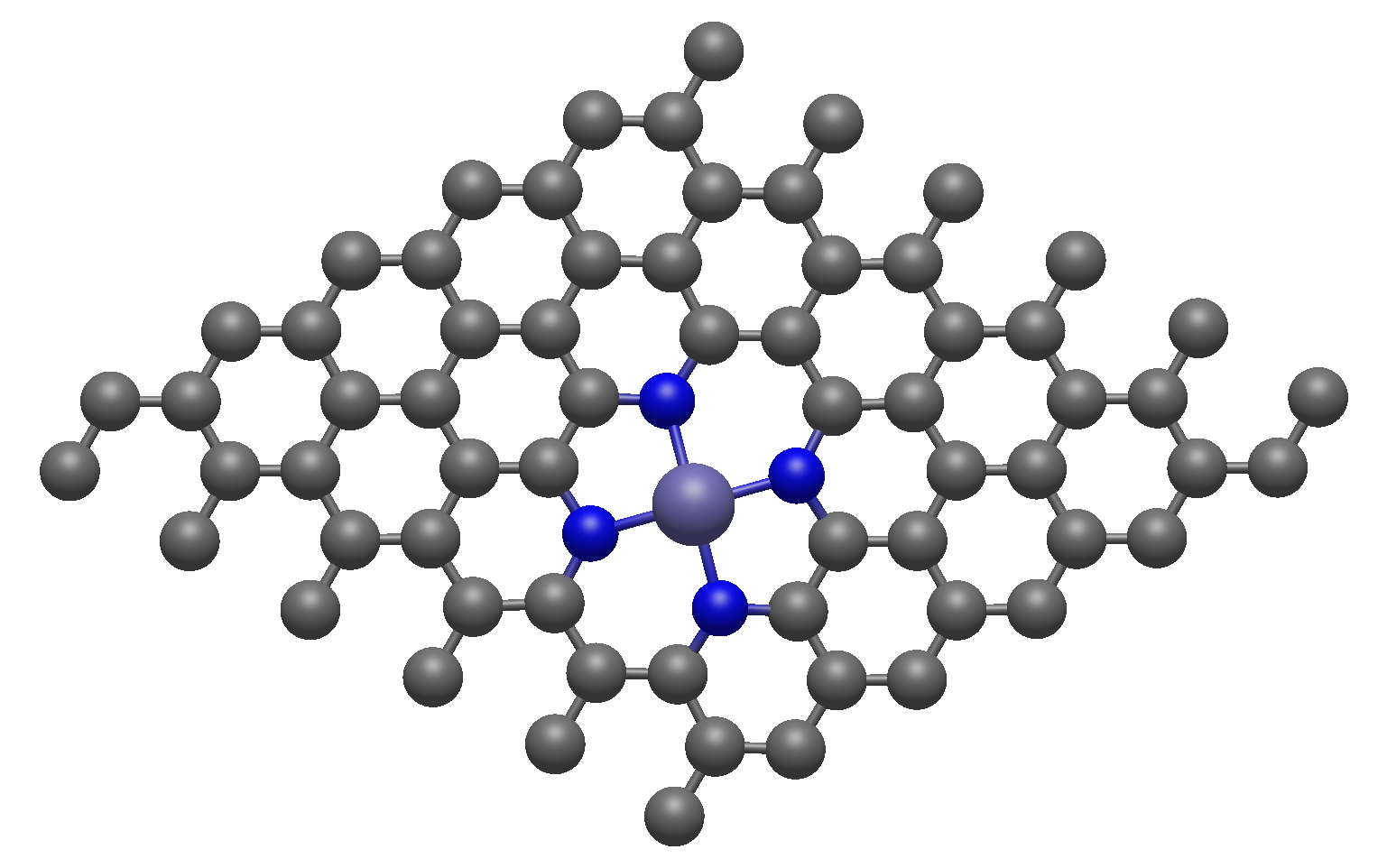}} \\
    \begin{flushleft} \hspace{0.9cm} \textbf{B} \end{flushleft} 
    \includegraphics[width=0.7\linewidth, trim=0cm 0cm 0cm 14cm,clip]{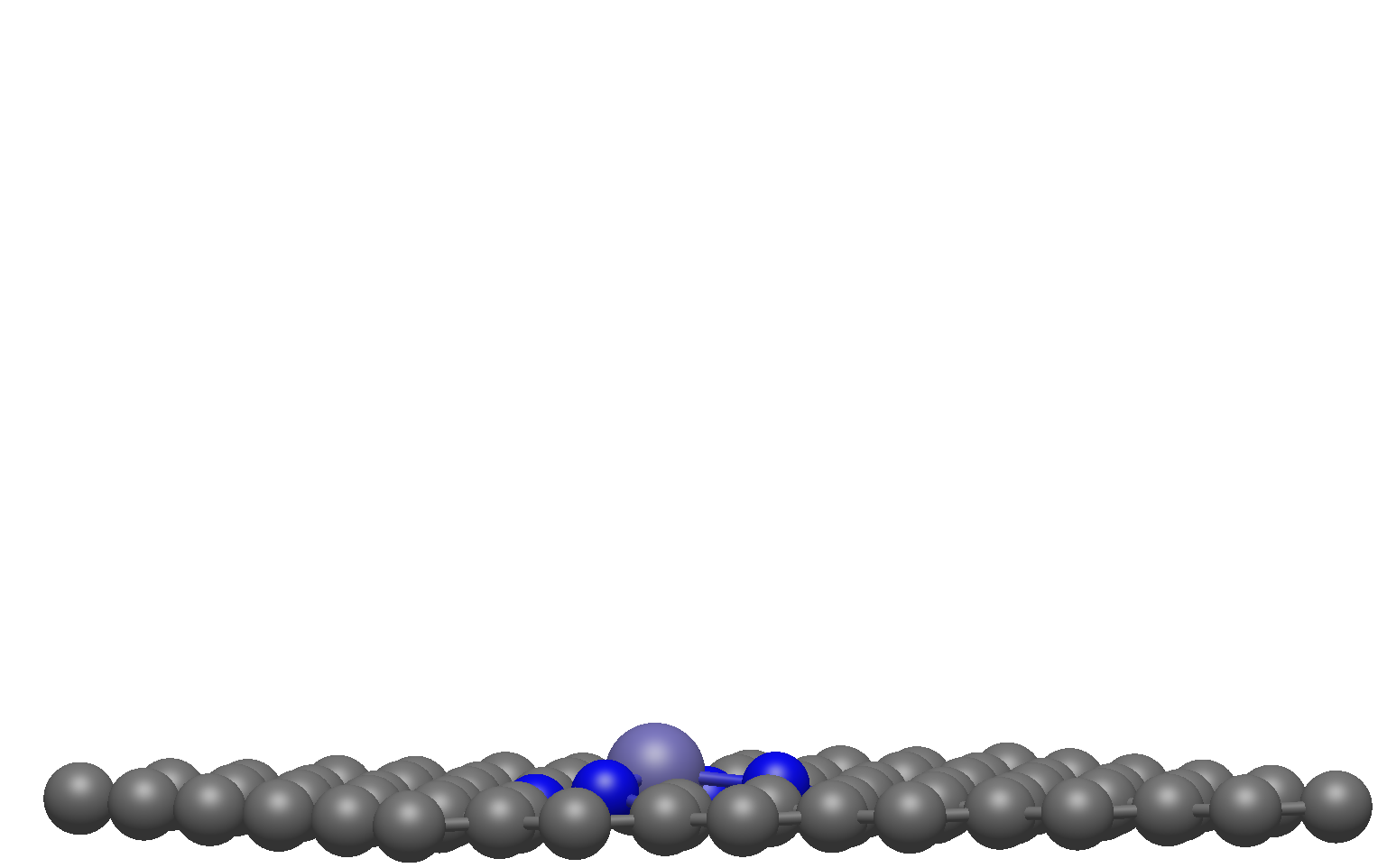} \\      
    \begin{flushleft} \hspace{0.9cm} \textbf{C} \end{flushleft} 
    \includegraphics[width=0.25\linewidth, trim=0cm 0cm 9cm 10cm,clip]{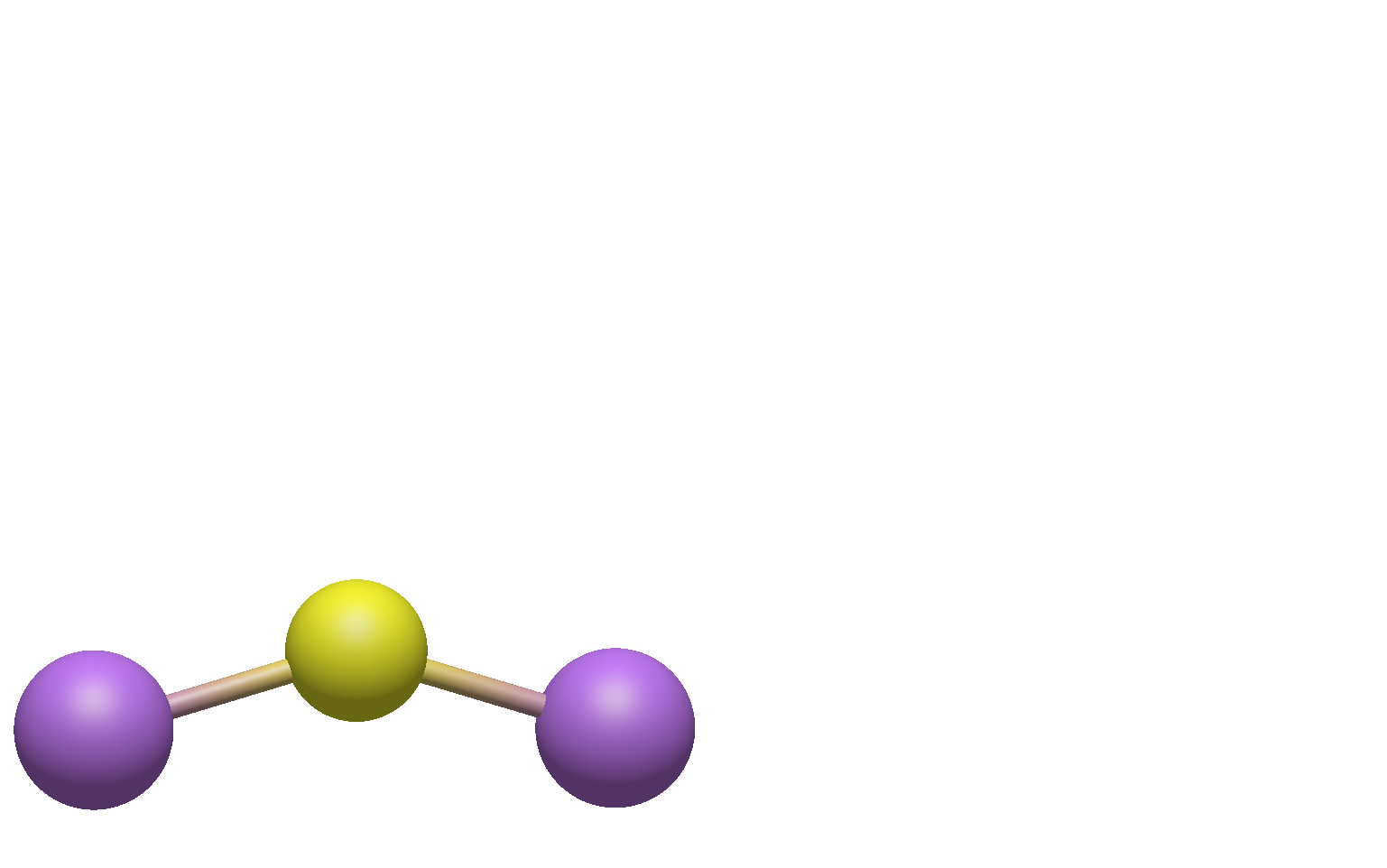}
    \includegraphics[width=0.25\linewidth, trim=0cm 0cm 9cm 10cm,clip]{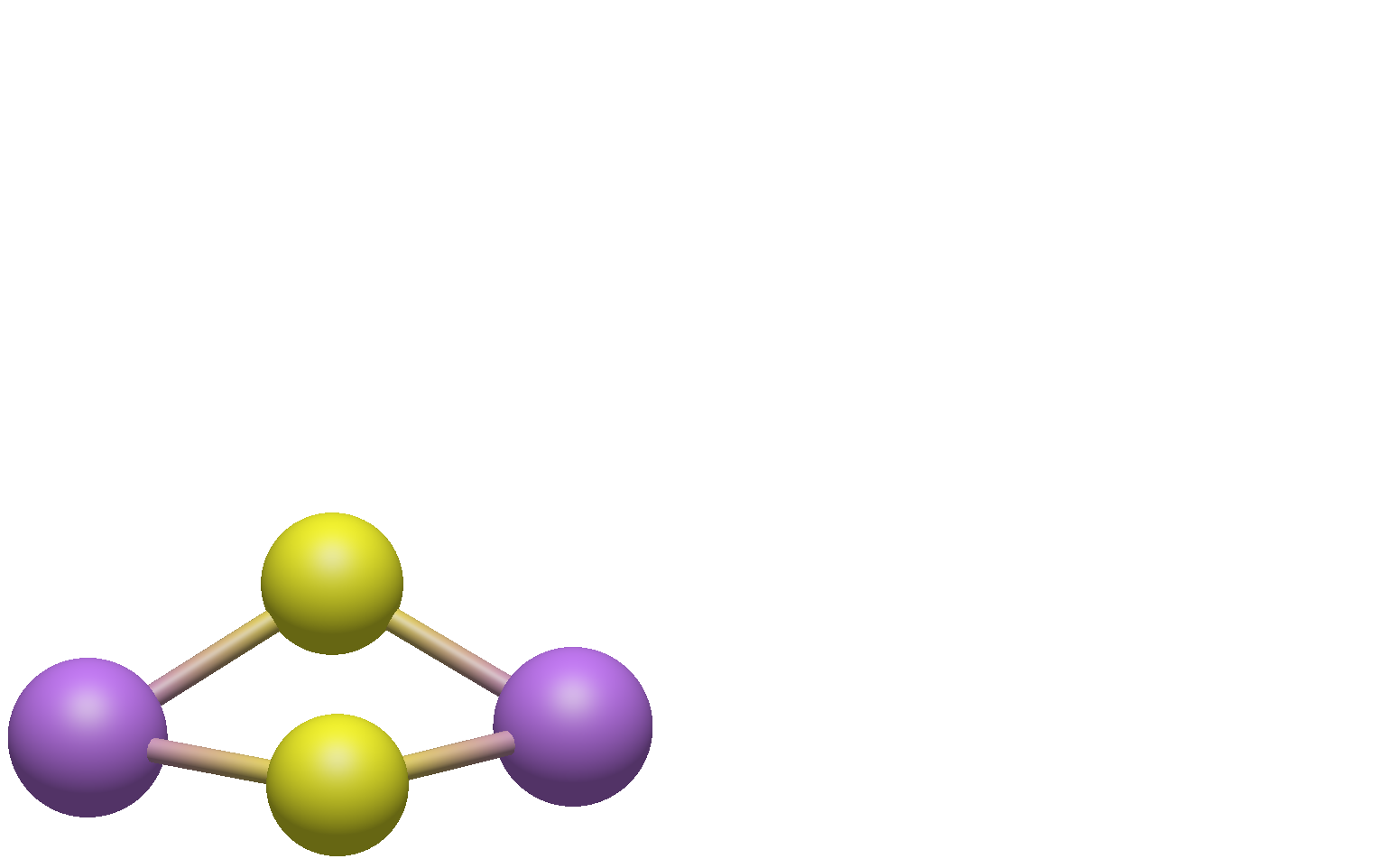} \\ 
    \includegraphics[width=0.25\linewidth, trim=0cm 0cm 9cm 6cm,clip]{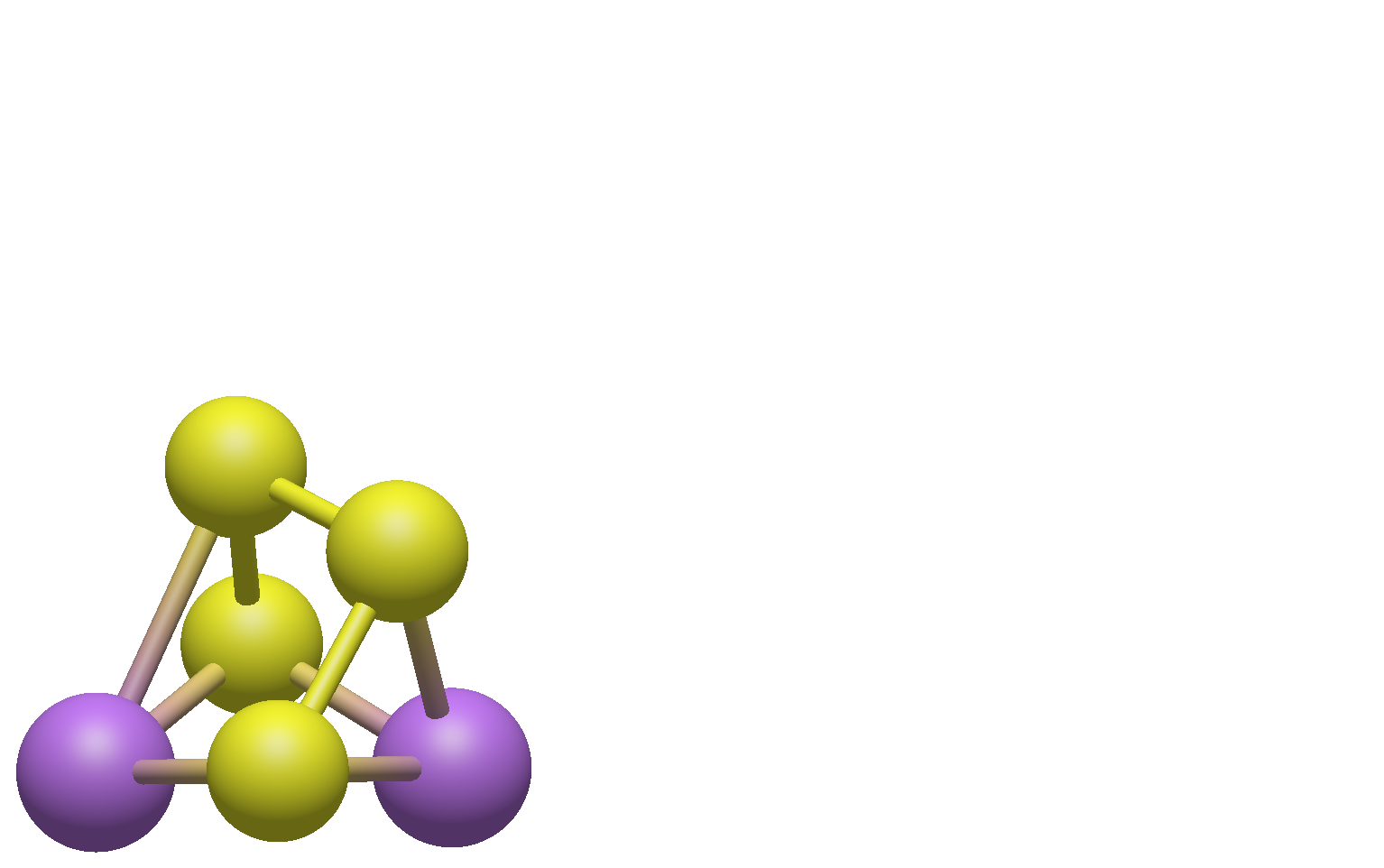} \\
    \includegraphics[width=0.25\linewidth, trim=0cm 0cm 9cm 0cm,clip]{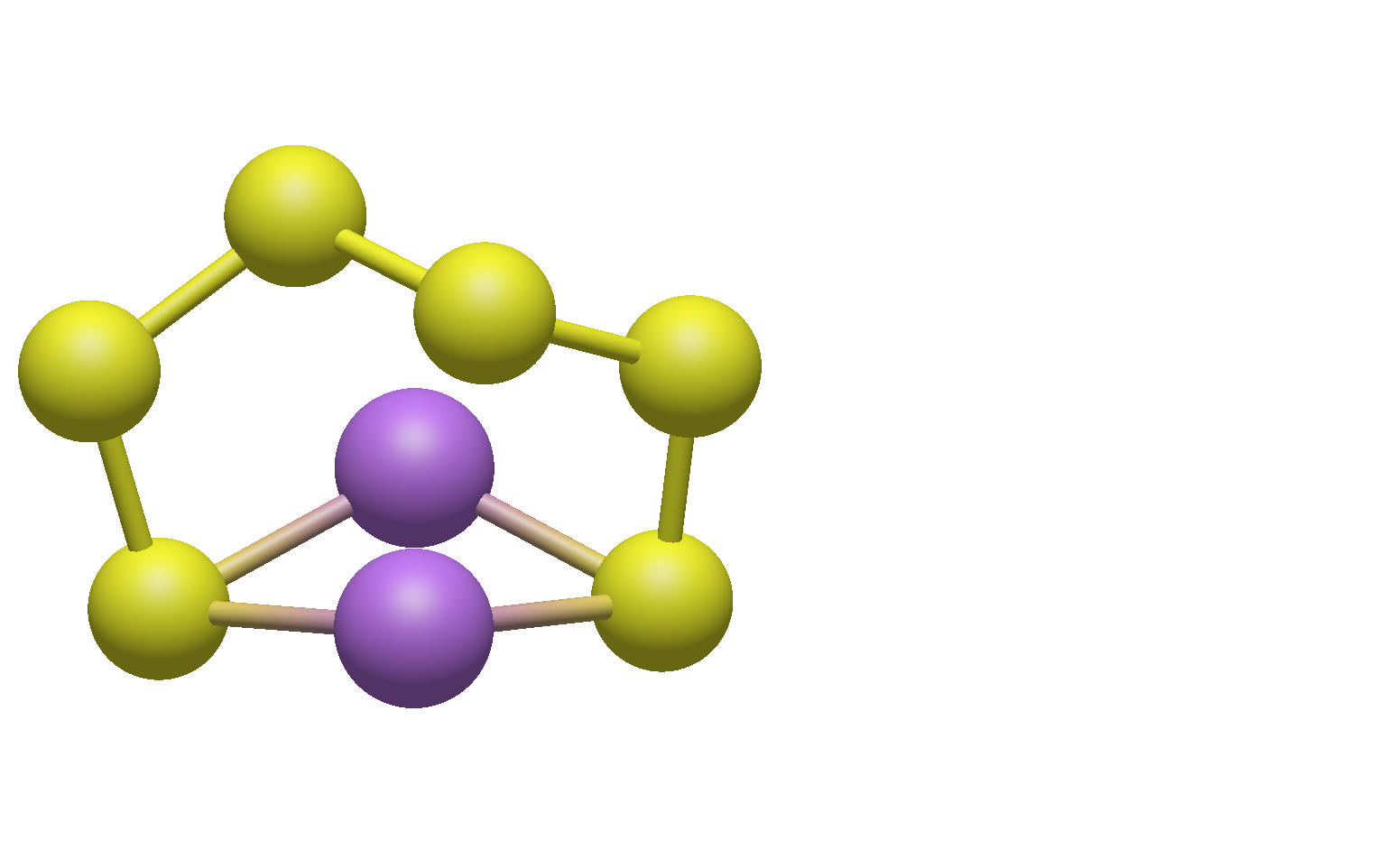}
    \includegraphics[width=0.25\linewidth, trim=0cm 0cm 9cm 0cm,clip]{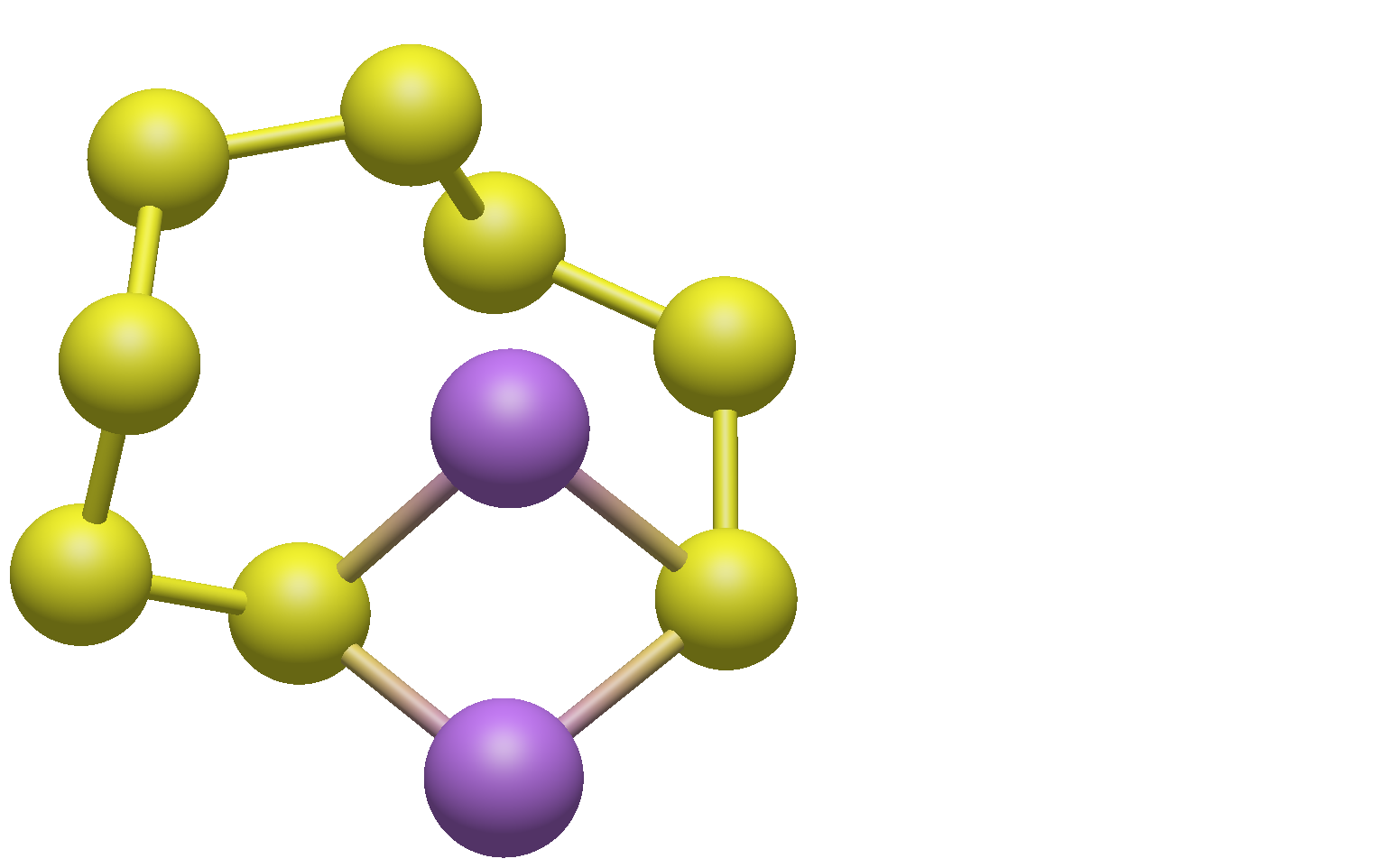}
    \caption{ Top (A) and side (B) views of \TMNC \ structure. 
    Geometry optimised Li$_2$S$_n$ ($n = 1,2, 4, 6, 8$) structures.
    Grey, blue, purple, magenta and yellow denote the C, N, TM, Li and S elements, respectively (C).}
    \label{fig:structures}
\end{figure}

In this study, we examine two SAC materials with \TMNC \ formation (TM = Fe, Zn) and five LiPS intermediates (Li$_2$S$_n$, \textit{n} = 1, 2, 4, 6, 8).
The \TMNC \ configurations, based on a modified 72-atom relaxed pristine graphene lattice, consist of 66 C atoms (92.96 at\%), four substitutional N atoms (5.63 at\%) and one interstitial TM atom (1.41 at\%).
The TM and two N atoms were manually shifted \textit{out of plane} by 0.5~\AA~ and 0.3~\AA, respectively, in order to realistically represent the previously observed geometry in SAC materials, when LiPSs are present \cite{Zeng2019_FeSAC, Zhou2020_SAC}.
The top and side views of the \TMNC, and the relaxed LiPS structures are shown in Figure \ref{fig:structures}.

\subsection{DFT set up} \label{sec:dftsetup}

We perform all spin-polarised DFT simulations with the CASTEP code \cite{CASTEP}. 
We calculate the exchange and correlation potential using the generalised gradient approximation (GGA) with the Perdew--Burke--Ernzerhof (PBE) functional, using plane-wave cut-off energy of 500 eV. 
We use energy and force convergence criteria of 10$^{-5}$ eV and $5\times10^{-5}$ eV/\AA~ respectively for the self-consistent field (SCF) for single-point and geometry optimisation simulations, and 10$^{-5}$ eV and 10$^{-3}$ eV/\AA~ respectively, for the geometry optimisation simulations.
The sampling of the Brillouin zone is carried on a Monkhorst-Pack grid, using a $3 \times 3 \times 1$ \textbf{k}-point mesh.
We include Van der Waals dispersion corrections as described in Grimme's empirical method. \cite{Grimme2006}
We consider periodic boundaries conditions in the simulation box and a vacuum separation distance of 18 \AA~ at the direction normal to the substrate's surface to avoid replica interaction.

We define the binding energy ($E_{\mathrm{b}}$) as:
\begin{equation}
E_{\mathrm{b}} = E_{\mathrm{mat}} - (E_{\mathrm{SAC}} + E_{\mathrm{LiPS}})
\end{equation}
where $E_{\mathrm{SAC}}$, $E_{\mathrm{LiPS}}$ and $E_{\mathrm{mat}}$ are the energies of the substrate, LiPS and substrate and LiPS together, respectively.
For reference, we report the total energies for the SACs and LiPSs as: 
$E_{\mathrm{tot}}^{\mathrm{Fe-N_4-C}} = -12383.41$ eV, $E_{\mathrm{tot}}^{\mathrm{Zn-N_4-C}} = -13307.78$ eV, $E_{\mathrm{tot}}^{\mathrm{Li_2S}} = -708.74$ eV, $E_{\mathrm{tot}}^{\mathrm{Li_2S_2}} = -1013.31$ eV, $E_{\mathrm{tot}}^{\mathrm{Li_2S_4}} = -1621.69$ eV, $E_{\mathrm{tot}}^{\mathrm{Li_2S_6}} = -2229.35$ eV, $E_{\mathrm{tot}}^{\mathrm{Li_2S_8}} = -2836.12$ eV.

\subsection{SOAP Representation} \label{SOAP}

Each configuration consisting of a LiPS and the SAC is described by means of $D$ features deriving from the Smooth Overlap of Atomic Orbitals representation \cite{SOAP}.
This representation is smooth, invariant to permutations, rotations, and translations of the system.


Within the SOAP representation, structures are encoded by means Gaussian smeared atomic densities expanded via orthonormal functions, based on spherical harmonics and radial basis functions, centred at atoms positions.
Operatively, the SOAP power spectrum is calculated as:
\begin{equation}
    {\bf p}_{n,n',l}^{Z1,Z2} = \pi \sqrt{\frac{8}{2l+1}} \sum_{m}{c_{n,l,m}^{Z_{1}} * c_{n',l,m}^{Z_{2}} }
\label{eq:1}
\end{equation}
where $n$ and $n'$ are the indices for the radial basis functions (up to $n_{max}$, $l$ labels the degree of the spherical harmonics (up to $l_{max}$) and $Z_{1}$ and $Z_{2}$   is a label associated to the atom species.
The coefficients $c_{nlmZ}$ in equation \ref{eq:1} are instead defined as (inner products):
\begin{equation}
    c_{n,l,m}^{Z} = \int \int \int_{\mathcal{R}} dV g_{n}(r) Y_{l,m} (\theta, \Phi) \rho^{Z} (r)
\end{equation}
where $\rho_{Z}(r)$, is the Gaussian smoothed atomic density for atoms with atomic number $Z$ defined as $Y_{l,m} (\theta, \Phi) $ are the real spherical harmonics, and $g_{n}(r)$ is the radial basis function (spherical Gaussian type orbitals here).

To calculate the SOAP power spectra we adopt the DSCRIBE \cite{dscribe} library, and utilise the following parameters: $n_{max}$~=~6 , $l_{max}$~=~7 , and $r_{cut}$~=~6.5$\AA$ .
After calculating the power spectrum associated to each atomic sites in a structure, we evaluate an average power spectrum which is associated to the atomic configuration. 

We utilise the elements in the averaged $\bf{p}$ power specture as the features in kernel ridge regression (KRR).
SOAP power spectrum coefficients,\cite{de2016,Bartok2018,Deringer2020,Deringer2021} and local-density expansion coefficients more in general,\cite{Zeni2020,Zeni2021,Lysogorskiy2021,kovacs2021}  have been largely successful features in kernel-based and linear machine learning models for structure classification and energy regression




\subsection{Kernel Ridge Regression} \label{sec:KRR}

KRR is a well-known and popular non-parametric regression approach. 
In particular, it consists of L2-norm regularised least squares fitting which exploits the kernel trick \cite{KRR},
where the latter allows for an efficient framework to transform and compare data in an high number dimensions.
Within the framework of kernel methods, data are represented by a matrix of pairwise similarity where each $(i,j)$ entry is defined by the kernel function $k(x_{i},x_{j})$. 
The latter returns the dot product of the features vectors associated to each $i$ and $j$ data point in the projected space, for any mapping $\phi: x \rightarrow R^{N}$.
The kernel trick thus allows for operations in a high dimensional space without the explicit need of calculating $\phi(x)$.
Given a set of covariate variables $x_i$, the KRR prediction $f(x_i)$ is written as a linear combination \cite{rasmussen_2006_gpr}:

\begin{equation}
f(x_i) = \sum_{i = 1}^{N} {\alpha}_{i}   ~ k(x_{i},x)
\end{equation}

and the vector of $\alpha_{i}$ weights is found by solving:

\begin{equation}
\boldsymbol{\alpha}  = (\mathbf{K} + \lambda \mathbf{I})^{-1} \cdot \mathbf{y}, 
\end{equation}

where $\mathbf{K}$ is the kernel matrix, $\lambda$ tunes the regularisation strength and $\mathbf{y}$ labels response variables in the training set.
In the reported case study we choose 
$\lambda$ to 0.025 and a radial basis function kernel to evaluate the ${K}_{i,j}$ elements of the $\mathbf{K}$ matrix.

\subsection{Generalized Convex Hull} \label{GCH}

The generalised convex hull (GCH) allows to evaluate the probabilities of structures being stabilised by general thermodynamic constraints. 
%
In a nutshell, the convex hull of a finite number of points is defined as the smallest convex set that contains them.

In materials science, the convex hull construction is used to identify structures that are stable.
Given a set of energies, as a function of one or more thermodynamic variables, only the configurations which lie at the vertexes of the convex hull which are lower in energy will be stable. 
If a structure does not lie on one of such vertexes it is unstable with respect to decomposition into two or more parent structures (assuming fixed thermodynamic conditions and disregarding possible stabilization due to kinetic effects).

The generalized convex hull construction adopts a geometric fingerprints $\boldsymbol{\Phi} = \{ \Phi_i \}$, to encode the full structural diversity of a dataset of configurations, rather than one or more explicit thermodynamic variables to predict stable phases.
The GCH framework has been successfully used in the past for predicting previously unseen ice phases \cite{Engel2018} and molecular crystals polymorphs \cite{Anelli2018}.
When moving from the bulk phase to a finite-size system as the one here considered, the assumption that unstable phases will decompose into two or more more stable phases is lost.
Nevertheless this elegant construction provides a rigorous framework to account on the same footing for both geometrically diversity and predicted energetic stability in the choice of diverse candidates for relaxation at DFT level.

\section{Workflow validation for {Fe}--N$_4$--C SAC}

\subsection{Li$_{2}$S on Fe--N$_4$--C}

\begin{figure}
    \centering
    \includegraphics[width=6cm]{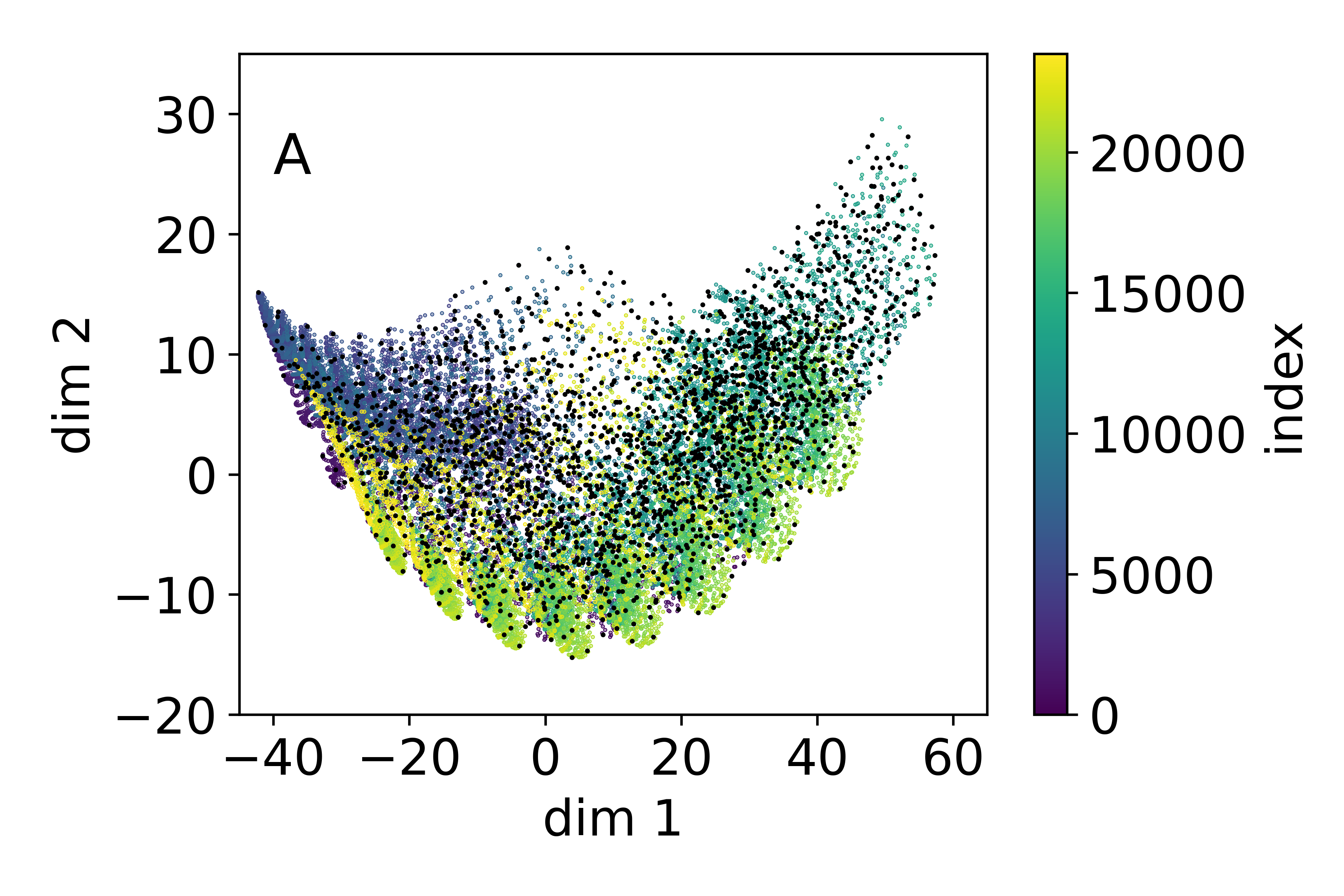}
    \includegraphics[width=6.5cm]{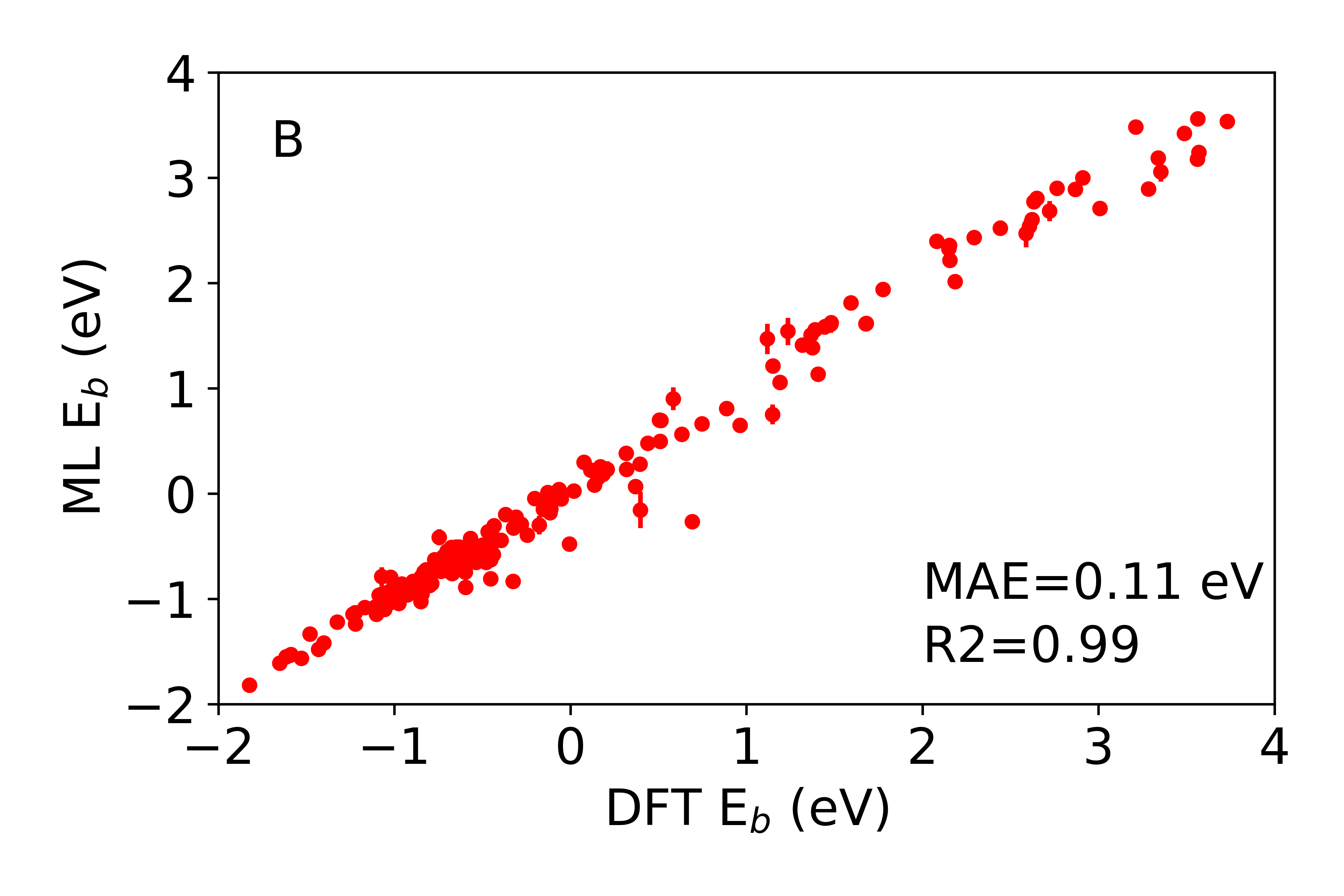}
    \includegraphics[width=6.5cm]{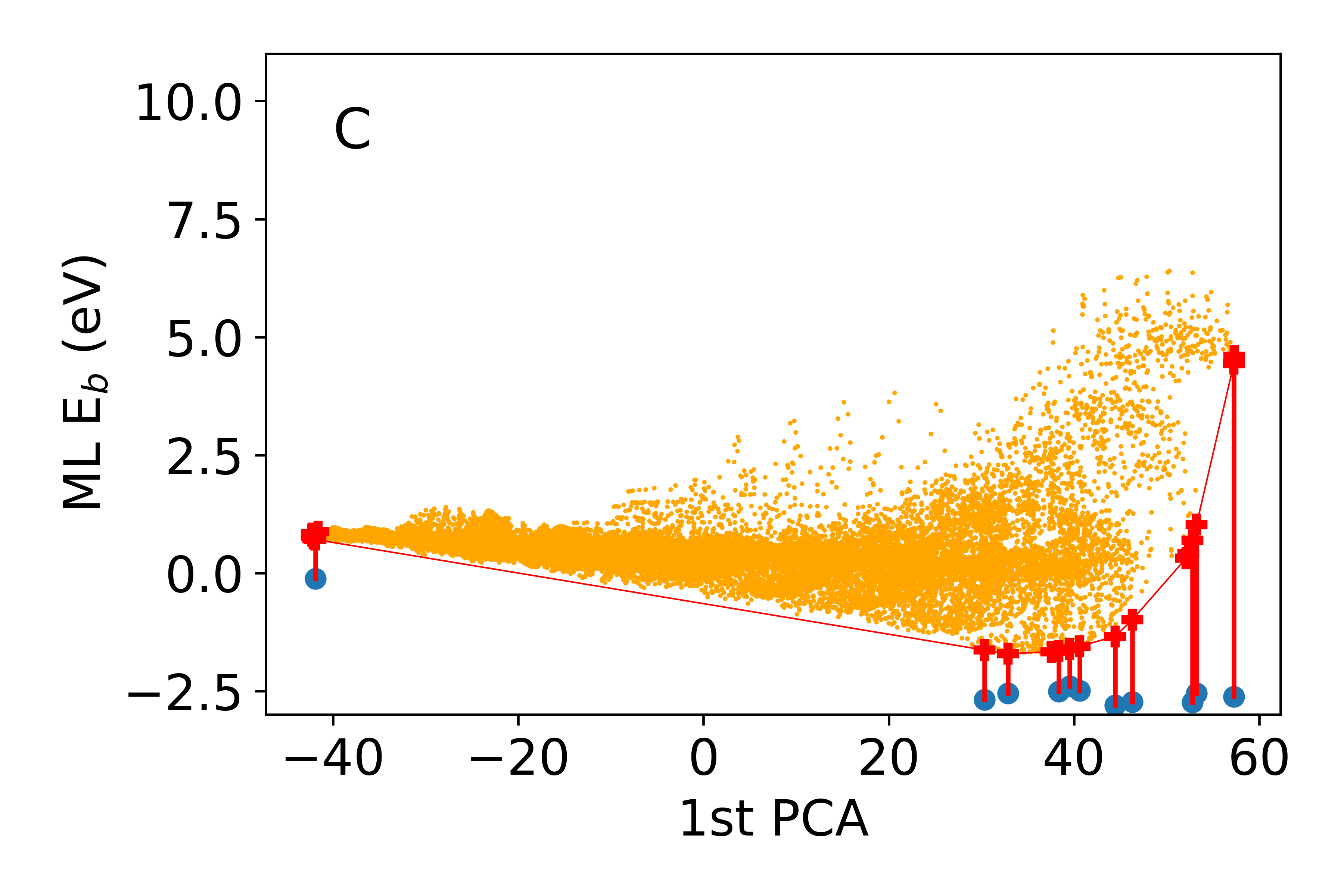}
    \caption{A) Visualisation of the SOAP features for initial full set of $\sim$ 23000 structure generated for \lst~ on Fe-based SAC, when projected in the space determined by their first two PCA components (adimensional). Points in the map are colour coded according to their structure index, while the $\sim$ 2000 structures selected by the FPS procedure are shown in black. Note their heterogeneous distribution in the PCA space, which also comprises outliers.
    B) Parity plot between the ML-predicted and DFT-calculated binding energies for the training, testing, and validation structures. Mean Absolute Error (MAE) and R2 score are reported for reference.
    C) Visualisation of the GCH construction when utilising the first SOAP feature's PCA component as the thermodynamic variable and binding energy as the energetic variables. Vertexes are highlighted with red crosses. The binding energy of structures found by relaxing each GCH vertex is shown in blue. A thick line acts as a guide to the eye in connecting each GCH vertex energy to its relaxed configuration binding energy.}   
    \label{fig:wflow2}
\end{figure} 

To showcase in a detailed manner the application of the workflow we discuss the case of Li$_{2}$ binding on a Fe-based SAC. 
Figure \ref{fig:wflow2}A shows a 2D map where the two axes are the first and second principal components of the SOAP features, associated to all structures generated for this system.
Each point in the map corresponds to one of those structures. 
The colour scale refers to the order with which structures have been generated during the systematic enumeration of LiPS roto-translations.
Points in black corresponds to structures selected via an heuristic FPS scheme where on the SOAP features for each structure and an Euclidean metric where considered to evaluate the distance between different structures.

We repeat the same process for the other LiPS molecules and run DFT-based single point calculation on the first $\sim10\%$ most different structures (i.e., circa 2720 out of 27,200 structures per each LiPS), and utilise the $\sim$80\% of that data as the training dataset in the Kernel Ridge regression fitting.
Features associated to each data point are normalised by removing the mean and scaling to unit variance before performing the KRR training.
$\sim$18\% of the data are used as the training test.
The obtained ML model exhibits a Mean Absolute Error (MAE) of 0.10~eV for training and testing, which is well within the acceptable error limits for this type of calculations.
Same MAE values are also obtained when performing an 8-fold cross-validation test \cite{Stone1974}.
To further evaluate our model's accuracy, we select the remaining $\sim$2\% most different structures ($\sim$50 structures per each LiPS) as our test set and calculate the binding energy from single point DFT calculations.
The parity plot between the ML predicted and DFT true binding energies for this test is illustrated in Figure \ref{fig:wflow2}B.
A MAE of 0.11 eV is registered for the test set, along with a very good correlation (R2 = 0.99) between the single-point predicted binding energy and the ground truth energy.

We utilise the KRR regressor to evaluate the binding energies for the whole set of Li$_2$S 27,216 structures, generated at the first stage of our minima exploration protocol.
We then map them on a Energy - First SOAP PCA dimension map, Figure \ref{fig:wflow2}C, where each LiPS roto-translated configuration corresponds to one orange dot.
The thin red edges and the 11 bold red crosses respectively identify the edges and vertexes of the convex hull construction associated to this dataset.
While a number of CH vertexes appear at very negative first PCA component values, we consider only one because of their small structural and energetic diversity.
The blue dots label the DFT-level Binding Energy found when relaxing each of the structure lying at a CH vertexes. 
We then analyse the intermolecular distances in the LiPS molecule and the intramolecular one between the LiPS and the Fe and N atoms in the support, and
that all lie in a distinct local minima.

\subsection{LiPSs on Fe--N$_4$--C}

\begin{figure*}[t]
    \centering
    \includegraphics[width=7cm]{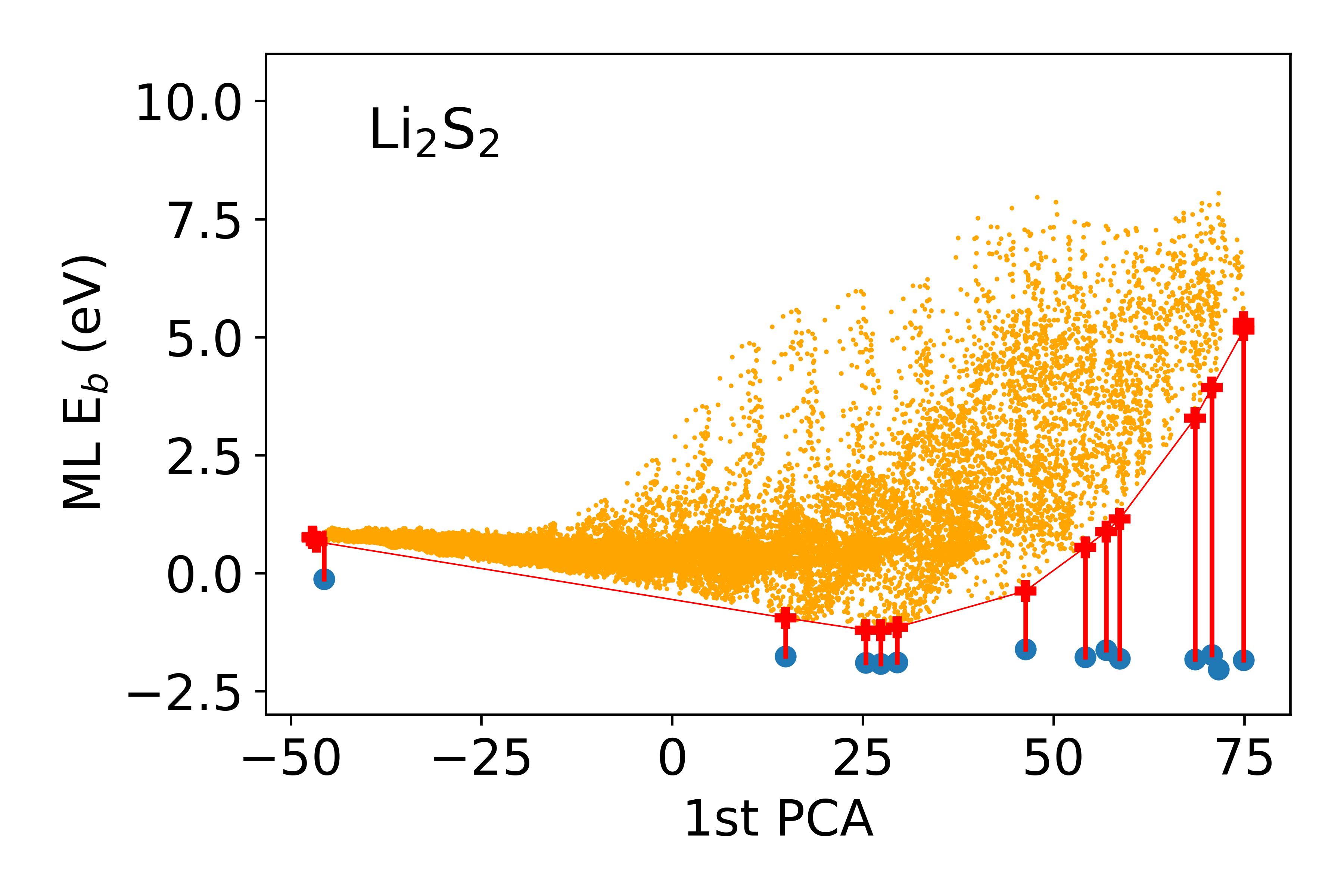}
    \includegraphics[width=7cm]{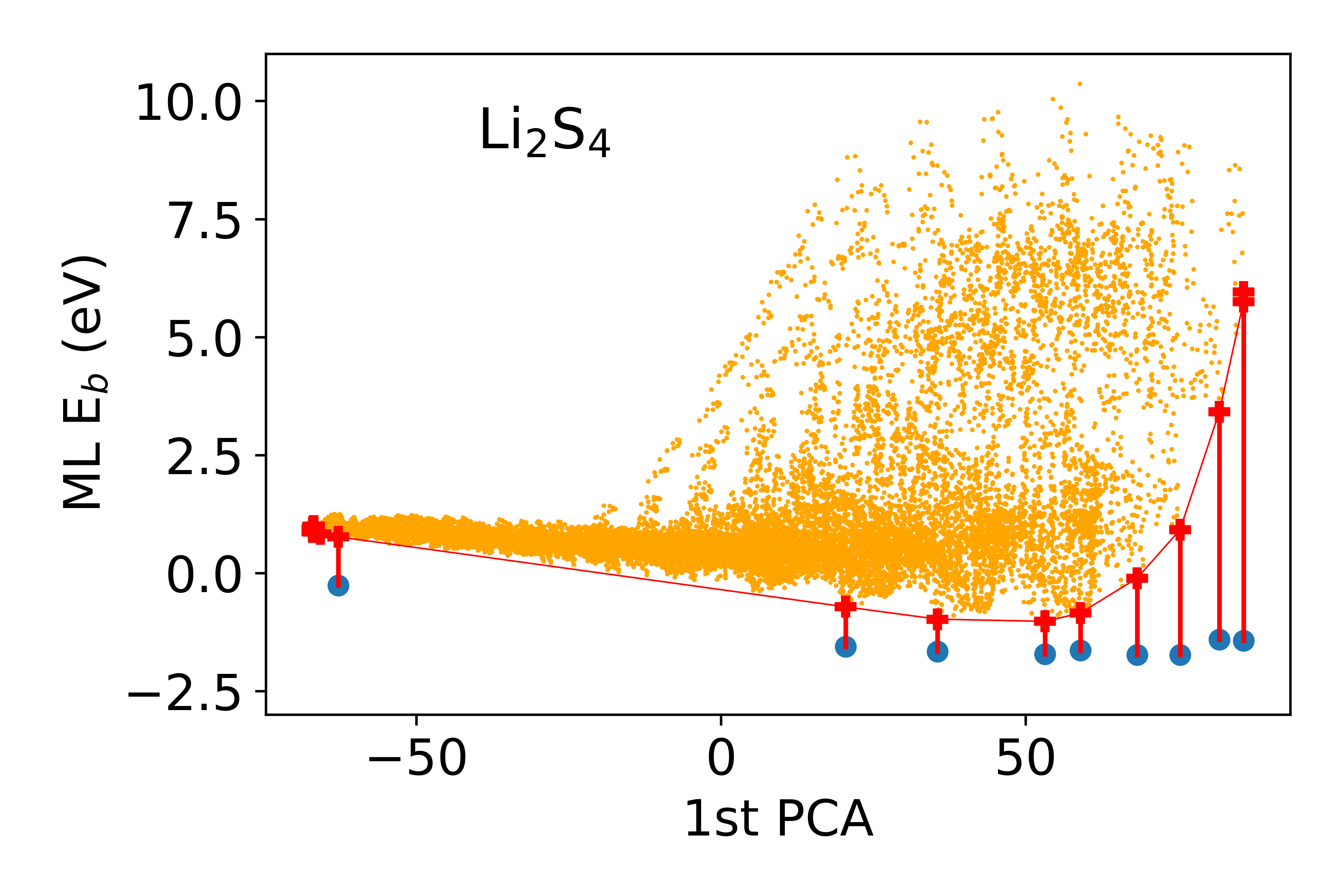}
    \includegraphics[width=7cm]{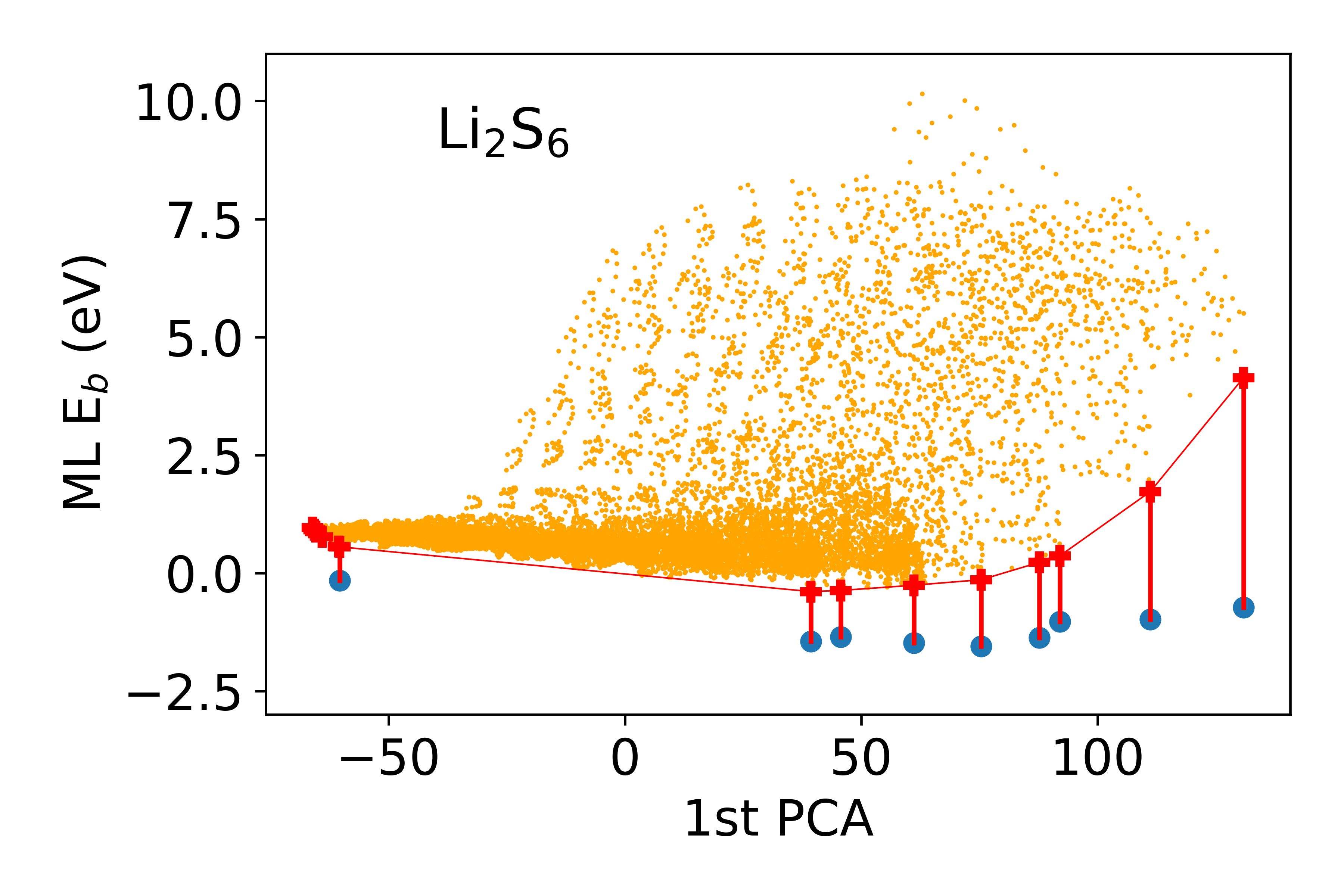}
    \includegraphics[width=7cm]{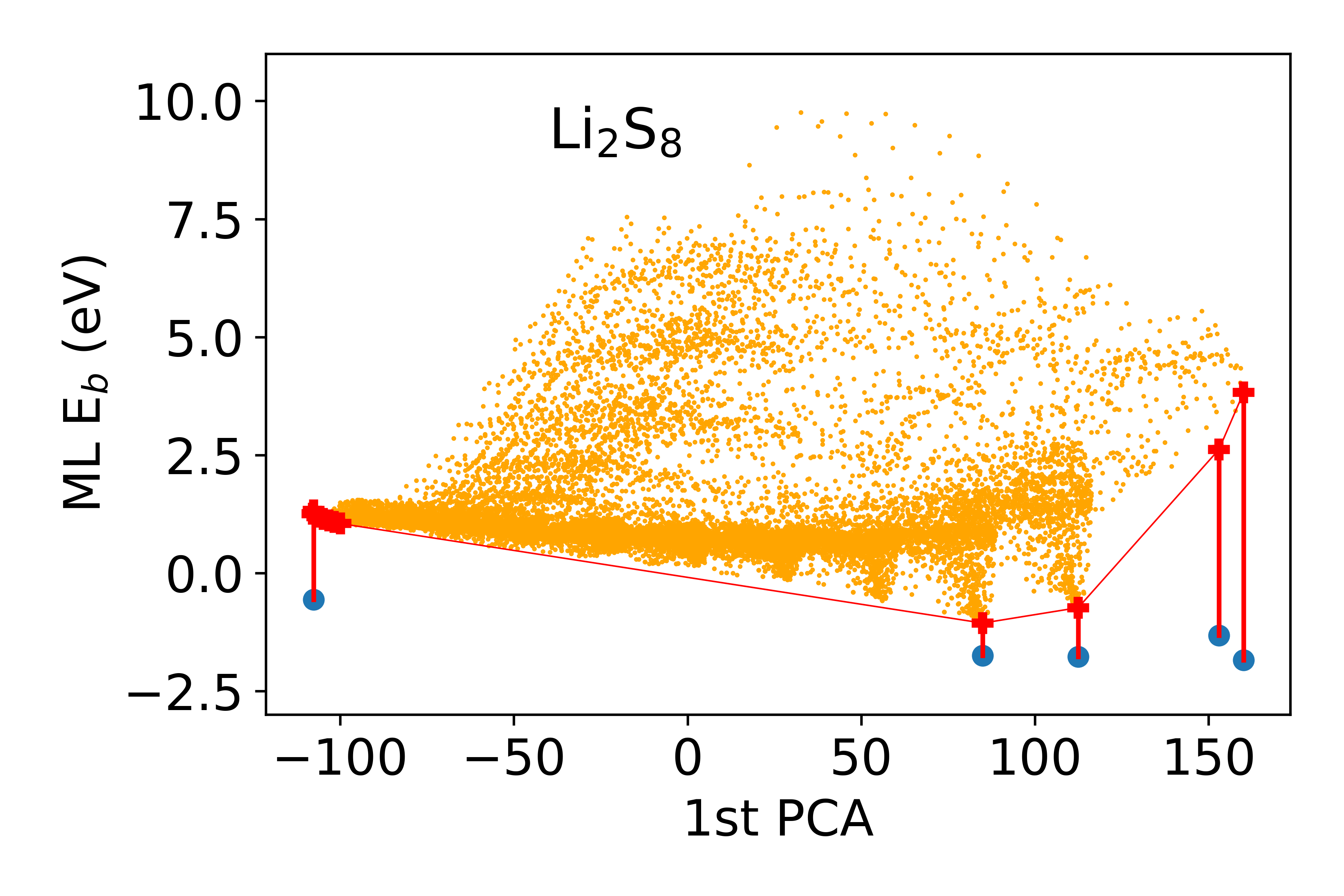}
    \caption{The surrogate GCH construction for Li$_2$S$_n$ ($n$ = 1, 4, 6, 8) structures adsorbed on \FeNC~ is illustrated with orange points. GCH vertexes are shown with red crosses. Blue dots show the binding Energy corresponding to structures sampled after relaxing the GCH vertexes.}
    \label{fig:gch2}
\end{figure*}

We follow the protocol described for Li$_{2}$S$_{2}$ to calculate the binding energy for the rest of the examined LiPSs (Li$_2$S$\mathrm{_n, n = 1, 4, 6, 8}$) when interacting with \FeNC.
Figure \ref{fig:gch2} illustrates the GCH construction resulting from the SOAP-KRR model.
Each LiPS configuration is reported as an orange dot, the GCH vertexes are highlighted with red crosses, and the edge connecting to vertexes is visualised with a red line.
The blue dots in Figure \ref{fig:gch2} instead report the DFT-level Binding Energy found when relaxing each GCH vertex.
For the four examined cases, the LiPS isomers found to give the lowest binding energies after geometry optimisation are GCH vertex with positive 1st PCA component and low single-point binding energy (lower right-hand side region of the plots).

First, we compare the binding energies found with our protocol with respect to those reported in the literature, and gather them in Table \ref{tab:eref}.
The reported binding energies are in good agreement with reported values from the literature, showing similar or stronger binding, but for one case were the binding energy is understimated by 0.1eV.
This highlights the predictive power of the framework under investigation, which is generally able to identify structures offering the strongest binding. 
We note that when we include higher energy Li$_{2}$S$_{2}$ CH vertexes at the relaxation stage, we identify a configuration with binding energy of -2.04eV, in line with what reported in the literature.
In this regards, the binding energy change trend among the various LiPSs is also in good agreement with our previous study \cite{Andritsos2021}, which adopts the exact same DFT calculation set up.

\begin{figure}[t]
    \centering
    \text{Top view} \hspace{2.5cm} \text{Side view} \\
    \includegraphics[width=0.4\linewidth]{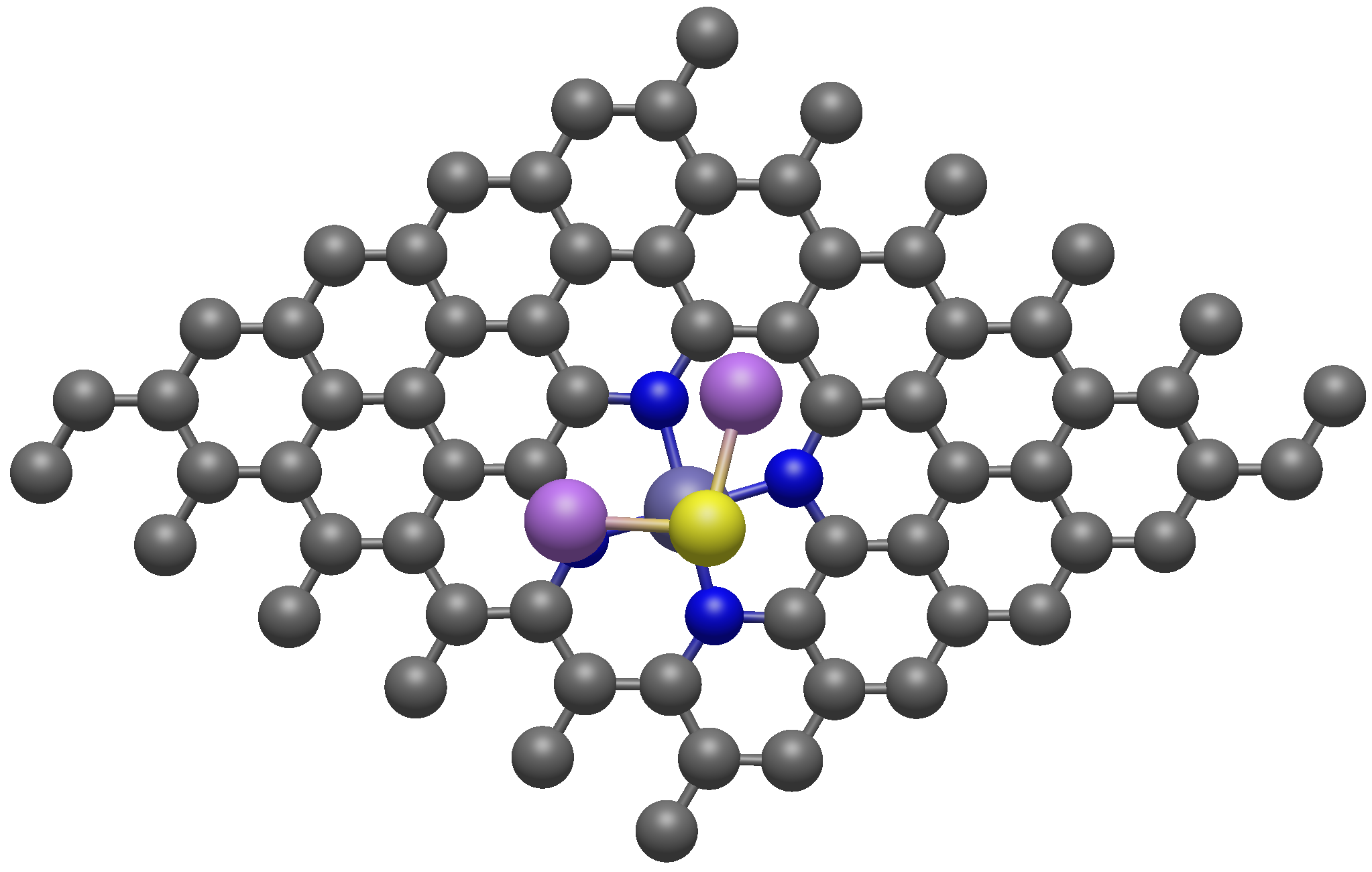}
    \includegraphics[width=0.4\linewidth]{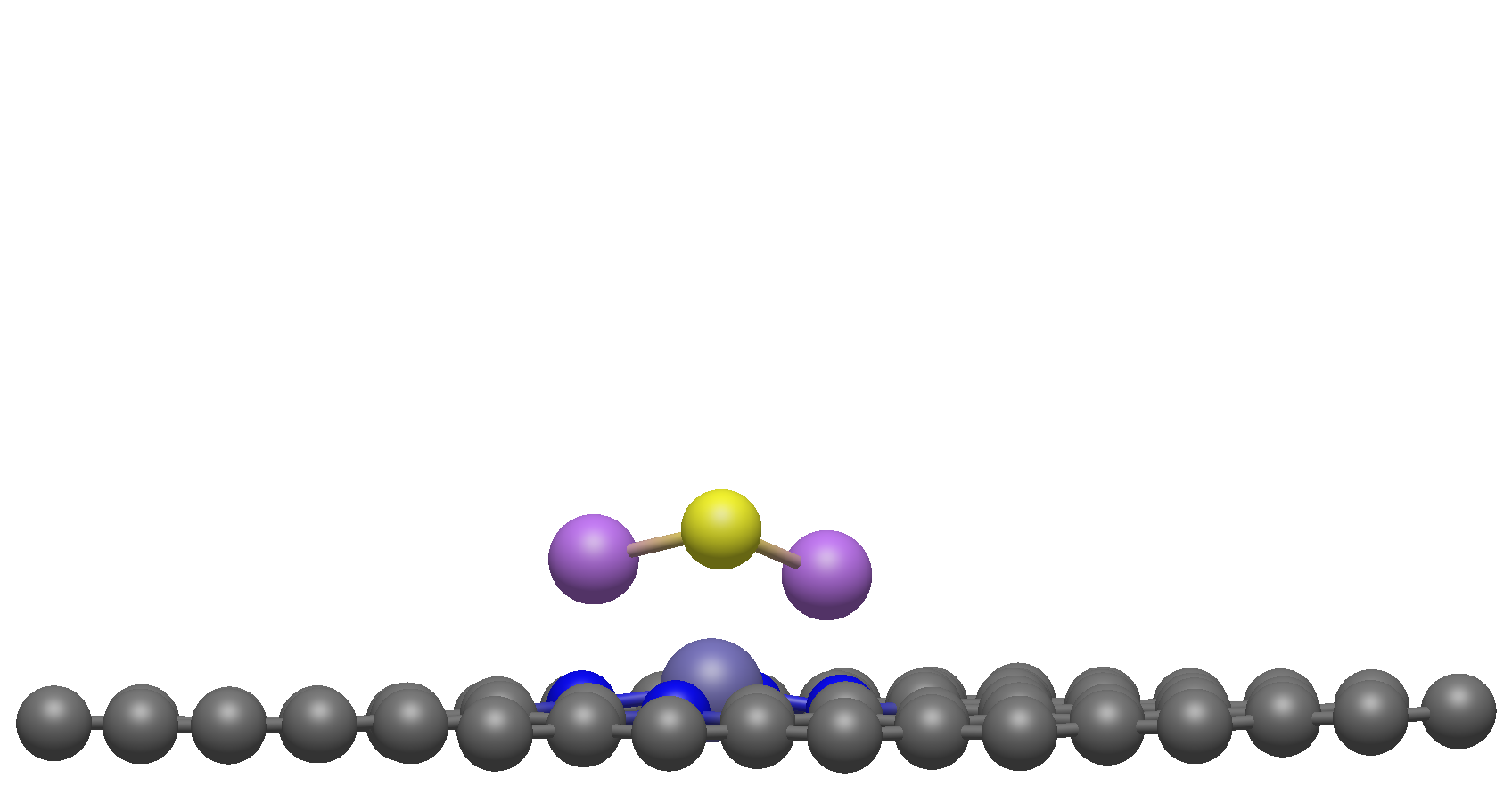} \\
    \includegraphics[width=0.4\linewidth]{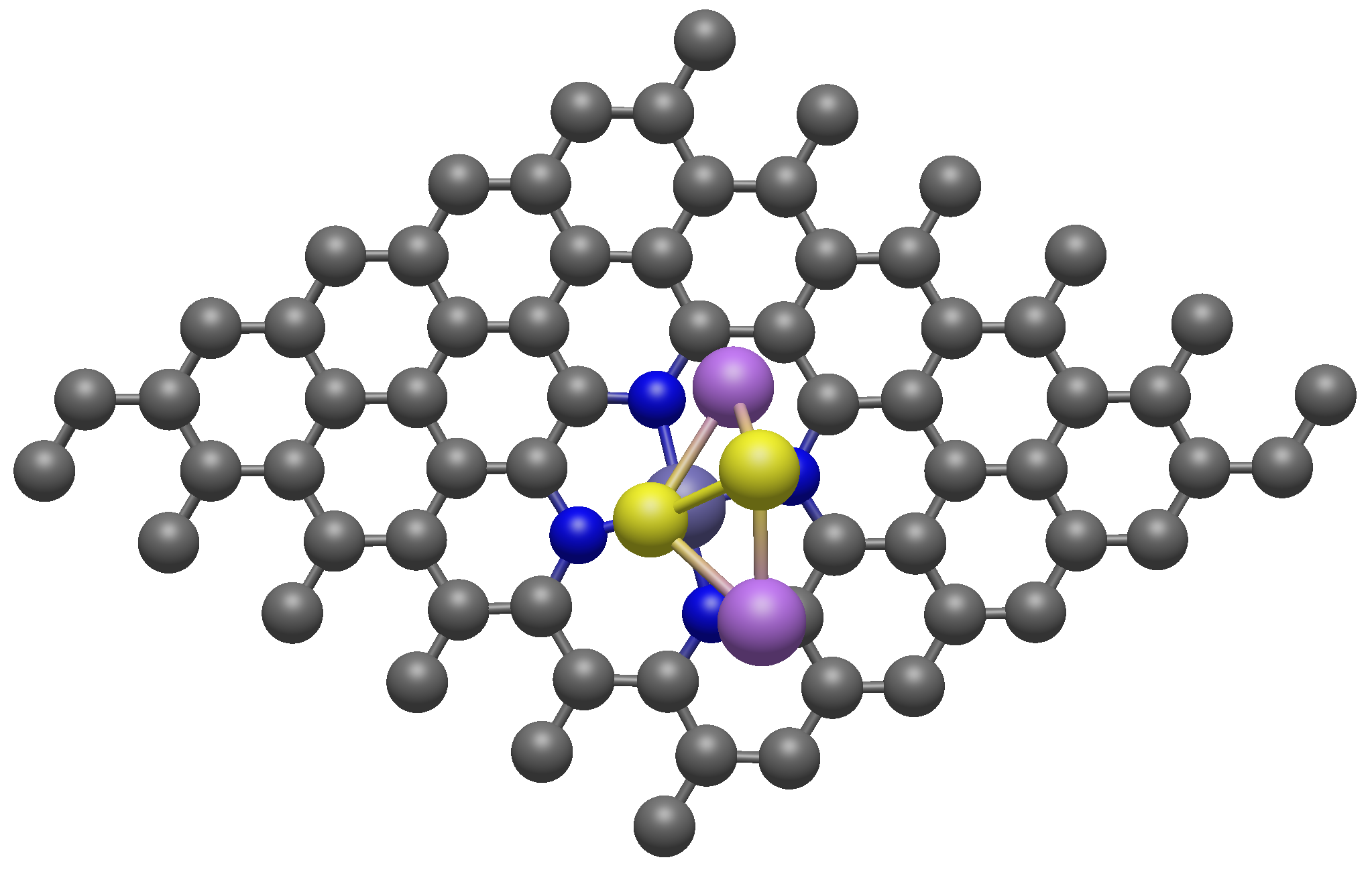}
    \includegraphics[width=0.4\linewidth]{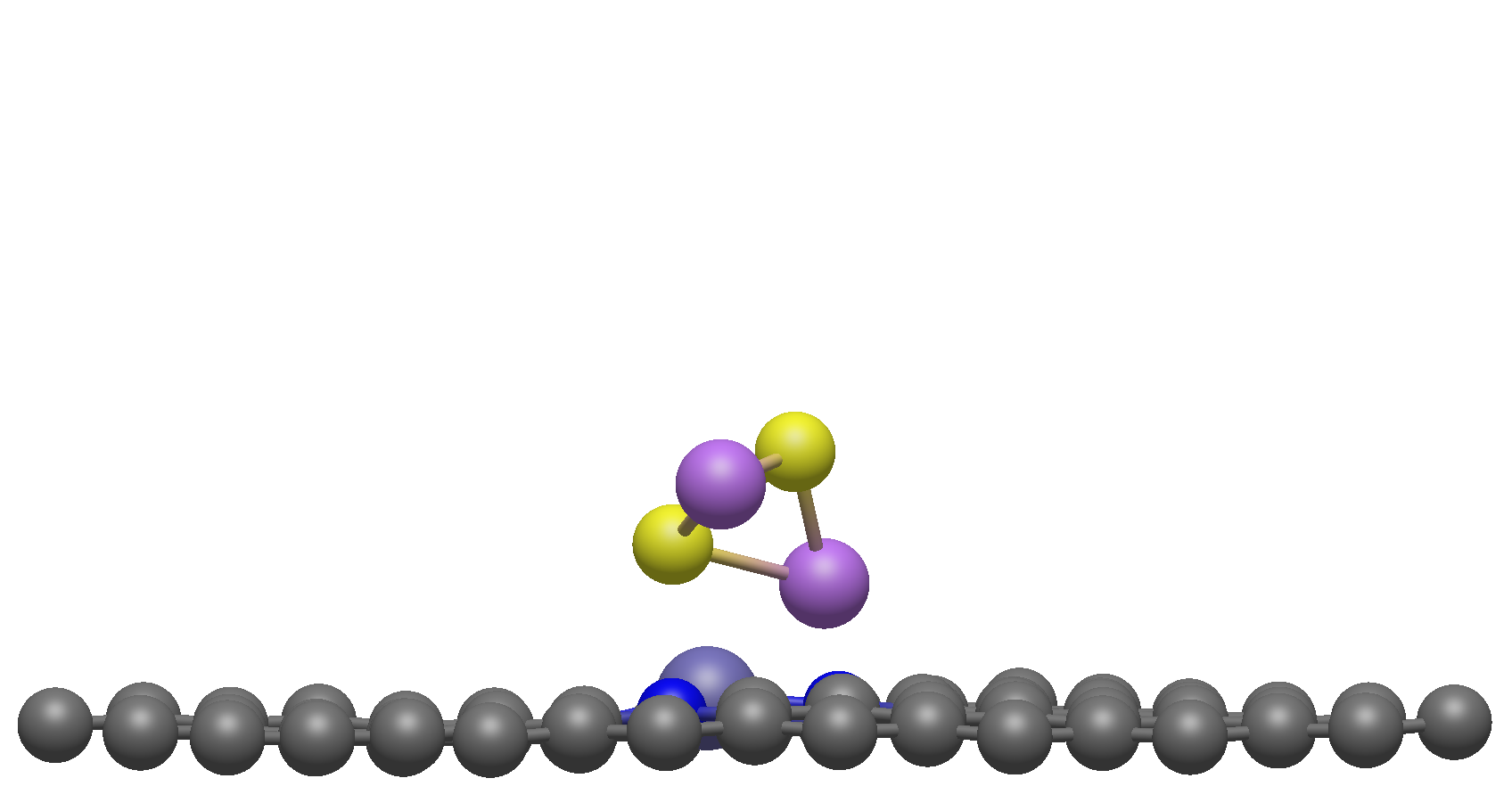} \\
    \includegraphics[width=0.4\linewidth]{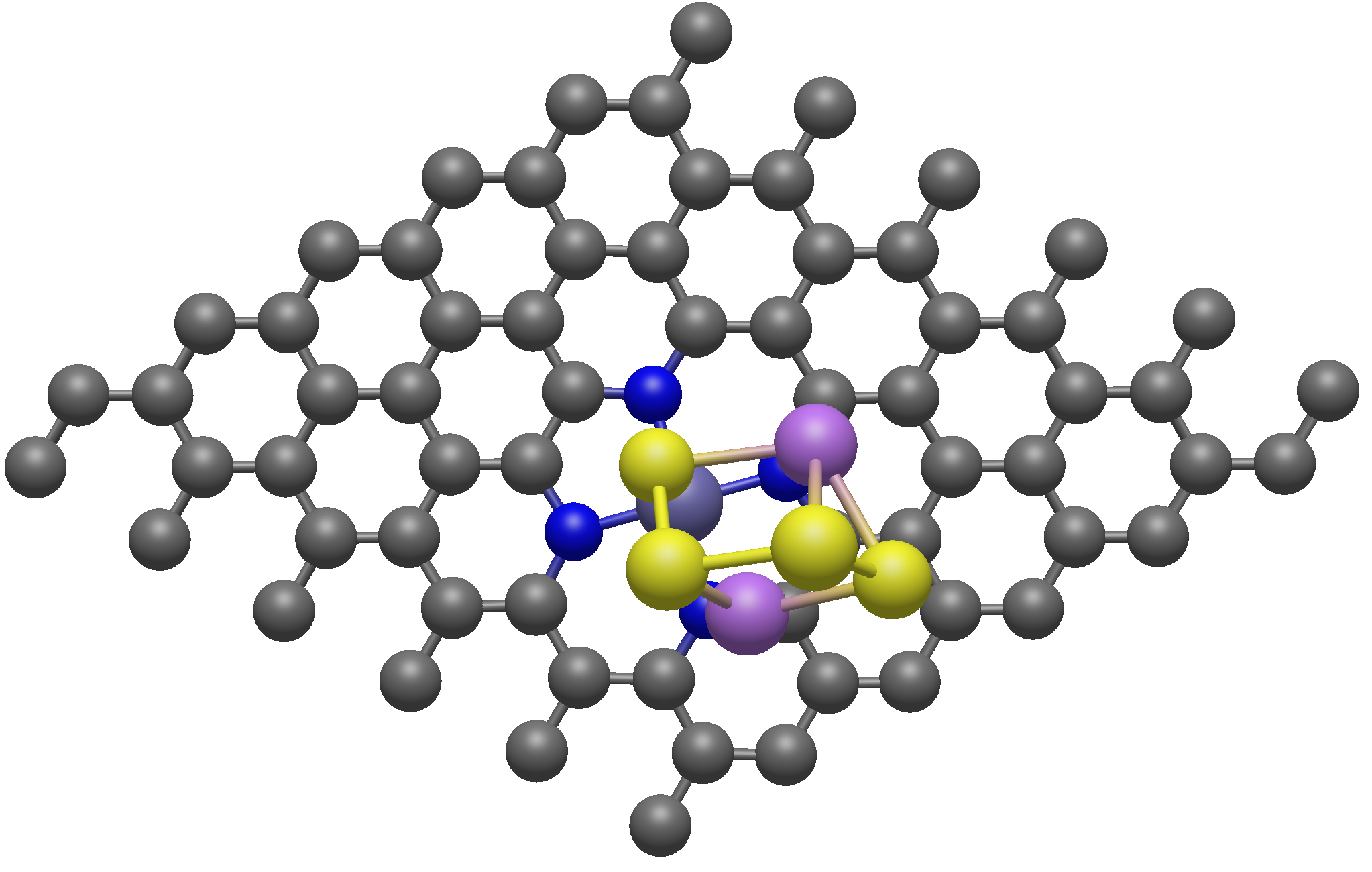}
    \includegraphics[width=0.4\linewidth]{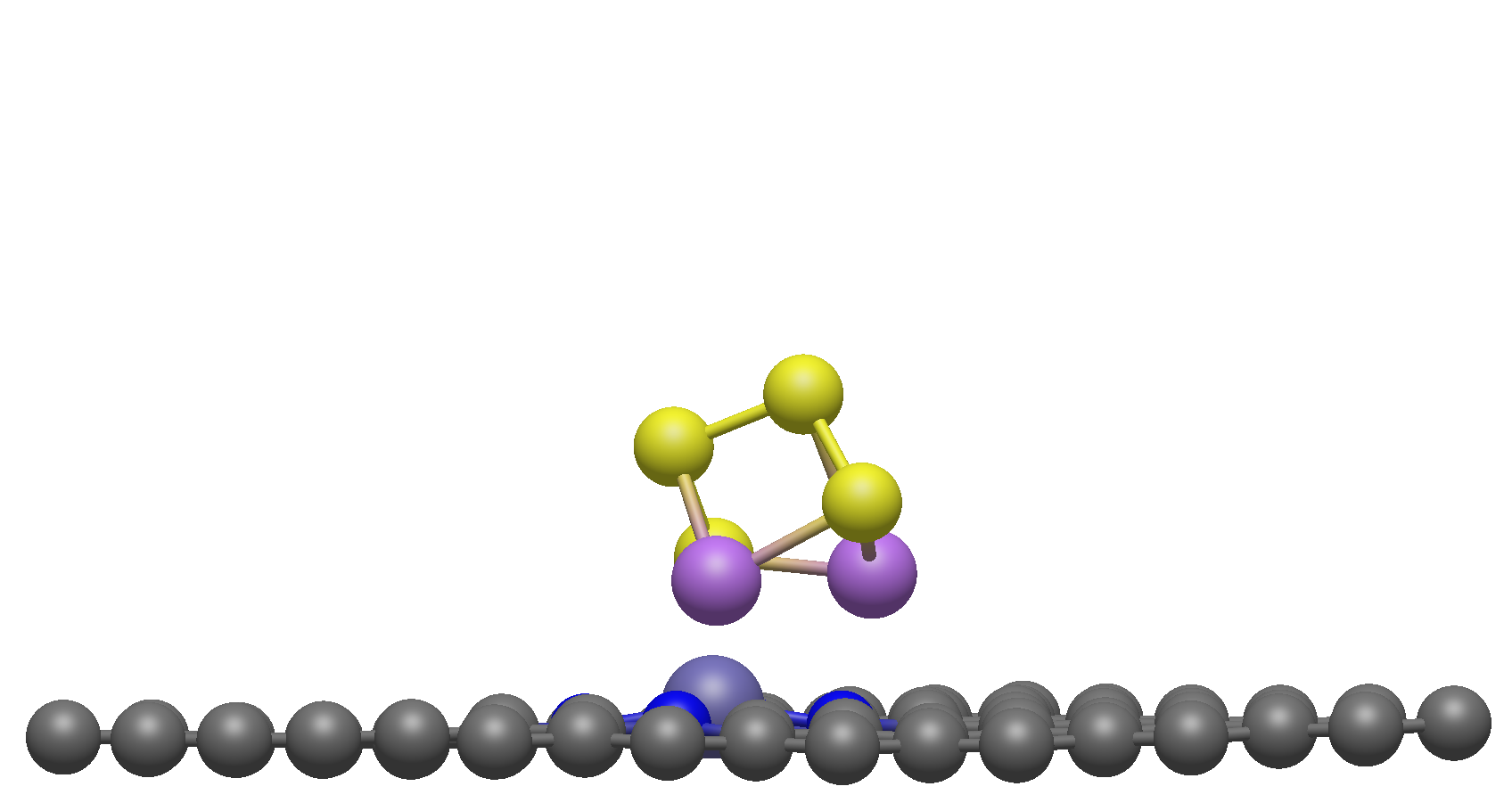} \\
    \includegraphics[width=0.4\linewidth]{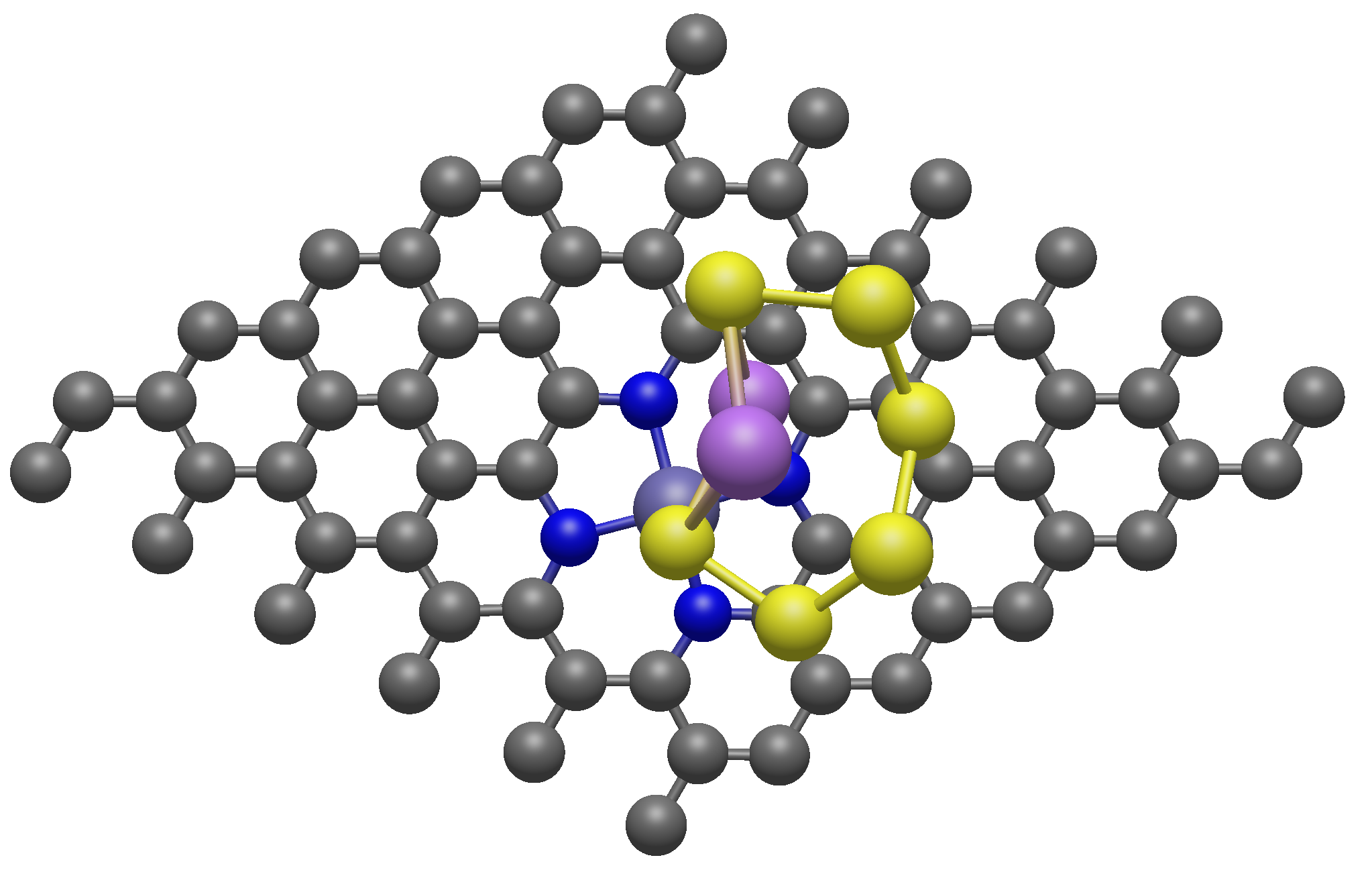}
    \includegraphics[width=0.4\linewidth]{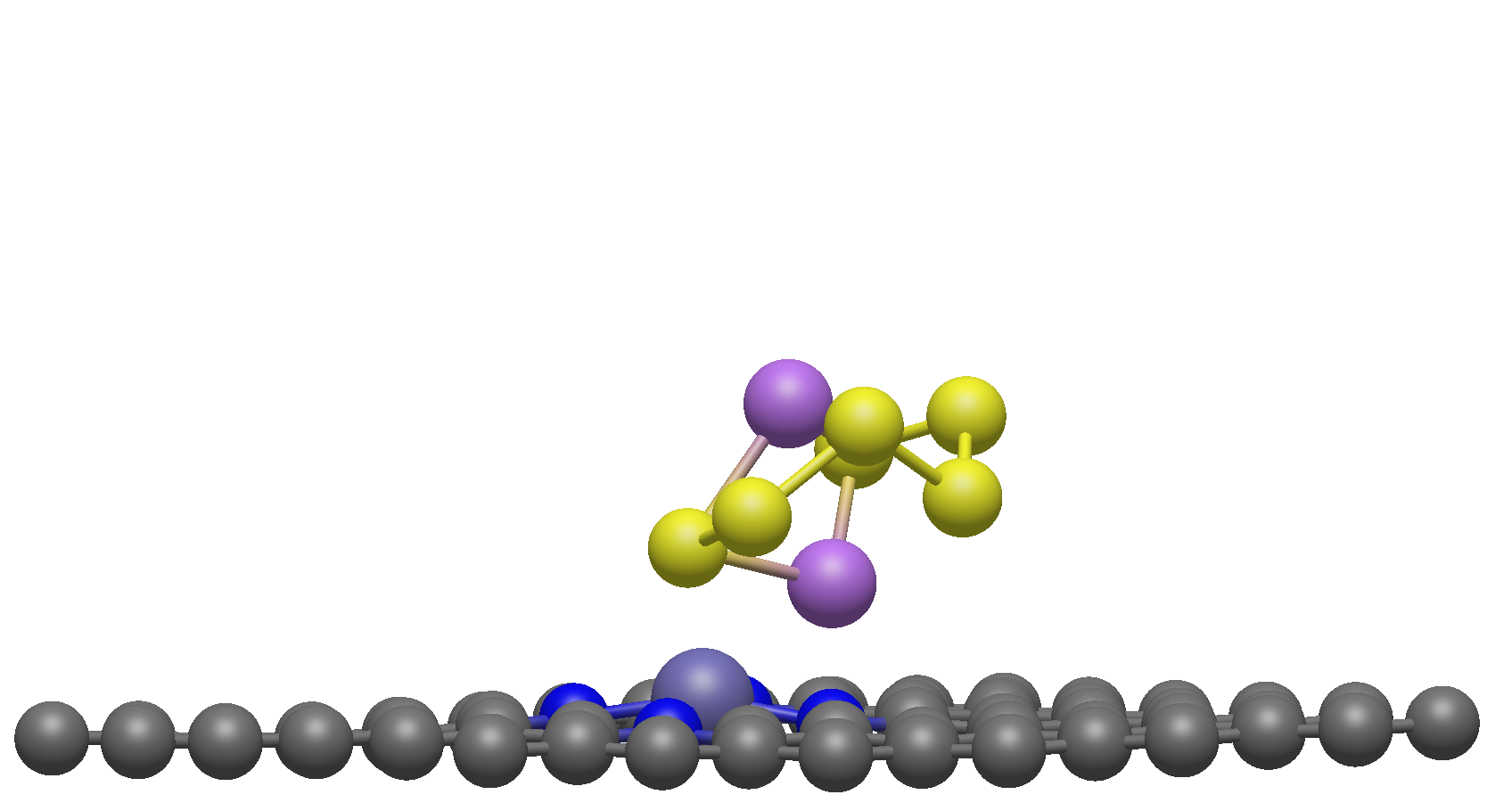} \\
    \includegraphics[width=0.4\linewidth]{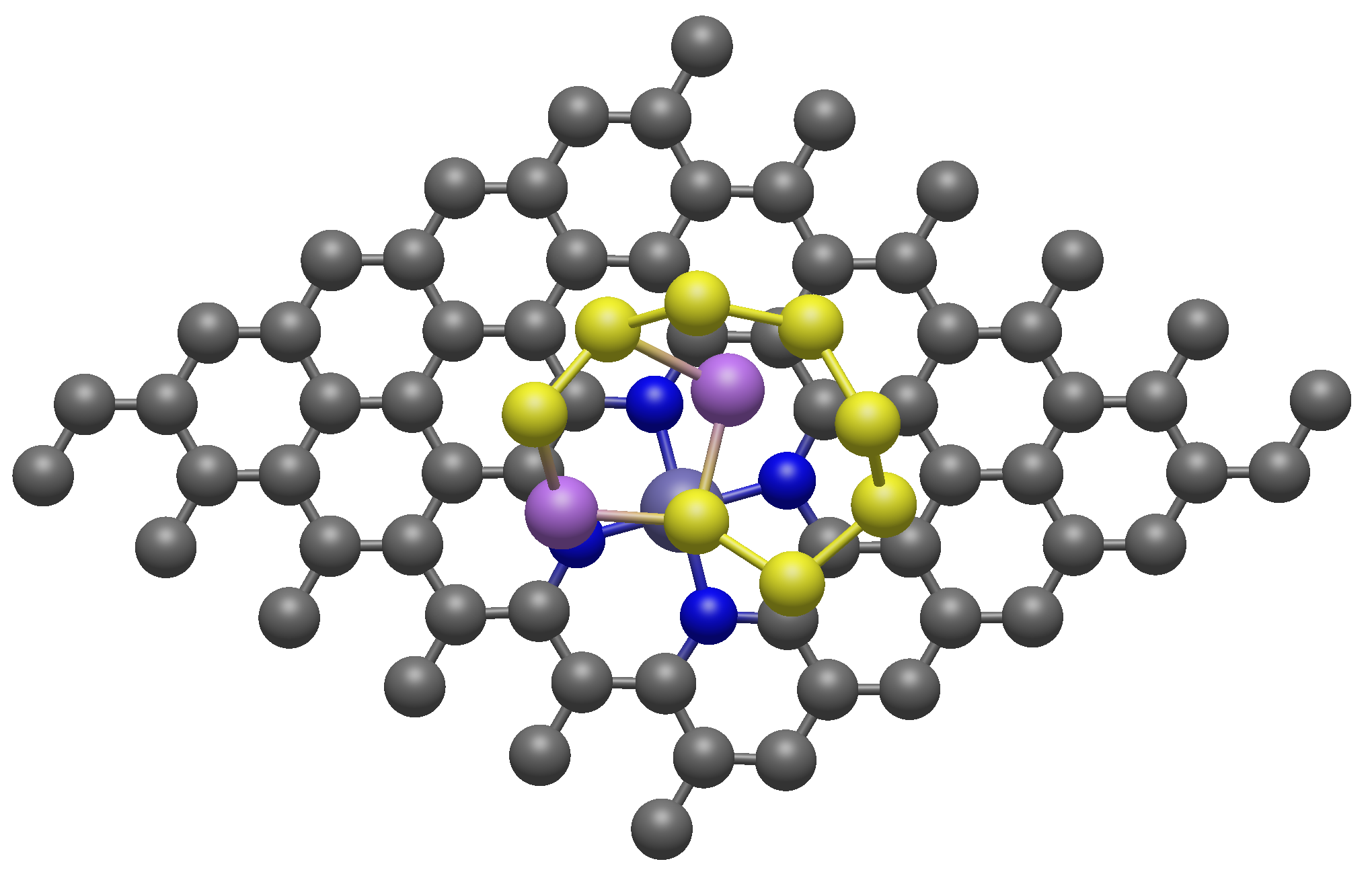}
    \includegraphics[width=0.4\linewidth]{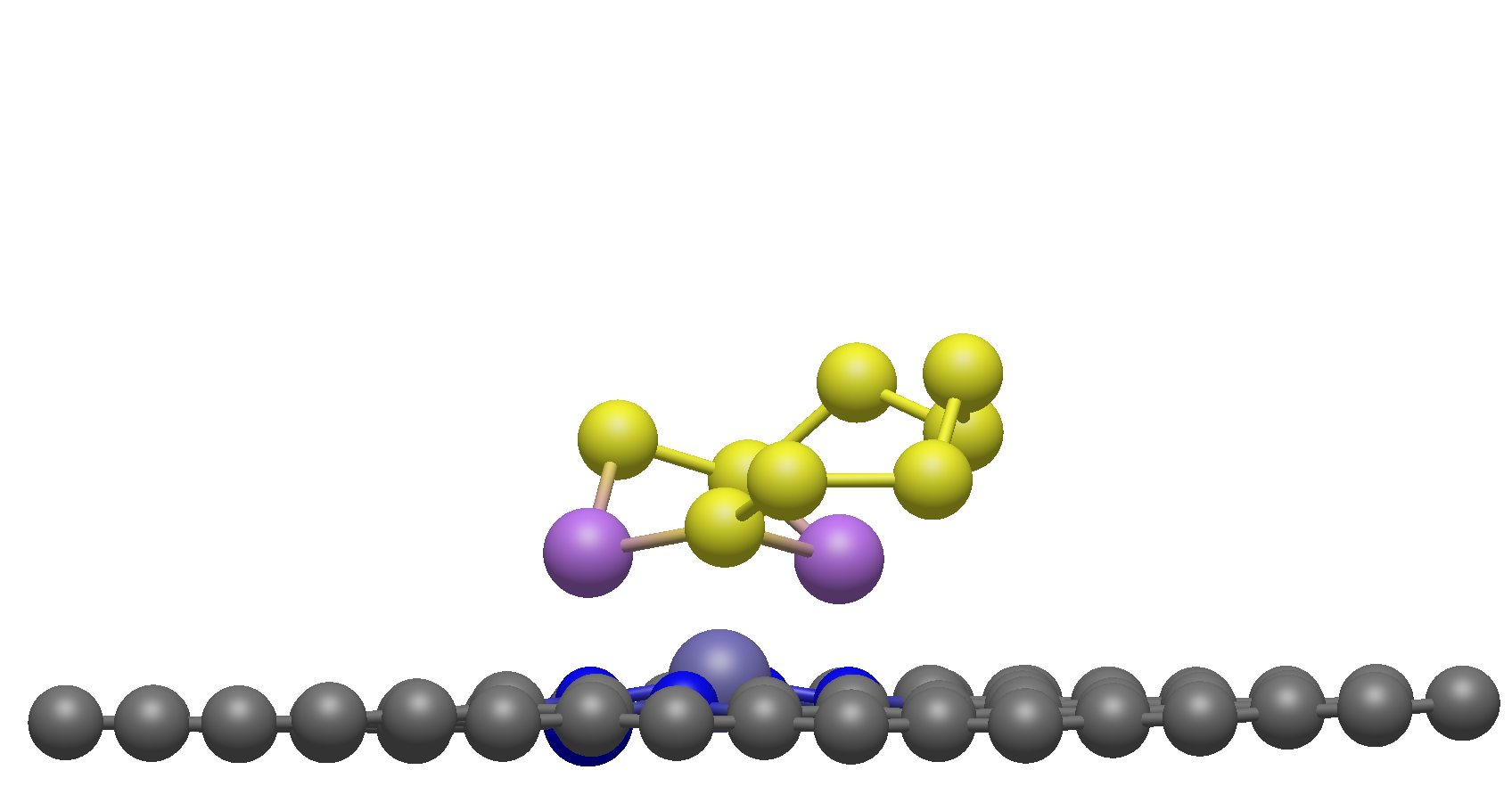} \\
    \caption{Top and side views of the lowest energy configurations for each LiPS (Li$_2$S$_n, n = 1, 2, 4, 6, 8$) on \FeNC, as reported in Table \ref{tab:EadsFe}. Colour code follows from Figure \ref{fig:structures}}
    \label{fig:Fe_str}
\end{figure}

Figure \ref{fig:Fe_str} shows the top and side views of LiPS structures lowest energy structures.
All relaxed LiPS structures have a S atom located on/near the top of the TM atom, while the Li atoms are located closer to the substrate. 
This is verified by an atomic distance analysis, reported in Table \ref{tab:EadsFe}, describing the shortest Li--TM and S--TM distances for each LiPS molecule corresponding to the lowest energy structure.
%
The distance between Li atoms and the TM is always smaller than the one between the S atoms and the TM.
The closest S--TM distance decreases for high order LiPSs, while the opposite is observed for the closest Li--TM distance.
The tendency of the TM in ``pulling'' the S atoms closer while ``pushing'' the Li atoms away is in agreement with our previous report in the literature \cite{Andritsos2021}.

\begin{table}[t]
    \centering
    \caption{Summary of the highest LiPS binding energies for \FeNC \ SAC calculated by geometry optimised DFT simulations found in ``This work'' or found in the literature.\cite{Andritsos2021,Zeng2019_FeSAC, Zhang2019Fe, Wang2019, Zhou2020_SAC} All datapoints are in eV.}
    \begin{tabular}{lcccccc}
        \hline
         & This work & Ref. \cite{Andritsos2021} & Ref. \cite{Zeng2019_FeSAC} & Ref. \cite{Zhang2019Fe} & Ref.\cite{Wang2019} & Ref. \cite{Zhou2020_SAC}  \\
         \hline
        Li$_{2}$S & -2.80 & $-$2.56 & $-$1.55 & $-$2.33 & $-$2.13 & NA\\
        Li$_{2}$S$_{2}$ & -1.94 & $-$2.02 & $-$1.30 & $-$2.04 & NA & NA\\
        Li$_{2}$S$_{4}$ & -1.74 & $-$1.60 & $-$0.65 & $-$ 1.40 & NA & NA\\
        Li$_{2}$S$_{6}$ & -1.57 & $-$1.52 & $-$0.95 & $-$1.42 &NA & $-$0.95 \\
        Li$_{2}$S$_{8}$ & -1.86 &  $-$1.77 & $-$0.80 &  $-$1.39 & NA & NA \\
        \hline
    \end{tabular}
    \label{tab:eref}
\end{table}

\begin{table}[h]
    \centering
    \renewcommand{\arraystretch}{1.2} 
    \caption{Summary of the highest LiPS binding energies for Fe-N$_{4}$-SAC, for structures predicted from the GCH vertexes, calculated by geometry optimised DFT simulations.
    $d_\mathrm{Li-TM}$ and $d_\mathrm{S-TM}$ refer to the distance to the TM of the closest Li and S atoms to the TM, respectively.
    Str. ID refers to the unique structure identification number based on the roto-translation of the LiPS with respect to the substrate, and its counting is consistent along the different SAC materials.}
    \label{tab:EadsFe}
    \begin{tabular}{lcccc}
    \hline
        Material  & $E_\mathrm{b}$ (eV) & $d_\mathrm{Li-TM}$ (\AA) & $d_\mathrm{S-TM}$ (\AA) & Str. ID\\
    \hline
        \FeNC/\lso & $-$2.80 & 2.57 & 2.29 & 9586 \\
        \FeNC/\lst & $-$1.94 & 2.61 & 2.22 & 19420 \\
        \FeNC/\lsf & $-$1.74 & 2.64 & 2.22 & 8637 \\
        \FeNC/\lss & $-$1.57 & 2.64 & 2.20 & 4092 \\
        \FeNC/\lse & $-$1.86 & 2.77 & 2.19 & 7335 \\
    \hline
    \end{tabular}
\end{table}

\section{Workflow Exploitation: \ZnNC \ SAC}

\begin{figure}[t]
    \centering
    \text{Top view} \hspace{2.5cm} \text{Side view} \\
    \includegraphics[width=0.4\linewidth]{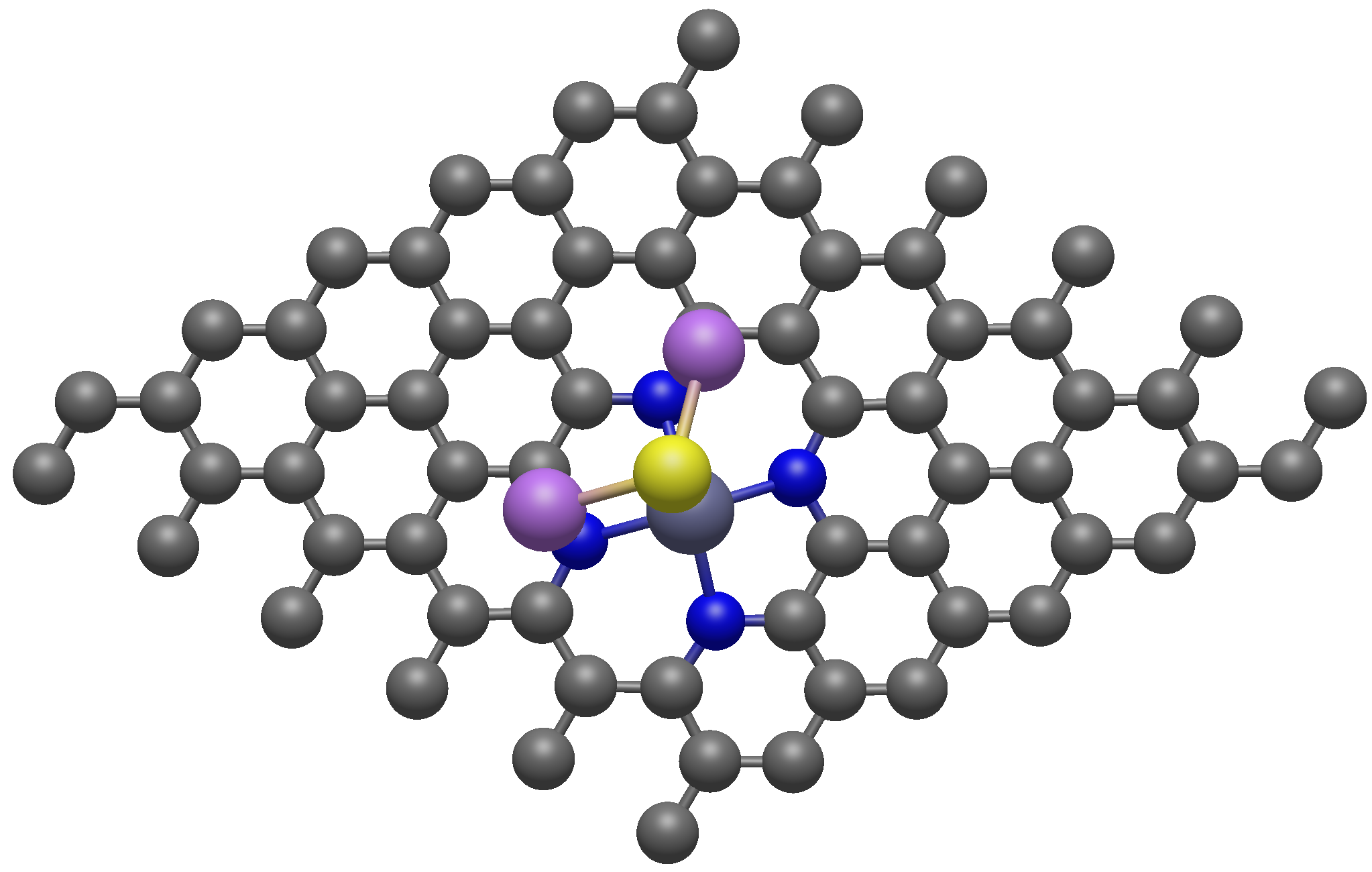}
    \includegraphics[width=0.4\linewidth]{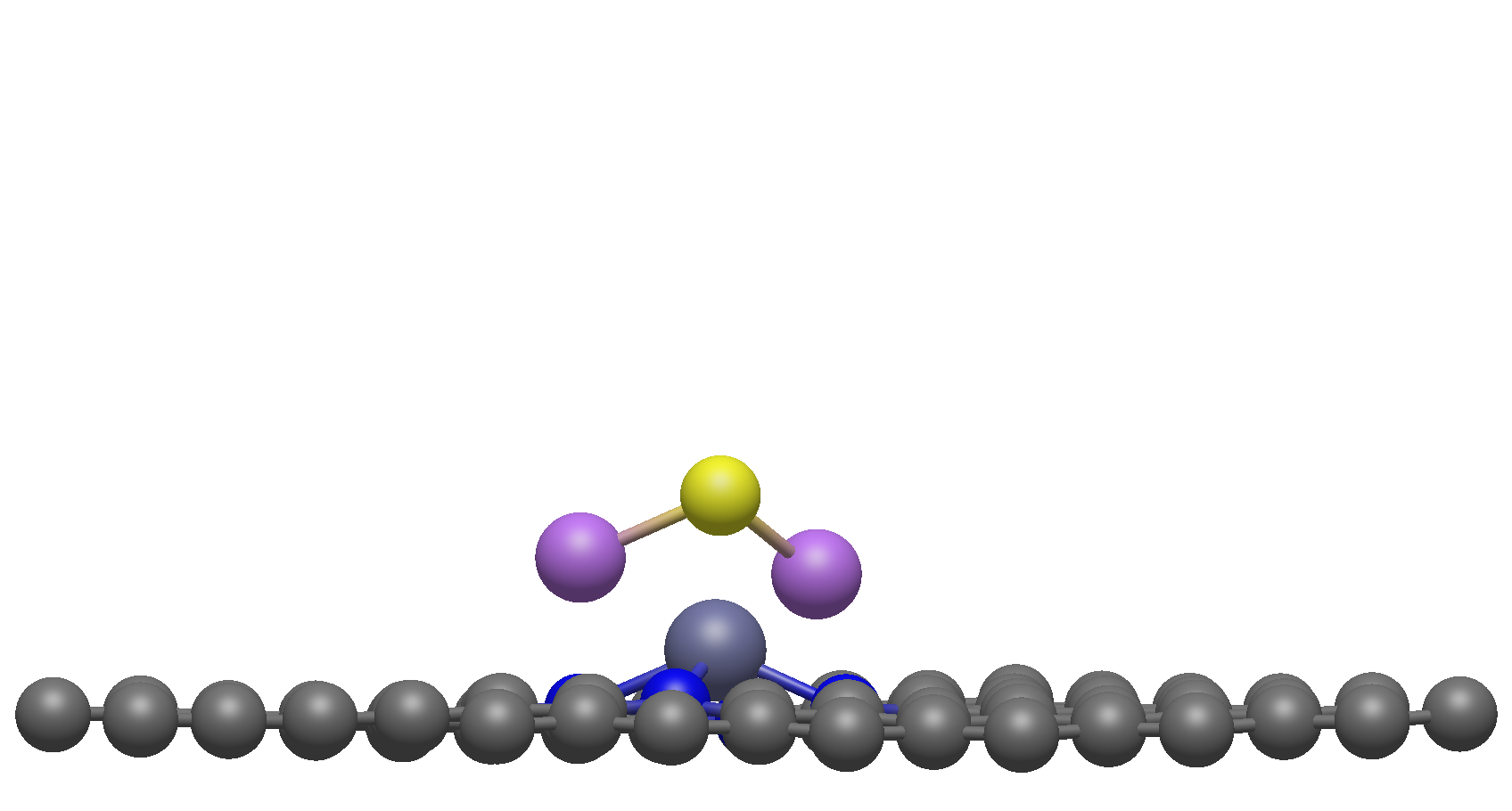} \\
    \includegraphics[width=0.4\linewidth]{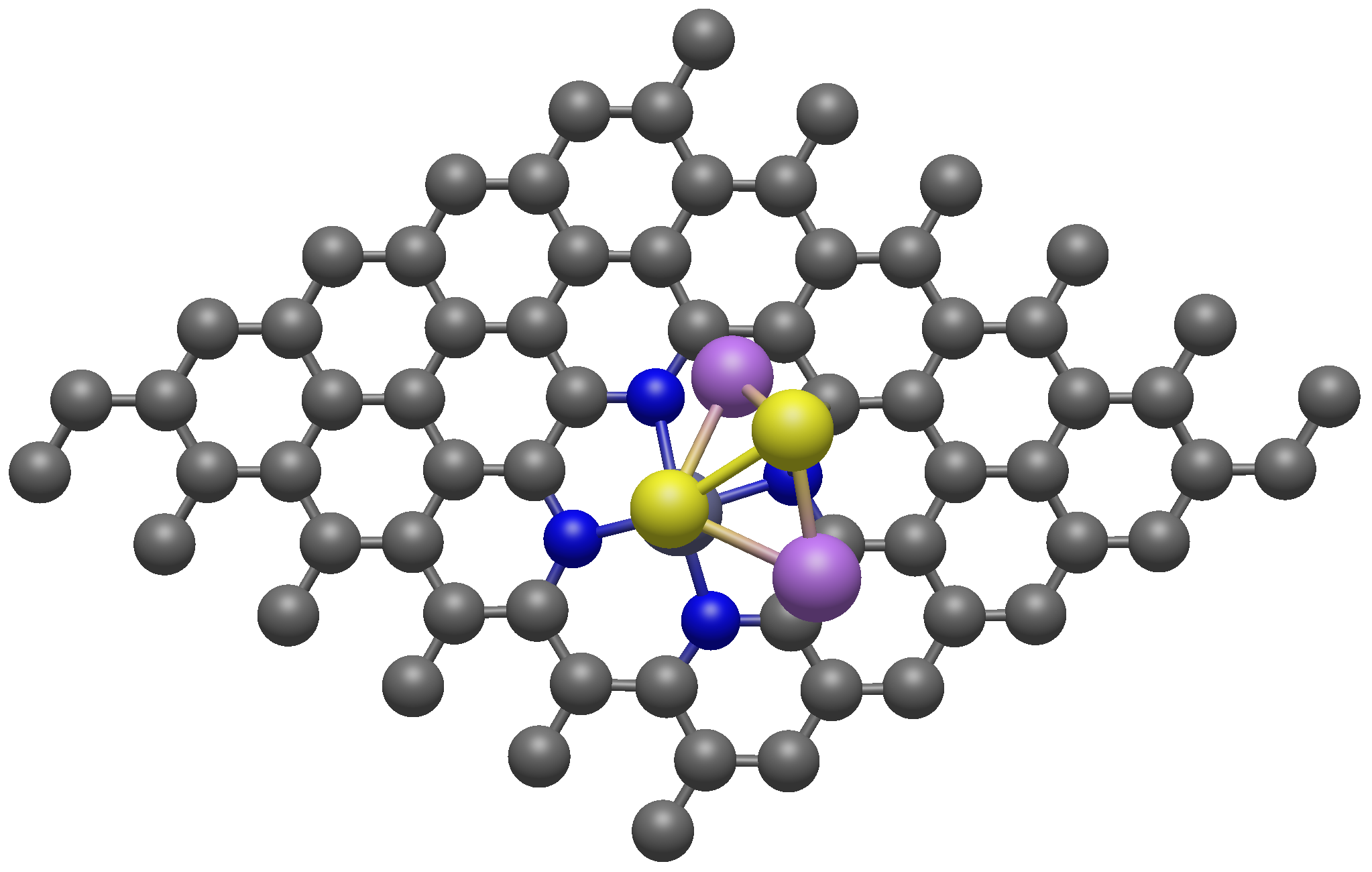}
    \includegraphics[width=0.4\linewidth]{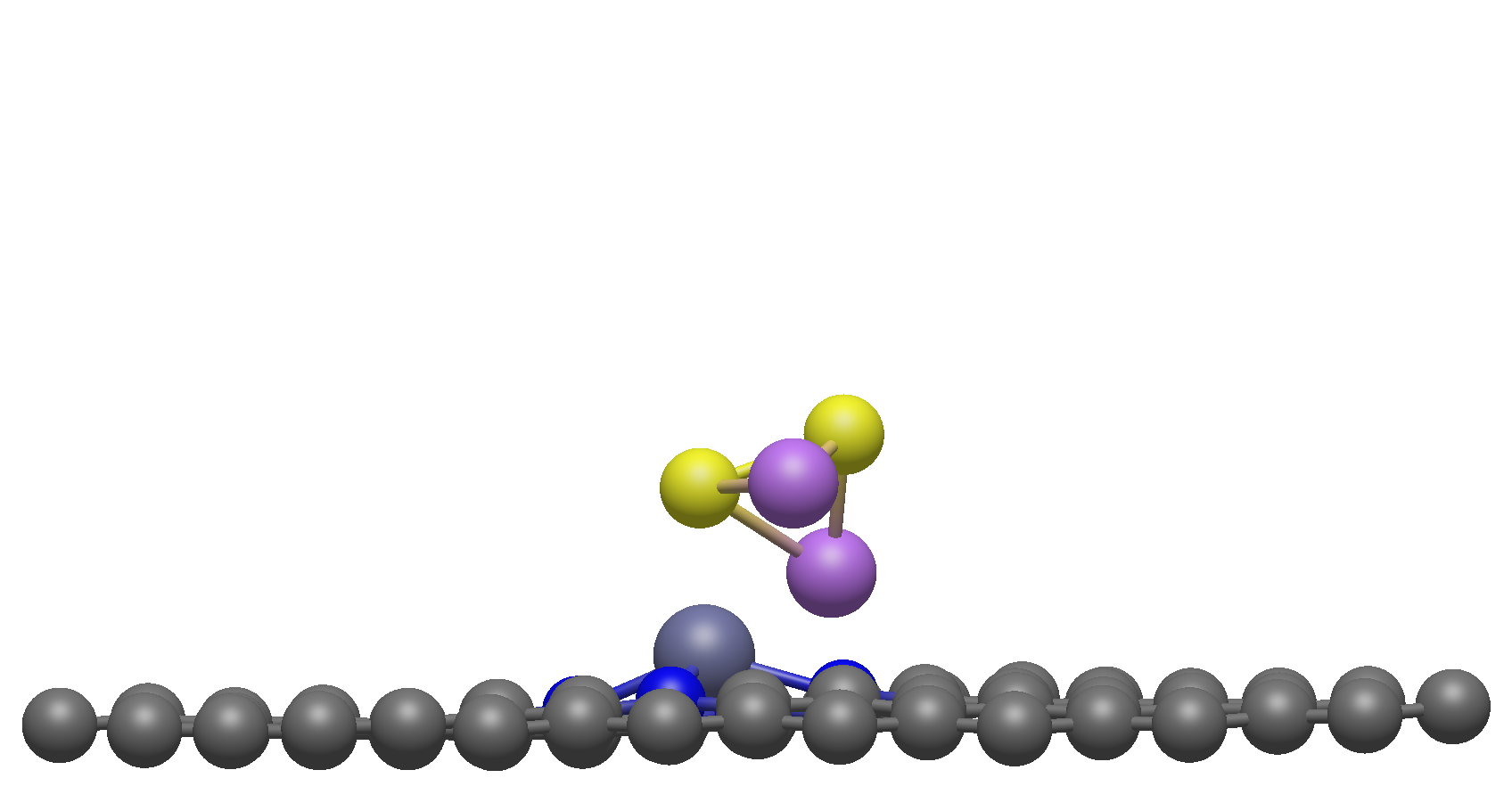} \\
    \includegraphics[width=0.4\linewidth]{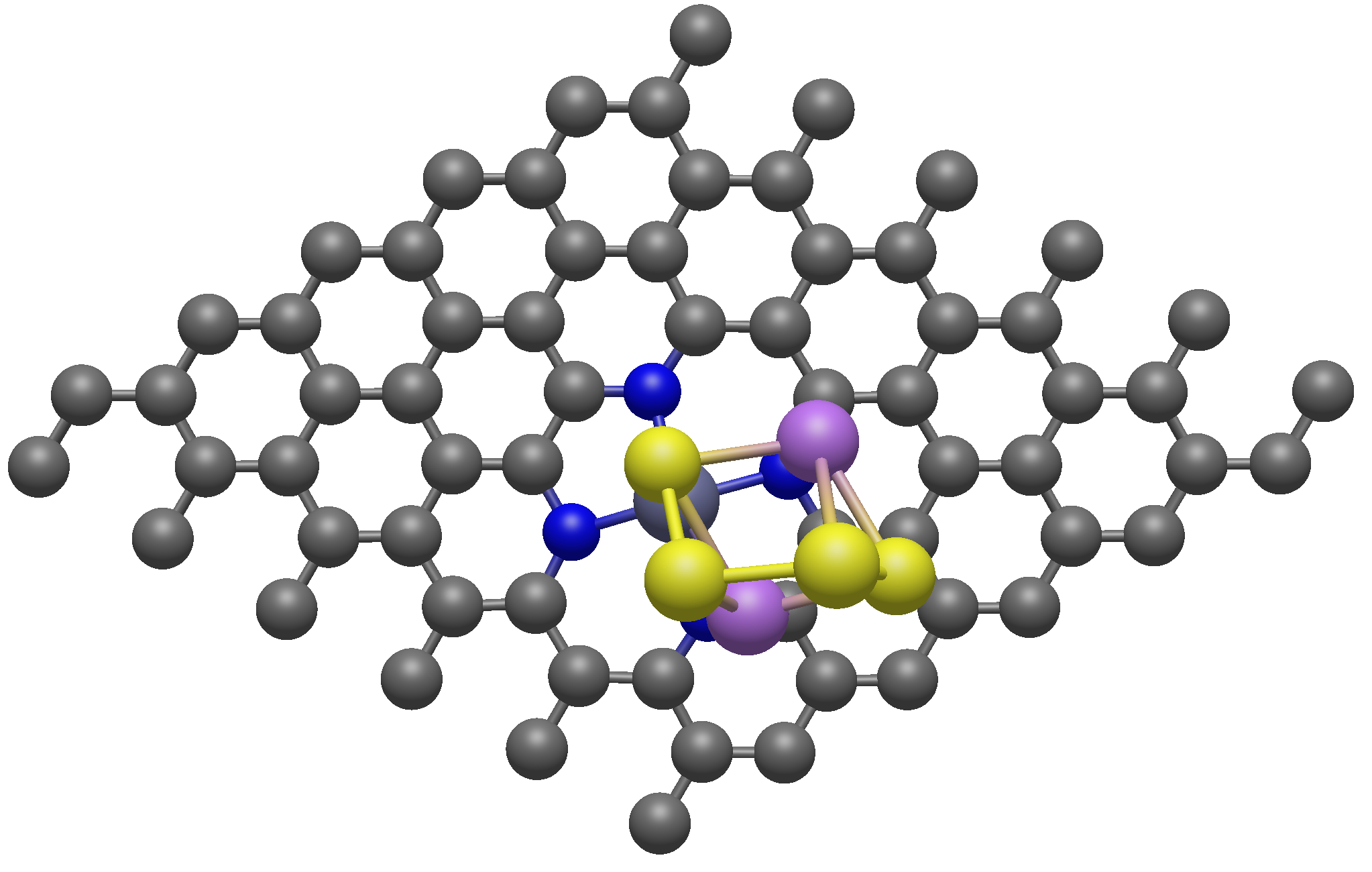}
    \includegraphics[width=0.4\linewidth]{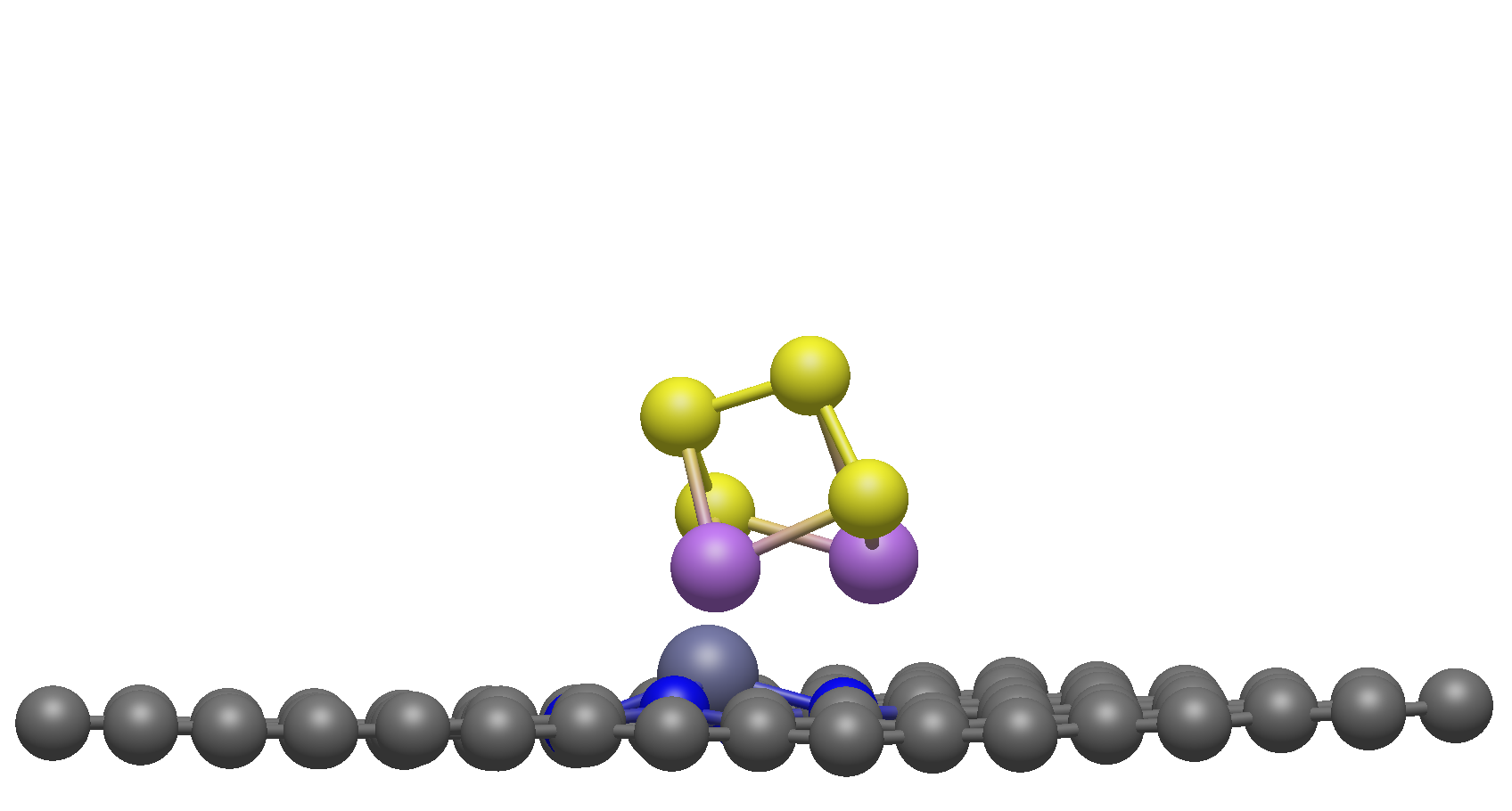} \\
    \includegraphics[width=0.4\linewidth]{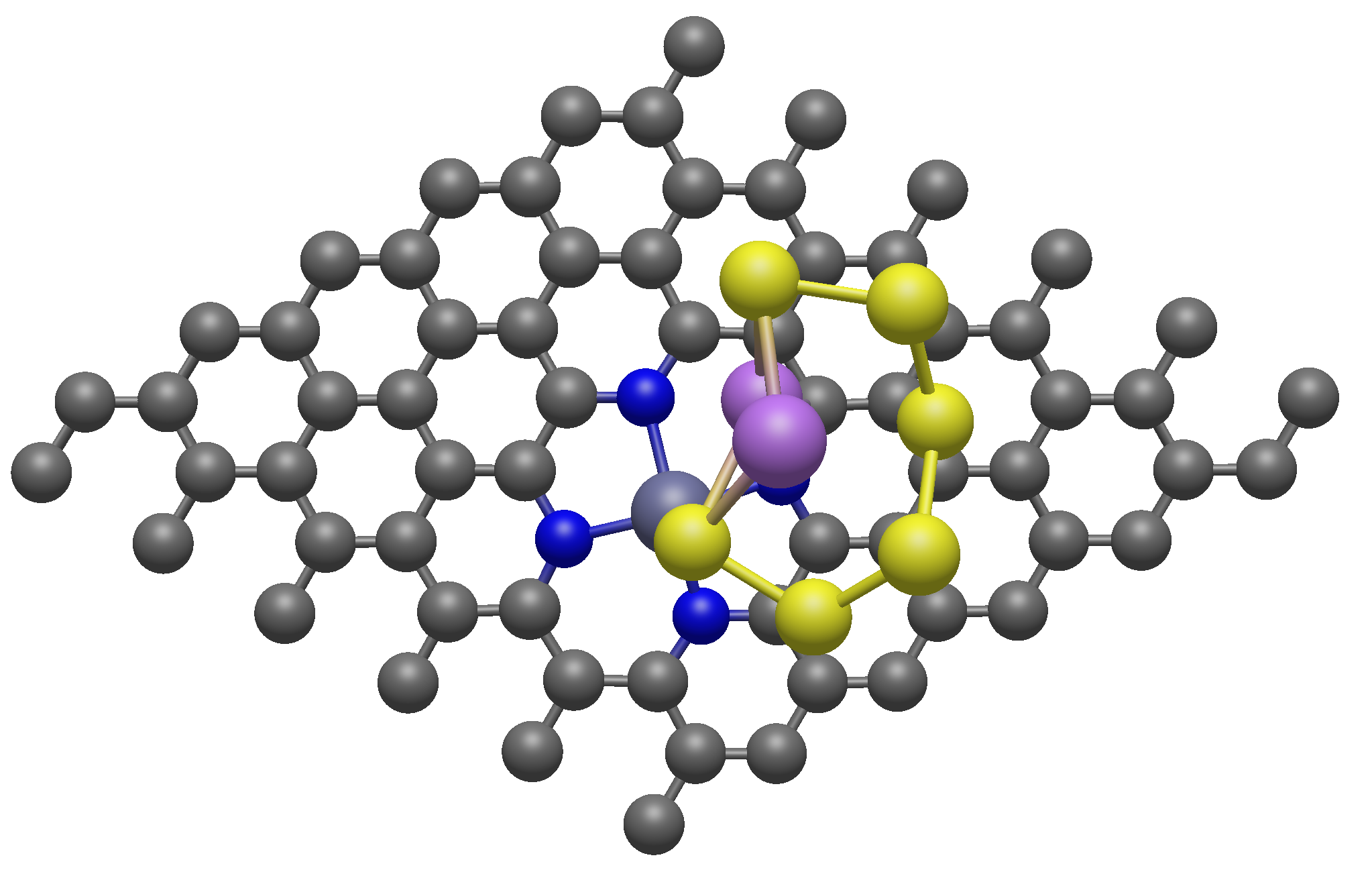}
    \includegraphics[width=0.4\linewidth]{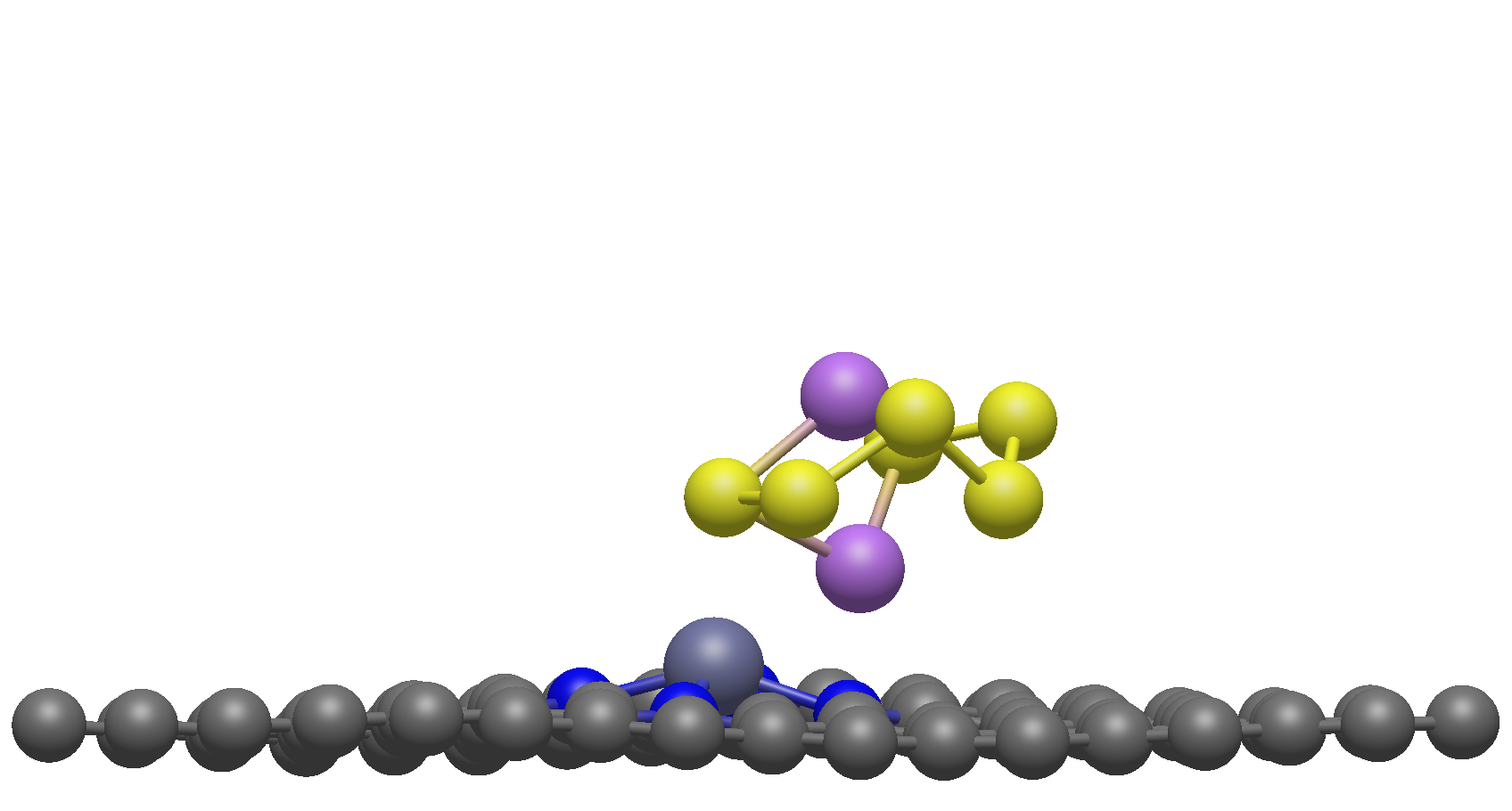} \\
    \includegraphics[width=0.4\linewidth]{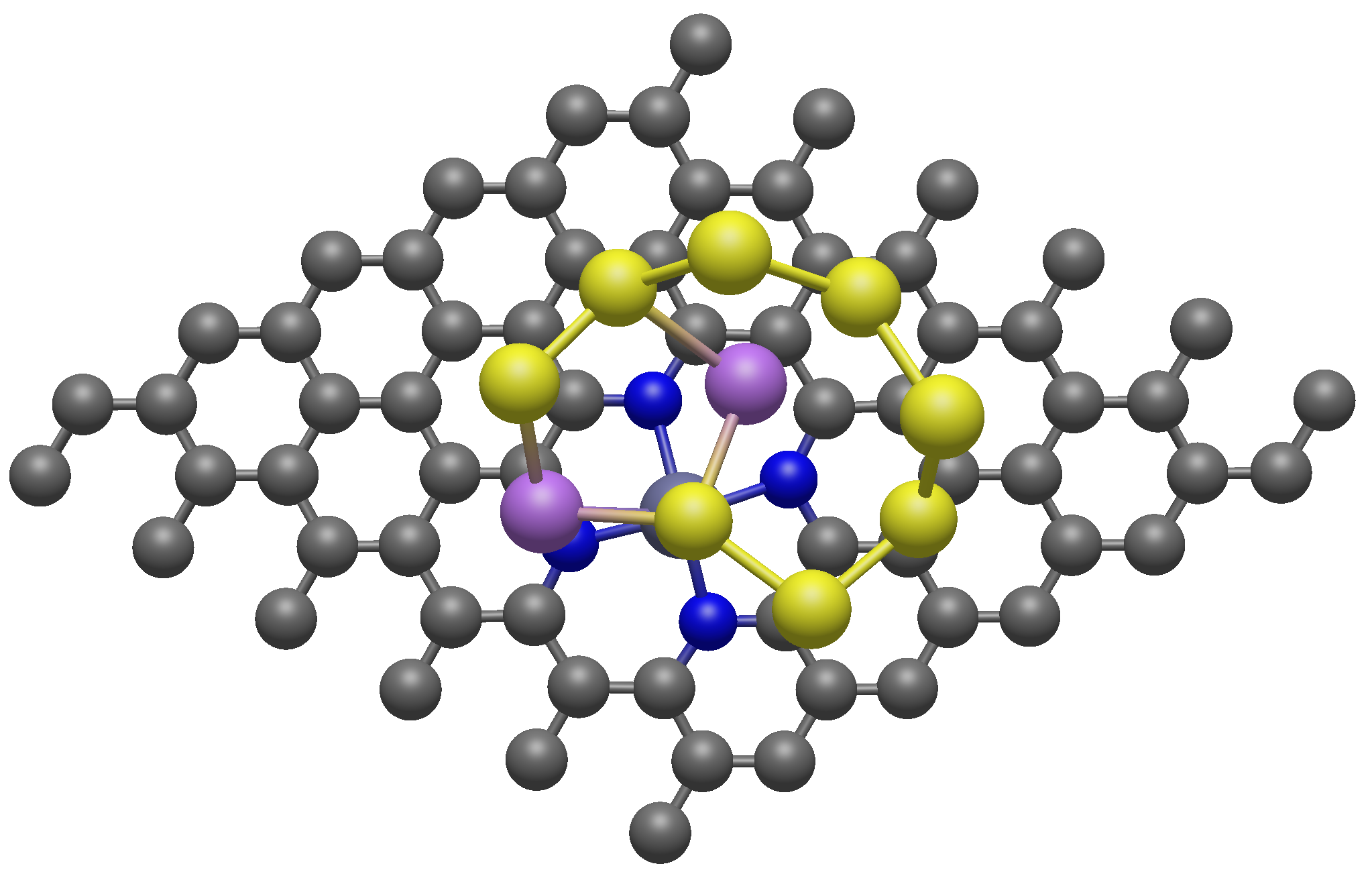}
    \includegraphics[width=0.4\linewidth]{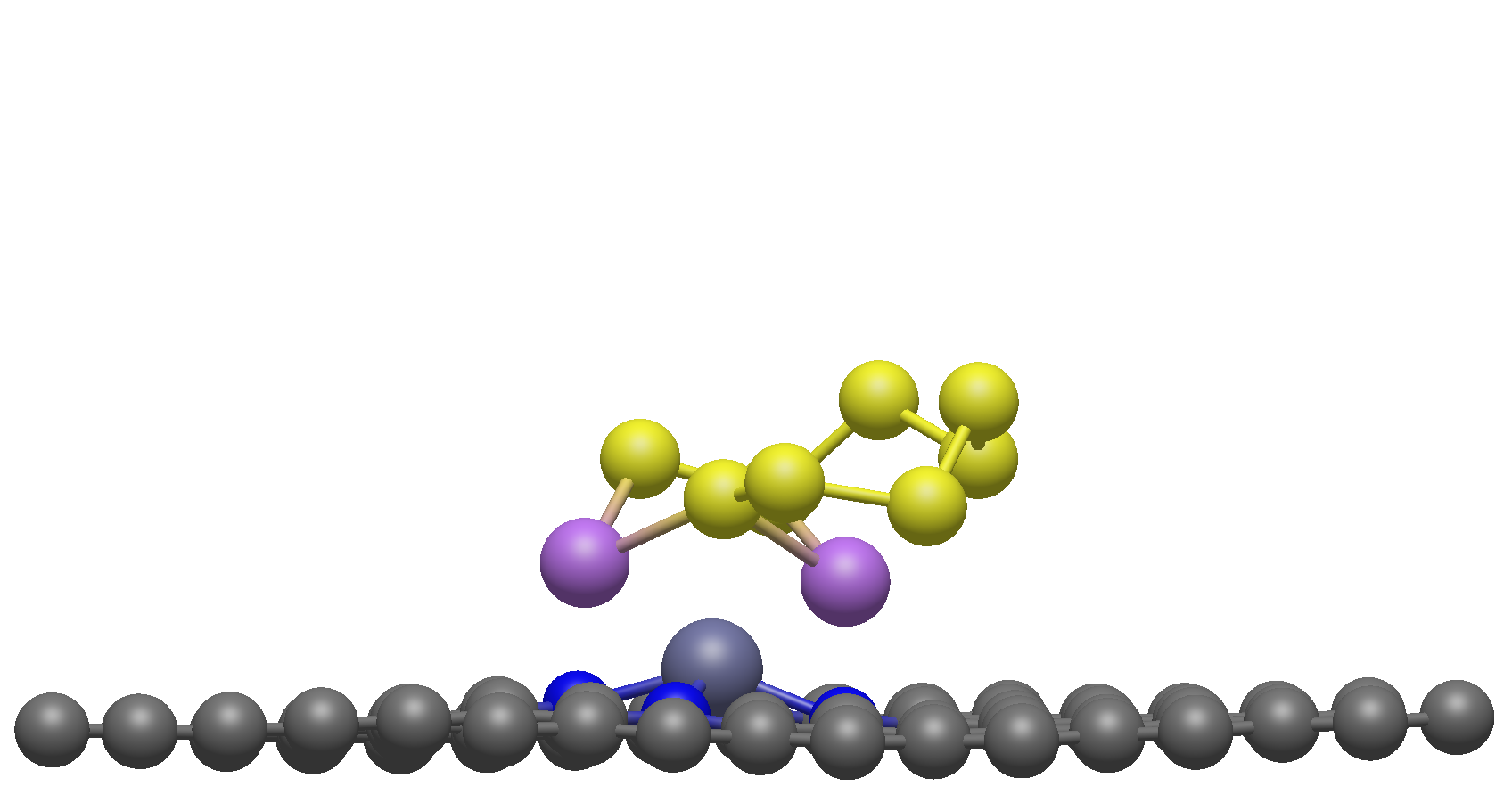} \\
    \caption{Top and side views of the lowest energy configurations for each LiPS (Li$_2$S$_n, n = 1, 2, 4, 6, 8$) on \ZnNC, as reported in Table \ref{tab:EadsZn}.Colour code follows from Figure \ref{fig:structures}}
    \label{fig:Zn_str}
\end{figure}

We move on to exploit the binding energy prediction workflow described in this study and identify the low-lying minima for the shuttle-effect cycle on a Zn-based SAC.
The system configuration and geometry of the Zn-based SAC structures were designed in accordance to the previously described Fe-based SAC, and are illustrated in Figure \ref{fig:Zn_str}.
We examine the binding energies for five LiPSs (Li$_2$S$_n, n = 1, 2, 4, 6, 8$), and compare LiPS energetics and geometries against the one found for Fe-based SAC.

\begin{table}[b]
    \centering
    \renewcommand{\arraystretch}{1.2} 
    \caption{Summary of the highest LiPS binding energies for Zn-N$_{4}$-SAC, for structures predicted from the GCH vertexes, calculated by geometry optimised DFT simulations.
    $d_\mathrm{Li-TM}$ and $d_\mathrm{S-TM}$ refer to the distance to the TM of the closest Li and S atoms to the TM, respectively.
    Str. ID refers to the unique structure identification number based on the roto-translation of the LiPS with respect to the substrate, and its counting is consistent along the different SAC materials.
    }
    \label{tab:EadsZn}
    \begin{tabular}{lcccc}
    \hline
        Material  & $E_\mathrm{b}$ (eV) & $d_\mathrm{Li-TM}$ (\AA) & $d_\mathrm{S-TM}$ (\AA) & Str. ID\\
    \hline
        \ZnNC/\lso & $-$2.51 & 2.57 & 2.27 & 10107  \\
        \ZnNC/\lst & $-$1.71 & 2.55 & 2.36 & 14382  \\
        \ZnNC/\lsf & $-$1.44 & 2.54 & 2.38 & 8637  \\
        \ZnNC/\lss & $-$1.27 & 2.71 & 2.46 & 4093  \\
        \ZnNC/\lse & $-$1.38 & 2.63 & 2.41 & 9518  \\
    \hline
    \end{tabular}
\end{table}

Table \ref{tab:EadsZn} shows the lowest identified binding energies for all examined structures lying on a GCH vertex.
Although the \ZnNC \ has not been extensively studied previously, the observed binding energies are in good agreement with previously reported values ($E_\mathrm{b}^\mathrm{Zn-N_4-C/Li_2S_6} = 1.02$ eV \cite{Zhou2020_SAC}).
Atomic binding on Zn-based SAC seems stronger than on Fe-based SAC, by 0.45 - 0.64 eV depending on the LiPS order.
The presence of Zn rather than FE in the SAC tends to affect more the lower order LiPSs comparing, a property that can be of help towards the reduction of the ``shuttle'' effect at the final catalytic steps of a LiPS. 
The shortest Li--TM distances behave similarly for either Zn and Fe-based SACs, although these are generally shorter when Zn is present.
In contrast, the shortest reported S--TM distances for the two systems follow an opposite trend.
When Zn is present the S--TM distance tends to increase as the LiPS order increases, and, generally, the distances are larger when comparing to the Fe-based SAC.
Based on the above, we can conclude that Zn tends to prefer Li bonding for high order LiPSs. 
This can potentially improve the conversion rate at that stage. \par

As a next step, we compare the lowest found energy configurations for the two different SACs, as reported in Tables \ref{tab:EadsFe} and \ref{tab:EadsZn} and as visualised in Figures \ref{fig:Fe_str} and \ref{fig:Zn_str}.
We note that some of the LiPSs which end up in the most competitive minima find the same, or almost the same, initial vertex in the GCH, comparing the Fe and Zn SAC cases.
For \lso, \lst and \lse, where the initial unrelaxed configurations are not lying in the same vertex for the two different SACs, the final relaxed configurations are minimising into similar configurations, as shown in Figures \ref{fig:Fe_str} and \ref{fig:Zn_str}.

It is a matter of debate whether best LiPS configurations found for a given substrate could be translated to other systems of similar nature. 
Our results suggest that relaxing a LiPS-SAC configuration which resulted in lower binding energy for the case of a given TM, is likely, yet not always, to lead to favourable configurations when substituting the TM itself. 
More robust approaches to identify putative minimum energy configurations, like the one here proposed, are thus necessary towards more accurate predictions.  \par

For completeness in the discussion on the computational costs associated with he discussed protocol, we display the number of DFT relaxation steps necessary in our workflow for each Zn-LiPS system in Table \ref{tab:MLefficiency}.
In particular, we report the total number of relaxation steps required in this workflow, including the single-point calculations necessary during the ``Reference Force-Energy calculations'' step, and the structure relaxation calculations of all GCH vertexes during the ``Minimise GCH Vertexes'' step.
We recognise that the computational cost of DFT-based calculations widely depends on multiple system and software dependent factors, hence effective cross-reference comparison is not trivial.
Additionally, the computational cost per DFT relaxation step varies when using self-consistent functionals, hence time comparison would not be immediately insightful.
However, it is heuristically observed that the total number of steps required for system relaxation, for structures with the same simulation parameters and configuration, can identify the level of disorder or system complexity of the initial structure.

\begin{table}
    \centering
    \renewcommand{\arraystretch}{1.2} 
   \caption{Reported relaxation steps for each material.
   "Single-point (S-P) number" refers to the calculations required during the ``Reference Force-Energy calculations'' to train the ML model, the label
   "Vertexes" refers the total number of geometry optimisations performed, and
   "Total steps" refers to the sum of the geometry optimisation steps for all vertexes.}
    \label{tab:MLefficiency}
    \begin{tabular}{lccc}
    \hline
    Material & S-P number & Vertexes & Total steps \\
    \hline
    \ZnNC/\lso & 537 & 9 & 3962 \\ 
    \ZnNC/\lst & 496 & 9 & 3922 \\ 
    \ZnNC/\lsf & 505 & 8 & 3139 \\ 
    \ZnNC/\lss & 404 & 7 & 8831 \\ 
    \ZnNC/\lse & 408 & 8 & 7700 \\ 
    \hline
    \end{tabular}
\end{table}

The standard approach of manually, and often arbitrarily, constructing the LiPS/SAC configuration is known to be ineffective and often might not correspond to the most energetically favourable structure, unless excessive testing is performed. 
On one hand we cannot exclude that other minima searches methods, e.g. via genetic algorithm approaches, will find minima structures competitive  with those found in this study.
On the other we note that these more extensive searches, involving hundreds or thousands structural relaxations, rather than tens of structural relaxations as in the proposed protocol, would result in much larger computational costs and practically unusable.
However, we argue that the workflow under examination proposes a good set of candidate structures and offers a fair balance between accuracy and computational efficiency.\par

\section{Discussion and Conclusions}

The binding energy of LiPSs on TM-N-SAC substrates is a useful descriptor to infer which SAC material is likely to provide an optimal performance in preventing shuttle effects in Li-S batteries.
Here, we propose an efficient protocol to identify a likely minimum energy configuration candidate, and thus the most informative binding energy prediction, via a combination of DFT single-point calculations, unsupervised and supervised data-driven algorithms, and DFT structure optimisations.
The protocol hinges on: I) the construction of a database of roto-translation of LiPS molecules deposited on a rigid TM-N-SAC substrate, II) the calculation of the binding energy for a carefully selected number of these, III) the training of a kernel ridge regressor to chart structure-binding energy relationships systematically, IV) the use of a generalised convex hull construction to isolate likely-stable and structurally diverse geometries, V) their final optimisation at DFT level. \par

We benchmark the protocol robustness for the case of LiPS on a Fe SAC and recover binding energies on-par or lower than the ones reported in the literature.
We move to analyse the LiPS adsorption properties on a Zn SAC and find that the latter systematically displays a strong interaction (i.e. lower binding energies).
From a structural perspective, we confirm that the minimum S-TM distances are always lower than the minimum Li-TM distances, for both systems.
We further find that, regardless of the TM nature, an increase in the nuclearity of the LiPS molecule corresponds to a decrease in the S-TM minimum distance and an increase in the Li-TM minimum distance.
Interestingly, we note that notwithstanding a change in the transition metal single atom catalysts, the lowest energy relaxed structure often, but not always, originates from a LiPS located and oriented in similar initial position and orientation w.r.t. the substrate TM. \par

Few upgrades could further benefit the efficiency and accuracy of the search strategy we propose.
If deemed necessary, the GCH vertex search can encompass not the first PCA component but also high order ones.
Further, adopting more complex statistical learning methods (e.g. Gaussian Process regression or Neural Network based) or other more recent local-density featurisation approaches (e.g. Atomic Cluster Expansion based \cite{Drautz2020, Zeni2021a, Lysogorskiy2021} approaches) could enable models which retain the same target accuracy but demand for a lesser number of training structures.
Second, co-regionalised \cite{Bonilla2009} or transfer learning  \cite{West2007} approaches may further benefit the making of more efficient and accurate models when tackling the study of LiPS adsorption on a substrates with many and diverse SACs transition metals.
Third, ML approaches could be also adopted to accelerate the minimisation \cite{Jacobsen2018, delrio2019, Yang2021} and reduce drastically the number of DFT force-and-energy calculations which are necessary at the "structure optimisation" stage following the surrogate generalised convex hull construction.
Currently, the proposed strategy does not aim to replace DFT-based calculations, but rather to guide and speed-up DFT-based searches of LiPS configurations for strong adsorption on SACs.
The ML acceleration of the structural relaxation step could further diminish the need of DFT single-point calculations by a factor of 2-10.

Finally, we remark that the method is general and transferable to study not only other LiPS-substrate interactions, but, in principle, also other molecular system - substrate interactions.
Here, we demonstrate our protocol's transferability by examining the \ZnNC \ SAC, and observing consistent results as in the \FeNC \ SAC.
Conversely we believe this work will nurture the advancement in minima searches protocols that investigate medium and large molecules adsorption on rigid and soft substrates of diverse nature.

\bibliography{references}

\begin{thebibliography}{49}%
\makeatletter
\providecommand \@ifxundefined [1]{%
 \@ifx{#1\undefined}
}%
\providecommand \@ifnum [1]{%
 \ifnum #1\expandafter \@firstoftwo
 \else \expandafter \@secondoftwo
 \fi
}%
\providecommand \@ifx [1]{%
 \ifx #1\expandafter \@firstoftwo
 \else \expandafter \@secondoftwo
 \fi
}%
\providecommand \natexlab [1]{#1}%
\providecommand \enquote  [1]{``#1''}%
\providecommand \bibnamefont  [1]{#1}%
\providecommand \bibfnamefont [1]{#1}%
\providecommand \citenamefont [1]{#1}%
\providecommand \href@noop [0]{\@secondoftwo}%
\providecommand \href [0]{\begingroup \@sanitize@url \@href}%
\providecommand \@href[1]{\@@startlink{#1}\@@href}%
\providecommand \@@href[1]{\endgroup#1\@@endlink}%
\providecommand \@sanitize@url [0]{\catcode `\\12\catcode `\$12\catcode
  `\&12\catcode `\#12\catcode `\^12\catcode `\_12\catcode `\%12\relax}%
\providecommand \@@startlink[1]{}%
\providecommand \@@endlink[0]{}%
\providecommand \url  [0]{\begingroup\@sanitize@url \@url }%
\providecommand \@url [1]{\endgroup\@href {#1}{\urlprefix }}%
\providecommand \urlprefix  [0]{URL }%
\providecommand \Eprint [0]{\href }%
\providecommand \doibase [0]{https://doi.org/}%
\providecommand \selectlanguage [0]{\@gobble}%
\providecommand \bibinfo  [0]{\@secondoftwo}%
\providecommand \bibfield  [0]{\@secondoftwo}%
\providecommand \translation [1]{[#1]}%
\providecommand \BibitemOpen [0]{}%
\providecommand \bibitemStop [0]{}%
\providecommand \bibitemNoStop [0]{.\EOS\space}%
\providecommand \EOS [0]{\spacefactor3000\relax}%
\providecommand \BibitemShut  [1]{\csname bibitem#1\endcsname}%
\let\auto@bib@innerbib\@empty
\bibitem [{\citenamefont {Ji}\ and\ \citenamefont {Nazar}(2010)}]{Ji2010rev}%
  \BibitemOpen
  \bibfield  {author} {\bibinfo {author} {\bibfnamefont {X.}~\bibnamefont
  {Ji}}\ and\ \bibinfo {author} {\bibfnamefont {L.~F.}\ \bibnamefont {Nazar}},\
  }\bibfield  {title} {\bibinfo {title} {Advances in li–s batteries},\
  }\href@noop {} {\bibfield  {journal} {\bibinfo  {journal} {J. Mater. Chem.}\
  }\textbf {\bibinfo {volume} {20}},\ \bibinfo {pages} {9821} (\bibinfo {year}
  {2010})}\BibitemShut {NoStop}%
\bibitem [{\citenamefont {Tao}\ \emph {et~al.}(2016)\citenamefont {Tao},
  \citenamefont {Wan}, \citenamefont {Liu}, \citenamefont {Wang}, \citenamefont
  {Yao}, \citenamefont {Zheng}, \citenamefont {Seh}, \citenamefont {Cai},
  \citenamefont {Li}, \citenamefont {Zhou}, \citenamefont {Zu},\ and\
  \citenamefont {Cui}}]{Tao2016}%
  \BibitemOpen
  \bibfield  {author} {\bibinfo {author} {\bibfnamefont {X.}~\bibnamefont
  {Tao}}, \bibinfo {author} {\bibfnamefont {J.}~\bibnamefont {Wan}}, \bibinfo
  {author} {\bibfnamefont {C.}~\bibnamefont {Liu}}, \bibinfo {author}
  {\bibfnamefont {H.}~\bibnamefont {Wang}}, \bibinfo {author} {\bibfnamefont
  {H.}~\bibnamefont {Yao}}, \bibinfo {author} {\bibfnamefont {G.}~\bibnamefont
  {Zheng}}, \bibinfo {author} {\bibfnamefont {Z.~W.}\ \bibnamefont {Seh}},
  \bibinfo {author} {\bibfnamefont {Q.}~\bibnamefont {Cai}}, \bibinfo {author}
  {\bibfnamefont {W.}~\bibnamefont {Li}}, \bibinfo {author} {\bibfnamefont
  {G.}~\bibnamefont {Zhou}}, \bibinfo {author} {\bibfnamefont {C.}~\bibnamefont
  {Zu}},\ and\ \bibinfo {author} {\bibfnamefont {Y.}~\bibnamefont {Cui}},\
  }\bibfield  {title} {\bibinfo {title} {Balancing surface adsorption and
  diffusion of lithium-polysulfides on nonconductive oxides for lithium-sulfur
  battery design},\ }\href {https://doi.org/10.1038/ncomms11203} {\bibfield
  {journal} {\bibinfo  {journal} {Nature Communications}\ }\textbf {\bibinfo
  {volume} {7}},\ \bibinfo {pages} {1} (\bibinfo {year} {2016})}\BibitemShut
  {NoStop}%
\bibitem [{\citenamefont {Ren}\ \emph {et~al.}(2019)\citenamefont {Ren},
  \citenamefont {Ma}, \citenamefont {Zhang},\ and\ \citenamefont
  {Tang}}]{Ren2019}%
  \BibitemOpen
  \bibfield  {author} {\bibinfo {author} {\bibfnamefont {W.}~\bibnamefont
  {Ren}}, \bibinfo {author} {\bibfnamefont {W.}~\bibnamefont {Ma}}, \bibinfo
  {author} {\bibfnamefont {S.}~\bibnamefont {Zhang}},\ and\ \bibinfo {author}
  {\bibfnamefont {B.}~\bibnamefont {Tang}},\ }\bibfield  {title} {\bibinfo
  {title} {Recent advances in shuttle effect inhibition for lithium sulfur
  batteries},\ }\href@noop {} {\bibfield  {journal} {\bibinfo  {journal}
  {Energy Storage Materials}\ }\textbf {\bibinfo {volume} {23}},\ \bibinfo
  {pages} {707 } (\bibinfo {year} {2019})}\BibitemShut {NoStop}%
\bibitem [{\citenamefont {Zhou}\ \emph
  {et~al.}(2020{\natexlab{a}})\citenamefont {Zhou}, \citenamefont {Danilov},
  \citenamefont {Eichel},\ and\ \citenamefont {Notten}}]{Zhou2020rev}%
  \BibitemOpen
  \bibfield  {author} {\bibinfo {author} {\bibfnamefont {L.}~\bibnamefont
  {Zhou}}, \bibinfo {author} {\bibfnamefont {D.~L.}\ \bibnamefont {Danilov}},
  \bibinfo {author} {\bibfnamefont {R.-A.}\ \bibnamefont {Eichel}},\ and\
  \bibinfo {author} {\bibfnamefont {P.~H.~L.}\ \bibnamefont {Notten}},\
  }\bibfield  {title} {\bibinfo {title} {Host materials anchoring polysulfides
  in li–s batteries reviewed},\ }\href@noop {} {\bibfield  {journal}
  {\bibinfo  {journal} {Advanced Energy Materials}\ ,\ \bibinfo {pages}
  {2001304}} (\bibinfo {year} {2020}{\natexlab{a}})}\BibitemShut {NoStop}%
\bibitem [{\citenamefont {Zhang}\ \emph {et~al.}(2019)\citenamefont {Zhang},
  \citenamefont {Liang}, \citenamefont {lei Man}, \citenamefont {Wang},
  \citenamefont {Huang}, \citenamefont {bo~Shu}, \citenamefont {gang Liu},\
  and\ \citenamefont {Wang}}]{Zhang2019Fe}%
  \BibitemOpen
  \bibfield  {author} {\bibinfo {author} {\bibfnamefont {L.}~\bibnamefont
  {Zhang}}, \bibinfo {author} {\bibfnamefont {P.}~\bibnamefont {Liang}},
  \bibinfo {author} {\bibfnamefont {X.}~\bibnamefont {lei Man}}, \bibinfo
  {author} {\bibfnamefont {D.}~\bibnamefont {Wang}}, \bibinfo {author}
  {\bibfnamefont {J.}~\bibnamefont {Huang}}, \bibinfo {author} {\bibfnamefont
  {H.}~\bibnamefont {bo~Shu}}, \bibinfo {author} {\bibfnamefont
  {Z.}~\bibnamefont {gang Liu}},\ and\ \bibinfo {author} {\bibfnamefont
  {L.}~\bibnamefont {Wang}},\ }\bibfield  {title} {\bibinfo {title} {Fe, n
  co-doped graphene as a multi-functional anchor material for lithium-sulfur
  battery},\ }\href@noop {} {\bibfield  {journal} {\bibinfo  {journal} {Journal
  of Physics and Chemistry of Solids}\ }\textbf {\bibinfo {volume} {126}},\
  \bibinfo {pages} {280 } (\bibinfo {year} {2019})}\BibitemShut {NoStop}%
\bibitem [{\citenamefont {Wang}\ \emph {et~al.}(2019)\citenamefont {Wang},
  \citenamefont {Jia}, \citenamefont {Zhong}, \citenamefont {Xiao},
  \citenamefont {Wang}, \citenamefont {Zang}, \citenamefont {Liu},
  \citenamefont {Zheng}, \citenamefont {Luo}, \citenamefont {Yang},
  \citenamefont {Fan}, \citenamefont {Duan}, \citenamefont {Wu}, \citenamefont
  {Lin},\ and\ \citenamefont {Zhang}}]{Wang2019}%
  \BibitemOpen
  \bibfield  {author} {\bibinfo {author} {\bibfnamefont {J.}~\bibnamefont
  {Wang}}, \bibinfo {author} {\bibfnamefont {L.}~\bibnamefont {Jia}}, \bibinfo
  {author} {\bibfnamefont {J.}~\bibnamefont {Zhong}}, \bibinfo {author}
  {\bibfnamefont {Q.}~\bibnamefont {Xiao}}, \bibinfo {author} {\bibfnamefont
  {C.}~\bibnamefont {Wang}}, \bibinfo {author} {\bibfnamefont {K.}~\bibnamefont
  {Zang}}, \bibinfo {author} {\bibfnamefont {H.}~\bibnamefont {Liu}}, \bibinfo
  {author} {\bibfnamefont {H.}~\bibnamefont {Zheng}}, \bibinfo {author}
  {\bibfnamefont {J.}~\bibnamefont {Luo}}, \bibinfo {author} {\bibfnamefont
  {J.}~\bibnamefont {Yang}}, \bibinfo {author} {\bibfnamefont {H.}~\bibnamefont
  {Fan}}, \bibinfo {author} {\bibfnamefont {W.}~\bibnamefont {Duan}}, \bibinfo
  {author} {\bibfnamefont {Y.}~\bibnamefont {Wu}}, \bibinfo {author}
  {\bibfnamefont {H.}~\bibnamefont {Lin}},\ and\ \bibinfo {author}
  {\bibfnamefont {Y.}~\bibnamefont {Zhang}},\ }\bibfield  {title} {\bibinfo
  {title} {Single-atom catalyst boosts electrochemical conversion reactions in
  batteries},\ }\href@noop {} {\bibfield  {journal} {\bibinfo  {journal}
  {Energy Storage Materials}\ }\textbf {\bibinfo {volume} {18}},\ \bibinfo
  {pages} {246 } (\bibinfo {year} {2019})}\BibitemShut {NoStop}%
\bibitem [{\citenamefont {Zeng}\ \emph {et~al.}(2019)\citenamefont {Zeng},
  \citenamefont {Hu}, \citenamefont {Chen},\ and\ \citenamefont
  {Shang}}]{Zeng2019_FeSAC}%
  \BibitemOpen
  \bibfield  {author} {\bibinfo {author} {\bibfnamefont {Q.-W.}\ \bibnamefont
  {Zeng}}, \bibinfo {author} {\bibfnamefont {R.-M.}\ \bibnamefont {Hu}},
  \bibinfo {author} {\bibfnamefont {Z.-B.}\ \bibnamefont {Chen}},\ and\
  \bibinfo {author} {\bibfnamefont {J.-X.}\ \bibnamefont {Shang}},\ }\bibfield
  {title} {\bibinfo {title} {Single-atom fe and n co-doped graphene for
  lithium-sulfur batteries: a density functional theory study},\ }\href@noop {}
  {\bibfield  {journal} {\bibinfo  {journal} {Materials Research Express}\
  }\textbf {\bibinfo {volume} {6}},\ \bibinfo {pages} {095620} (\bibinfo {year}
  {2019})}\BibitemShut {NoStop}%
\bibitem [{\citenamefont {Du}\ \emph {et~al.}(2019)\citenamefont {Du},
  \citenamefont {Chen}, \citenamefont {Hu}, \citenamefont {Chuang},
  \citenamefont {Xie}, \citenamefont {Hu}, \citenamefont {Yan}, \citenamefont
  {Kong}, \citenamefont {Wu}, \citenamefont {Ji},\ and\ \citenamefont
  {Wan}}]{Du2019}%
  \BibitemOpen
  \bibfield  {author} {\bibinfo {author} {\bibfnamefont {Z.}~\bibnamefont
  {Du}}, \bibinfo {author} {\bibfnamefont {X.}~\bibnamefont {Chen}}, \bibinfo
  {author} {\bibfnamefont {W.}~\bibnamefont {Hu}}, \bibinfo {author}
  {\bibfnamefont {C.}~\bibnamefont {Chuang}}, \bibinfo {author} {\bibfnamefont
  {S.}~\bibnamefont {Xie}}, \bibinfo {author} {\bibfnamefont {A.}~\bibnamefont
  {Hu}}, \bibinfo {author} {\bibfnamefont {W.}~\bibnamefont {Yan}}, \bibinfo
  {author} {\bibfnamefont {X.}~\bibnamefont {Kong}}, \bibinfo {author}
  {\bibfnamefont {X.}~\bibnamefont {Wu}}, \bibinfo {author} {\bibfnamefont
  {H.}~\bibnamefont {Ji}},\ and\ \bibinfo {author} {\bibfnamefont {L.-J.}\
  \bibnamefont {Wan}},\ }\bibfield  {title} {\bibinfo {title} {Cobalt in
  nitrogen-doped graphene as single-atom catalyst for high-sulfur content
  lithium–sulfur batteries},\ }\href {https://doi.org/10.1021/jacs.8b12973}
  {\bibfield  {journal} {\bibinfo  {journal} {Journal of the American Chemical
  Society}\ }\textbf {\bibinfo {volume} {141}},\ \bibinfo {pages} {3977}
  (\bibinfo {year} {2019})},\ \bibinfo {note} {pMID: 30764605}\BibitemShut
  {NoStop}%
\bibitem [{\citenamefont {Zhou}\ \emph
  {et~al.}(2020{\natexlab{b}})\citenamefont {Zhou}, \citenamefont {Zhao},
  \citenamefont {Wang}, \citenamefont {Yang}, \citenamefont {Johannessen},
  \citenamefont {Chen}, \citenamefont {Liu}, \citenamefont {Ye}, \citenamefont
  {Wu}, \citenamefont {Peng}, \citenamefont {Liu}, \citenamefont {Jiang},
  \citenamefont {Zhang},\ and\ \citenamefont {Cui}}]{Zhou2020_SAC}%
  \BibitemOpen
  \bibfield  {author} {\bibinfo {author} {\bibfnamefont {G.}~\bibnamefont
  {Zhou}}, \bibinfo {author} {\bibfnamefont {S.}~\bibnamefont {Zhao}}, \bibinfo
  {author} {\bibfnamefont {T.}~\bibnamefont {Wang}}, \bibinfo {author}
  {\bibfnamefont {S.-Z.}\ \bibnamefont {Yang}}, \bibinfo {author}
  {\bibfnamefont {B.}~\bibnamefont {Johannessen}}, \bibinfo {author}
  {\bibfnamefont {H.}~\bibnamefont {Chen}}, \bibinfo {author} {\bibfnamefont
  {C.}~\bibnamefont {Liu}}, \bibinfo {author} {\bibfnamefont {Y.}~\bibnamefont
  {Ye}}, \bibinfo {author} {\bibfnamefont {Y.}~\bibnamefont {Wu}}, \bibinfo
  {author} {\bibfnamefont {Y.}~\bibnamefont {Peng}}, \bibinfo {author}
  {\bibfnamefont {C.}~\bibnamefont {Liu}}, \bibinfo {author} {\bibfnamefont
  {S.~P.}\ \bibnamefont {Jiang}}, \bibinfo {author} {\bibfnamefont
  {Q.}~\bibnamefont {Zhang}},\ and\ \bibinfo {author} {\bibfnamefont
  {Y.}~\bibnamefont {Cui}},\ }\bibfield  {title} {\bibinfo {title} {Theoretical
  calculation guided design of single-atom catalysts toward fast kinetic and
  long-life li–s batteries},\ }\href
  {https://doi.org/10.1021/acs.nanolett.9b04719} {\bibfield  {journal}
  {\bibinfo  {journal} {Nano Letters}\ }\textbf {\bibinfo {volume} {20}},\
  \bibinfo {pages} {1252} (\bibinfo {year} {2020}{\natexlab{b}})},\ \bibinfo
  {note} {pMID: 31887051}\BibitemShut {NoStop}%
\bibitem [{\citenamefont {Wang}\ \emph {et~al.}(2020)\citenamefont {Wang},
  \citenamefont {Song}, \citenamefont {Yu}, \citenamefont {Ullah},
  \citenamefont {Guan}, \citenamefont {Chu}, \citenamefont {Zhang},
  \citenamefont {Zhao}, \citenamefont {Li},\ and\ \citenamefont
  {Liu}}]{WangC2020}%
  \BibitemOpen
  \bibfield  {author} {\bibinfo {author} {\bibfnamefont {C.}~\bibnamefont
  {Wang}}, \bibinfo {author} {\bibfnamefont {H.}~\bibnamefont {Song}}, \bibinfo
  {author} {\bibfnamefont {C.}~\bibnamefont {Yu}}, \bibinfo {author}
  {\bibfnamefont {Z.}~\bibnamefont {Ullah}}, \bibinfo {author} {\bibfnamefont
  {Z.}~\bibnamefont {Guan}}, \bibinfo {author} {\bibfnamefont {R.}~\bibnamefont
  {Chu}}, \bibinfo {author} {\bibfnamefont {Y.}~\bibnamefont {Zhang}}, \bibinfo
  {author} {\bibfnamefont {L.}~\bibnamefont {Zhao}}, \bibinfo {author}
  {\bibfnamefont {Q.}~\bibnamefont {Li}},\ and\ \bibinfo {author}
  {\bibfnamefont {L.}~\bibnamefont {Liu}},\ }\bibfield  {title} {\bibinfo
  {title} {Iron single-atom catalyst anchored on nitrogen-rich mof-derived
  carbon nanocage to accelerate polysulfide redox conversion for lithium sulfur
  batteries},\ }\href {https://doi.org/10.1039/C9TA11680J} {\bibfield
  {journal} {\bibinfo  {journal} {J. Mater. Chem. A}\ }\textbf {\bibinfo
  {volume} {8}},\ \bibinfo {pages} {3421} (\bibinfo {year} {2020})}\BibitemShut
  {NoStop}%
\bibitem [{\citenamefont {Li}\ \emph {et~al.}(2020)\citenamefont {Li},
  \citenamefont {Wu}, \citenamefont {Zhang}, \citenamefont {Wang},
  \citenamefont {Zhang}, \citenamefont {Seh}, \citenamefont {Zhang},
  \citenamefont {Sun}, \citenamefont {Huang}, \citenamefont {Jiang},
  \citenamefont {Zhou},\ and\ \citenamefont {Sun}}]{Li2020}%
  \BibitemOpen
  \bibfield  {author} {\bibinfo {author} {\bibfnamefont {Y.}~\bibnamefont
  {Li}}, \bibinfo {author} {\bibfnamefont {J.}~\bibnamefont {Wu}}, \bibinfo
  {author} {\bibfnamefont {B.}~\bibnamefont {Zhang}}, \bibinfo {author}
  {\bibfnamefont {W.}~\bibnamefont {Wang}}, \bibinfo {author} {\bibfnamefont
  {G.}~\bibnamefont {Zhang}}, \bibinfo {author} {\bibfnamefont {Z.~W.}\
  \bibnamefont {Seh}}, \bibinfo {author} {\bibfnamefont {N.}~\bibnamefont
  {Zhang}}, \bibinfo {author} {\bibfnamefont {J.}~\bibnamefont {Sun}}, \bibinfo
  {author} {\bibfnamefont {L.}~\bibnamefont {Huang}}, \bibinfo {author}
  {\bibfnamefont {J.}~\bibnamefont {Jiang}}, \bibinfo {author} {\bibfnamefont
  {J.}~\bibnamefont {Zhou}},\ and\ \bibinfo {author} {\bibfnamefont
  {Y.}~\bibnamefont {Sun}},\ }\bibfield  {title} {\bibinfo {title} {Fast
  conversion and controlled deposition of lithium (poly)sulfides in
  lithium-sulfur batteries using high-loading cobalt single atoms},\
  }\href@noop {} {\bibfield  {journal} {\bibinfo  {journal} {Energy Storage
  Materials}\ }\textbf {\bibinfo {volume} {30}},\ \bibinfo {pages} {250 }
  (\bibinfo {year} {2020})}\BibitemShut {NoStop}%
\bibitem [{\citenamefont {Shao}\ \emph {et~al.}(2020)\citenamefont {Shao},
  \citenamefont {Xu}, \citenamefont {Guo}, \citenamefont {Su},\ and\
  \citenamefont {Chen}}]{Shao2020}%
  \BibitemOpen
  \bibfield  {author} {\bibinfo {author} {\bibfnamefont {Q.}~\bibnamefont
  {Shao}}, \bibinfo {author} {\bibfnamefont {L.}~\bibnamefont {Xu}}, \bibinfo
  {author} {\bibfnamefont {D.}~\bibnamefont {Guo}}, \bibinfo {author}
  {\bibfnamefont {Y.}~\bibnamefont {Su}},\ and\ \bibinfo {author}
  {\bibfnamefont {J.}~\bibnamefont {Chen}},\ }\bibfield  {title} {\bibinfo
  {title} {Atomic level design of single iron atom embedded mesoporous hollow
  carbon spheres as multi-effect nanoreactors for advanced lithium–sulfur
  batteries},\ }\href@noop {} {\bibfield  {journal} {\bibinfo  {journal} {J.
  Mater. Chem. A}\ }\textbf {\bibinfo {volume} {8}},\ \bibinfo {pages} {23772}
  (\bibinfo {year} {2020})}\BibitemShut {NoStop}%
\bibitem [{\citenamefont {Andritsos}\ \emph {et~al.}(2021)\citenamefont
  {Andritsos}, \citenamefont {Lekakou},\ and\ \citenamefont
  {Cai}}]{Andritsos2021}%
  \BibitemOpen
  \bibfield  {author} {\bibinfo {author} {\bibfnamefont {E.~I.}\ \bibnamefont
  {Andritsos}}, \bibinfo {author} {\bibfnamefont {C.}~\bibnamefont {Lekakou}},\
  and\ \bibinfo {author} {\bibfnamefont {Q.}~\bibnamefont {Cai}},\ }\bibfield
  {title} {\bibinfo {title} {Single-atom catalysts as promising cathode
  materials for lithium–sulfur batteries},\ }\href
  {https://doi.org/10.1021/acs.jpcc.1c04491} {\bibfield  {journal} {\bibinfo
  {journal} {The Journal of Physical Chemistry C}\ }\textbf {\bibinfo {volume}
  {125}},\ \bibinfo {pages} {18108} (\bibinfo {year} {2021})}\BibitemShut
  {NoStop}%
\bibitem [{\citenamefont {Jand}\ \emph {et~al.}(2016)\citenamefont {Jand},
  \citenamefont {Chen},\ and\ \citenamefont {Kaghazchi}}]{Jand2016}%
  \BibitemOpen
  \bibfield  {author} {\bibinfo {author} {\bibfnamefont {S.~P.}\ \bibnamefont
  {Jand}}, \bibinfo {author} {\bibfnamefont {Y.}~\bibnamefont {Chen}},\ and\
  \bibinfo {author} {\bibfnamefont {P.}~\bibnamefont {Kaghazchi}},\ }\bibfield
  {title} {\bibinfo {title} {Comparative theoretical study of adsorption of
  lithium polysulfides (li2sx) on pristine and defective graphene},\
  }\href@noop {} {\bibfield  {journal} {\bibinfo  {journal} {Journal of Power
  Sources}\ }\textbf {\bibinfo {volume} {308}},\ \bibinfo {pages} {166 }
  (\bibinfo {year} {2016})}\BibitemShut {NoStop}%
\bibitem [{\citenamefont {Chen}\ \emph {et~al.}(2020)\citenamefont {Chen},
  \citenamefont {Zuo}, \citenamefont {Ye}, \citenamefont {Li}, \citenamefont
  {Deng},\ and\ \citenamefont {Ong}}]{Chi2020_rev}%
  \BibitemOpen
  \bibfield  {author} {\bibinfo {author} {\bibfnamefont {C.}~\bibnamefont
  {Chen}}, \bibinfo {author} {\bibfnamefont {Y.}~\bibnamefont {Zuo}}, \bibinfo
  {author} {\bibfnamefont {W.}~\bibnamefont {Ye}}, \bibinfo {author}
  {\bibfnamefont {X.}~\bibnamefont {Li}}, \bibinfo {author} {\bibfnamefont
  {Z.}~\bibnamefont {Deng}},\ and\ \bibinfo {author} {\bibfnamefont {S.~P.}\
  \bibnamefont {Ong}},\ }\bibfield  {title} {\bibinfo {title} {A critical
  review of machine learning of energy materials},\ }\href@noop {} {\bibfield
  {journal} {\bibinfo  {journal} {Advanced Energy Materials}\ }\textbf
  {\bibinfo {volume} {10}},\ \bibinfo {pages} {1903242} (\bibinfo {year}
  {2020})}\BibitemShut {NoStop}%
\bibitem [{\citenamefont {Liu}\ \emph {et~al.}(2021)\citenamefont {Liu},
  \citenamefont {Esan}, \citenamefont {Pan},\ and\ \citenamefont
  {An}}]{Liu2021_rev}%
  \BibitemOpen
  \bibfield  {author} {\bibinfo {author} {\bibfnamefont {Y.}~\bibnamefont
  {Liu}}, \bibinfo {author} {\bibfnamefont {O.~C.}\ \bibnamefont {Esan}},
  \bibinfo {author} {\bibfnamefont {Z.}~\bibnamefont {Pan}},\ and\ \bibinfo
  {author} {\bibfnamefont {L.}~\bibnamefont {An}},\ }\bibfield  {title}
  {\bibinfo {title} {Machine learning for advanced energy materials},\
  }\href@noop {} {\bibfield  {journal} {\bibinfo  {journal} {Energy and AI}\
  }\textbf {\bibinfo {volume} {3}},\ \bibinfo {pages} {100049} (\bibinfo {year}
  {2021})}\BibitemShut {NoStop}%
\bibitem [{\citenamefont {Rossi}\ \emph {et~al.}(2020)\citenamefont {Rossi},
  \citenamefont {Asara},\ and\ \citenamefont {Baletto}}]{Rossi2020}%
  \BibitemOpen
  \bibfield  {author} {\bibinfo {author} {\bibfnamefont {K.}~\bibnamefont
  {Rossi}}, \bibinfo {author} {\bibfnamefont {G.}~\bibnamefont {Asara}},\ and\
  \bibinfo {author} {\bibfnamefont {F.}~\bibnamefont {Baletto}},\ }\bibfield
  {title} {\bibinfo {title} {Structural screening and design of platinum
  nanosamples for oxygen reduction},\ }\href@noop {} {\bibfield  {journal}
  {\bibinfo  {journal} {Acs Catalysis}\ }\textbf {\bibinfo {volume} {10 (6)}},\
  \bibinfo {pages} {3911} (\bibinfo {year} {2020})}\BibitemShut {NoStop}%
\bibitem [{\citenamefont {Tran}\ and\ \citenamefont {Ulissi}(2018)}]{Tran2018}%
  \BibitemOpen
  \bibfield  {author} {\bibinfo {author} {\bibfnamefont {K.}~\bibnamefont
  {Tran}}\ and\ \bibinfo {author} {\bibfnamefont {Z.~W.}\ \bibnamefont
  {Ulissi}},\ }\bibfield  {title} {\bibinfo {title} {Active learning across
  intermetallics to guide discovery of electrocatalysts for co 2 reduction and
  h 2 evolution},\ }\href@noop {} {\bibfield  {journal} {\bibinfo  {journal}
  {Nature Catalysis}\ }\textbf {\bibinfo {volume} {1 (9)}},\ \bibinfo {pages}
  {696} (\bibinfo {year} {2018})}\BibitemShut {NoStop}%
\bibitem [{\citenamefont {Gu}\ \emph {et~al.}(2018)\citenamefont {Gu},
  \citenamefont {Plechac},\ and\ \citenamefont {Vlachos}}]{Gu2018}%
  \BibitemOpen
  \bibfield  {author} {\bibinfo {author} {\bibfnamefont {G.~H.}\ \bibnamefont
  {Gu}}, \bibinfo {author} {\bibfnamefont {P.}~\bibnamefont {Plechac}},\ and\
  \bibinfo {author} {\bibfnamefont {D.~G.}\ \bibnamefont {Vlachos}},\
  }\bibfield  {title} {\bibinfo {title} {Thermochemistry of gas-phase and
  surface species via lasso-assisted subgraph selection},\ }\href
  {https://doi.org/10.1039/C7RE00210F} {\bibfield  {journal} {\bibinfo
  {journal} {React. Chem. Eng.}\ }\textbf {\bibinfo {volume} {3}},\ \bibinfo
  {pages} {454} (\bibinfo {year} {2018})}\BibitemShut {NoStop}%
\bibitem [{\citenamefont {Robinson}\ \emph {et~al.}(2021)\citenamefont
  {Robinson}, \citenamefont {Xi}, \citenamefont {Kumar}, \citenamefont
  {Ferrari}, \citenamefont {Au}, \citenamefont {Titirici}, \citenamefont
  {Parra-Puerto}, \citenamefont {Kucernak}, \citenamefont {Fitch},
  \citenamefont {Garcia-Araez}, \citenamefont {Brown}, \citenamefont {Pasta},
  \citenamefont {Furness}, \citenamefont {Kibler}, \citenamefont {Walsh},
  \citenamefont {Johnson}, \citenamefont {Holc}, \citenamefont {Newton},
  \citenamefont {Champness}, \citenamefont {Markoulidis}, \citenamefont
  {Crean}, \citenamefont {Slade}, \citenamefont {Andritsos}, \citenamefont
  {Cai}, \citenamefont {Babar}, \citenamefont {Zhang}, \citenamefont {Lekakou},
  \citenamefont {Kulkarni}, \citenamefont {Rettie}, \citenamefont {Jervis},
  \citenamefont {Cornish}, \citenamefont {Marinescu}, \citenamefont {Offer},
  \citenamefont {Li}, \citenamefont {Bird}, \citenamefont {Grey}, \citenamefont
  {Chhowalla}, \citenamefont {Lecce}, \citenamefont {Owen}, \citenamefont
  {Miller}, \citenamefont {Brett}, \citenamefont {Liatard}, \citenamefont
  {Ainsworth},\ and\ \citenamefont {Shearing}}]{AndritsosRoadmap}%
  \BibitemOpen
  \bibfield  {author} {\bibinfo {author} {\bibfnamefont {J.~B.}\ \bibnamefont
  {Robinson}}, \bibinfo {author} {\bibfnamefont {K.}~\bibnamefont {Xi}},
  \bibinfo {author} {\bibfnamefont {R.~V.}\ \bibnamefont {Kumar}}, \bibinfo
  {author} {\bibfnamefont {A.~C.}\ \bibnamefont {Ferrari}}, \bibinfo {author}
  {\bibfnamefont {H.}~\bibnamefont {Au}}, \bibinfo {author} {\bibfnamefont
  {M.-M.}\ \bibnamefont {Titirici}}, \bibinfo {author} {\bibfnamefont
  {A.}~\bibnamefont {Parra-Puerto}}, \bibinfo {author} {\bibfnamefont
  {A.}~\bibnamefont {Kucernak}}, \bibinfo {author} {\bibfnamefont {S.~D.~S.}\
  \bibnamefont {Fitch}}, \bibinfo {author} {\bibfnamefont {N.}~\bibnamefont
  {Garcia-Araez}}, \bibinfo {author} {\bibfnamefont {Z.~L.}\ \bibnamefont
  {Brown}}, \bibinfo {author} {\bibfnamefont {M.}~\bibnamefont {Pasta}},
  \bibinfo {author} {\bibfnamefont {L.}~\bibnamefont {Furness}}, \bibinfo
  {author} {\bibfnamefont {A.~J.}\ \bibnamefont {Kibler}}, \bibinfo {author}
  {\bibfnamefont {D.~A.}\ \bibnamefont {Walsh}}, \bibinfo {author}
  {\bibfnamefont {L.~R.}\ \bibnamefont {Johnson}}, \bibinfo {author}
  {\bibfnamefont {C.}~\bibnamefont {Holc}}, \bibinfo {author} {\bibfnamefont
  {G.~N.}\ \bibnamefont {Newton}}, \bibinfo {author} {\bibfnamefont {N.~R.}\
  \bibnamefont {Champness}}, \bibinfo {author} {\bibfnamefont {F.}~\bibnamefont
  {Markoulidis}}, \bibinfo {author} {\bibfnamefont {C.}~\bibnamefont {Crean}},
  \bibinfo {author} {\bibfnamefont {R.~C.~T.}\ \bibnamefont {Slade}}, \bibinfo
  {author} {\bibfnamefont {E.~I.}\ \bibnamefont {Andritsos}}, \bibinfo {author}
  {\bibfnamefont {Q.}~\bibnamefont {Cai}}, \bibinfo {author} {\bibfnamefont
  {S.}~\bibnamefont {Babar}}, \bibinfo {author} {\bibfnamefont
  {T.}~\bibnamefont {Zhang}}, \bibinfo {author} {\bibfnamefont
  {C.}~\bibnamefont {Lekakou}}, \bibinfo {author} {\bibfnamefont
  {N.}~\bibnamefont {Kulkarni}}, \bibinfo {author} {\bibfnamefont {A.~J.~E.}\
  \bibnamefont {Rettie}}, \bibinfo {author} {\bibfnamefont {R.}~\bibnamefont
  {Jervis}}, \bibinfo {author} {\bibfnamefont {M.}~\bibnamefont {Cornish}},
  \bibinfo {author} {\bibfnamefont {M.}~\bibnamefont {Marinescu}}, \bibinfo
  {author} {\bibfnamefont {G.}~\bibnamefont {Offer}}, \bibinfo {author}
  {\bibfnamefont {Z.}~\bibnamefont {Li}}, \bibinfo {author} {\bibfnamefont
  {L.}~\bibnamefont {Bird}}, \bibinfo {author} {\bibfnamefont {C.~P.}\
  \bibnamefont {Grey}}, \bibinfo {author} {\bibfnamefont {M.}~\bibnamefont
  {Chhowalla}}, \bibinfo {author} {\bibfnamefont {D.~D.}\ \bibnamefont
  {Lecce}}, \bibinfo {author} {\bibfnamefont {R.~E.}\ \bibnamefont {Owen}},
  \bibinfo {author} {\bibfnamefont {T.~S.}\ \bibnamefont {Miller}}, \bibinfo
  {author} {\bibfnamefont {D.~J.~L.}\ \bibnamefont {Brett}}, \bibinfo {author}
  {\bibfnamefont {S.}~\bibnamefont {Liatard}}, \bibinfo {author} {\bibfnamefont
  {D.}~\bibnamefont {Ainsworth}},\ and\ \bibinfo {author} {\bibfnamefont
  {P.~R.}\ \bibnamefont {Shearing}},\ }\bibfield  {title} {\bibinfo {title}
  {2021 roadmap on lithium sulfur batteries},\ }\href@noop {} {\bibfield
  {journal} {\bibinfo  {journal} {Journal of Physics: Energy}\ }\textbf
  {\bibinfo {volume} {3}},\ \bibinfo {pages} {031501} (\bibinfo {year}
  {2021})}\BibitemShut {NoStop}%
\bibitem [{\citenamefont {Kilic}\ \emph {et~al.}(2020)\citenamefont {Kilic},
  \citenamefont {Çağla Odabaşı}, \citenamefont {Yildirim},\ and\
  \citenamefont {Eroglu}}]{Kilic2020}%
  \BibitemOpen
  \bibfield  {author} {\bibinfo {author} {\bibfnamefont {A.}~\bibnamefont
  {Kilic}}, \bibinfo {author} {\bibnamefont {Çağla Odabaşı}}, \bibinfo
  {author} {\bibfnamefont {R.}~\bibnamefont {Yildirim}},\ and\ \bibinfo
  {author} {\bibfnamefont {D.}~\bibnamefont {Eroglu}},\ }\bibfield  {title}
  {\bibinfo {title} {Assessment of critical materials and cell design factors
  for high performance lithium-sulfur batteries using machine learning},\
  }\href@noop {} {\bibfield  {journal} {\bibinfo  {journal} {Chemical
  Engineering Journal}\ }\textbf {\bibinfo {volume} {390}},\ \bibinfo {pages}
  {124117} (\bibinfo {year} {2020})}\BibitemShut {NoStop}%
\bibitem [{\citenamefont {Zhang}\ \emph {et~al.}(2021)\citenamefont {Zhang},
  \citenamefont {Wang}, \citenamefont {Ren}, \citenamefont {Liu},\ and\
  \citenamefont {Li}}]{Zhang2021}%
  \BibitemOpen
  \bibfield  {author} {\bibinfo {author} {\bibfnamefont {H.}~\bibnamefont
  {Zhang}}, \bibinfo {author} {\bibfnamefont {Z.}~\bibnamefont {Wang}},
  \bibinfo {author} {\bibfnamefont {J.}~\bibnamefont {Ren}}, \bibinfo {author}
  {\bibfnamefont {J.}~\bibnamefont {Liu}},\ and\ \bibinfo {author}
  {\bibfnamefont {J.}~\bibnamefont {Li}},\ }\bibfield  {title} {\bibinfo
  {title} {Ultra-fast and accurate binding energy prediction of shuttle
  effect-suppressive sulfur hosts for lithium-sulfur batteries using machine
  learning},\ }\href@noop {} {\bibfield  {journal} {\bibinfo  {journal} {Energy
  Storage Materials}\ }\textbf {\bibinfo {volume} {35}},\ \bibinfo {pages} {88}
  (\bibinfo {year} {2021})}\BibitemShut {NoStop}%
\bibitem [{\citenamefont {Lian}\ \emph {et~al.}(2021)\citenamefont {Lian},
  \citenamefont {Yang}, \citenamefont {Jan},\ and\ \citenamefont
  {Li}}]{Lian2021}%
  \BibitemOpen
  \bibfield  {author} {\bibinfo {author} {\bibfnamefont {Z.}~\bibnamefont
  {Lian}}, \bibinfo {author} {\bibfnamefont {M.}~\bibnamefont {Yang}}, \bibinfo
  {author} {\bibfnamefont {F.}~\bibnamefont {Jan}},\ and\ \bibinfo {author}
  {\bibfnamefont {B.}~\bibnamefont {Li}},\ }\bibfield  {title} {\bibinfo
  {title} {Machine learning derived blueprint for rational design of the
  effective single-atom cathode catalyst of the lithium–sulfur battery},\
  }\href {https://doi.org/10.1021/acs.jpclett.1c00927} {\bibfield  {journal}
  {\bibinfo  {journal} {The Journal of Physical Chemistry Letters}\ }\textbf
  {\bibinfo {volume} {12}},\ \bibinfo {pages} {7053} (\bibinfo {year}
  {2021})},\ \bibinfo {note} {pMID: 34291938}\BibitemShut {NoStop}%
\bibitem [{\citenamefont {Bart\'ok}\ \emph {et~al.}(2013)\citenamefont
  {Bart\'ok}, \citenamefont {Kondor},\ and\ \citenamefont {Cs\'anyi}}]{SOAP}%
  \BibitemOpen
  \bibfield  {author} {\bibinfo {author} {\bibfnamefont {A.~P.}\ \bibnamefont
  {Bart\'ok}}, \bibinfo {author} {\bibfnamefont {R.}~\bibnamefont {Kondor}},\
  and\ \bibinfo {author} {\bibfnamefont {G.}~\bibnamefont {Cs\'anyi}},\
  }\bibfield  {title} {\bibinfo {title} {On representing chemical
  environments},\ }\href@noop {} {\bibfield  {journal} {\bibinfo  {journal}
  {Phys. Rev. B}\ }\textbf {\bibinfo {volume} {87}},\ \bibinfo {pages} {184115}
  (\bibinfo {year} {2013})}\BibitemShut {NoStop}%
\bibitem [{\citenamefont {Imbalzano}\ \emph {et~al.}(2018)\citenamefont
  {Imbalzano}, \citenamefont {Anelli}, \citenamefont {Giofré}, \citenamefont
  {Klees}, \citenamefont {Behler},\ and\ \citenamefont
  {Ceriotti}}]{Imbalzano2018}%
  \BibitemOpen
  \bibfield  {author} {\bibinfo {author} {\bibfnamefont {G.}~\bibnamefont
  {Imbalzano}}, \bibinfo {author} {\bibfnamefont {A.}~\bibnamefont {Anelli}},
  \bibinfo {author} {\bibfnamefont {D.}~\bibnamefont {Giofré}}, \bibinfo
  {author} {\bibfnamefont {S.}~\bibnamefont {Klees}}, \bibinfo {author}
  {\bibfnamefont {J.}~\bibnamefont {Behler}},\ and\ \bibinfo {author}
  {\bibfnamefont {M.}~\bibnamefont {Ceriotti}},\ }\bibfield  {title} {\bibinfo
  {title} {Automatic selection of atomic fingerprints and reference
  configurations for machine-learning potentials},\ }\bibfield  {journal}
  {\bibinfo  {journal} {Journal of Chemical Physics}\ }\textbf {\bibinfo
  {volume} {148}},\ \href {https://doi.org/10.1063/1.5024611}
  {10.1063/1.5024611} (\bibinfo {year} {2018})\BibitemShut {NoStop}%
\bibitem [{\citenamefont {Vovk}(2013)}]{KRR}%
  \BibitemOpen
  \bibfield  {author} {\bibinfo {author} {\bibfnamefont {V.}~\bibnamefont
  {Vovk}},\ }\bibinfo {title} {Kernel ridge regression},\ in\ \href@noop {}
  {\emph {\bibinfo {booktitle} {Empirical Inference: Festschrift in Honor of
  Vladimir N. Vapnik}}}\ (\bibinfo  {publisher} {Springer},\ \bibinfo {year}
  {2013})\ pp.\ \bibinfo {pages} {105--116}\BibitemShut {NoStop}%
\bibitem [{\citenamefont {Pearson}(192)}]{Pearson1901}%
  \BibitemOpen
  \bibfield  {author} {\bibinfo {author} {\bibfnamefont {K.}~\bibnamefont
  {Pearson}},\ }\bibfield  {title} {\bibinfo {title} {n lines and planes of
  closest fit to systems of points in space},\ }\href@noop {} {\bibfield
  {journal} {\bibinfo  {journal} {Philosophical Magazine}\ }\textbf {\bibinfo
  {volume} {2 (11)}},\ \bibinfo {pages} {559 } (\bibinfo {year}
  {192})}\BibitemShut {NoStop}%
\bibitem [{\citenamefont {Anelli}\ \emph {et~al.}(2018)\citenamefont {Anelli},
  \citenamefont {Engel}, \citenamefont {Pickard},\ and\ \citenamefont
  {Ceriotti}}]{Anelli2018}%
  \BibitemOpen
  \bibfield  {author} {\bibinfo {author} {\bibfnamefont {A.}~\bibnamefont
  {Anelli}}, \bibinfo {author} {\bibfnamefont {E.~A.}\ \bibnamefont {Engel}},
  \bibinfo {author} {\bibfnamefont {C.~J.}\ \bibnamefont {Pickard}},\ and\
  \bibinfo {author} {\bibfnamefont {M.}~\bibnamefont {Ceriotti}},\ }\bibfield
  {title} {\bibinfo {title} {Generalized convex hull construction for materials
  discovery},\ }\bibfield  {journal} {\bibinfo  {journal} {Physical Review
  Materials}\ }\textbf {\bibinfo {volume} {2}},\ \href
  {https://doi.org/10.1103/PhysRevMaterials.2.103804}
  {10.1103/PhysRevMaterials.2.103804} (\bibinfo {year} {2018})\BibitemShut
  {NoStop}%
\bibitem [{\citenamefont {Clark}\ \emph {et~al.}(2005)\citenamefont {Clark},
  \citenamefont {Segall}, \citenamefont {Pickard}, \citenamefont {Hasnip},
  \citenamefont {Probert}, \citenamefont {Refson},\ and\ \citenamefont
  {Payne}}]{CASTEP}%
  \BibitemOpen
  \bibfield  {author} {\bibinfo {author} {\bibfnamefont {S.~J.}\ \bibnamefont
  {Clark}}, \bibinfo {author} {\bibfnamefont {M.~D.}\ \bibnamefont {Segall}},
  \bibinfo {author} {\bibfnamefont {C.~J.}\ \bibnamefont {Pickard}}, \bibinfo
  {author} {\bibfnamefont {P.~J.}\ \bibnamefont {Hasnip}}, \bibinfo {author}
  {\bibfnamefont {M.~I.~J.}\ \bibnamefont {Probert}}, \bibinfo {author}
  {\bibfnamefont {K.}~\bibnamefont {Refson}},\ and\ \bibinfo {author}
  {\bibfnamefont {M.~C.}\ \bibnamefont {Payne}},\ }\bibfield  {title} {\bibinfo
  {title} {First principles methods using castep},\ }\href@noop {} {\bibfield
  {journal} {\bibinfo  {journal} {Zeitschrift für Kristallographie -
  Crystalline Materials}\ }\textbf {\bibinfo {volume} {220}},\ \bibinfo {pages}
  {567 } (\bibinfo {year} {2005})}\BibitemShut {NoStop}%
\bibitem [{\citenamefont {Grimme}(2006)}]{Grimme2006}%
  \BibitemOpen
  \bibfield  {author} {\bibinfo {author} {\bibfnamefont {S.}~\bibnamefont
  {Grimme}},\ }\bibfield  {title} {\bibinfo {title} {Semiempirical gga-type
  density functional constructed with a long-range dispersion correction},\
  }\href@noop {} {\bibfield  {journal} {\bibinfo  {journal} {Journal of
  Computational Chemistry}\ }\textbf {\bibinfo {volume} {27}},\ \bibinfo
  {pages} {1787} (\bibinfo {year} {2006})}\BibitemShut {NoStop}%
\bibitem [{\citenamefont {Himanen}\ \emph {et~al.}(2020)\citenamefont
  {Himanen}, \citenamefont {J{\"a}ger}, \citenamefont {Morooka}, \citenamefont
  {Federici~Canova}, \citenamefont {Ranawat}, \citenamefont {Gao},
  \citenamefont {Rinke},\ and\ \citenamefont {Foster}}]{dscribe}%
  \BibitemOpen
  \bibfield  {author} {\bibinfo {author} {\bibfnamefont {L.}~\bibnamefont
  {Himanen}}, \bibinfo {author} {\bibfnamefont {M.~O.~J.}\ \bibnamefont
  {J{\"a}ger}}, \bibinfo {author} {\bibfnamefont {E.~V.}\ \bibnamefont
  {Morooka}}, \bibinfo {author} {\bibfnamefont {F.}~\bibnamefont
  {Federici~Canova}}, \bibinfo {author} {\bibfnamefont {Y.~S.}\ \bibnamefont
  {Ranawat}}, \bibinfo {author} {\bibfnamefont {D.~Z.}\ \bibnamefont {Gao}},
  \bibinfo {author} {\bibfnamefont {P.}~\bibnamefont {Rinke}},\ and\ \bibinfo
  {author} {\bibfnamefont {A.~S.}\ \bibnamefont {Foster}},\ }\bibfield  {title}
  {\bibinfo {title} {{DScribe: Library of descriptors for machine learning in
  materials science}},\ }\href {https://doi.org/10.1016/j.cpc.2019.106949}
  {\bibfield  {journal} {\bibinfo  {journal} {Computer Physics Communications}\
  }\textbf {\bibinfo {volume} {247}},\ \bibinfo {pages} {106949} (\bibinfo
  {year} {2020})}\BibitemShut {NoStop}%
\bibitem [{\citenamefont {De}\ \emph {et~al.}(2016)\citenamefont {De},
  \citenamefont {Bartók}, \citenamefont {Csányi},\ and\ \citenamefont
  {Ceriotti}}]{de2016}%
  \BibitemOpen
  \bibfield  {author} {\bibinfo {author} {\bibfnamefont {S.}~\bibnamefont
  {De}}, \bibinfo {author} {\bibfnamefont {A.~P.}\ \bibnamefont {Bartók}},
  \bibinfo {author} {\bibfnamefont {G.}~\bibnamefont {Csányi}},\ and\ \bibinfo
  {author} {\bibfnamefont {M.}~\bibnamefont {Ceriotti}},\ }\bibfield  {title}
  {\bibinfo {title} {Comparing molecules and solids across structural and
  alchemical space},\ }\href {https://doi.org/10.1039/C6CP00415F} {\bibfield
  {journal} {\bibinfo  {journal} {Phys. Chem. Chem. Phys.}\ }\textbf {\bibinfo
  {volume} {18}},\ \bibinfo {pages} {13754} (\bibinfo {year}
  {2016})}\BibitemShut {NoStop}%
\bibitem [{\citenamefont {Bartok}\ \emph {et~al.}(2017)\citenamefont {Bartok},
  \citenamefont {De}, \citenamefont {Poelking}, \citenamefont {Bernstein},
  \citenamefont {Kermode}, \citenamefont {Csanyi},\ and\ \citenamefont
  {Ceriotti}}]{Bartok2018}%
  \BibitemOpen
  \bibfield  {author} {\bibinfo {author} {\bibfnamefont {A.~P.}\ \bibnamefont
  {Bartok}}, \bibinfo {author} {\bibfnamefont {S.}~\bibnamefont {De}}, \bibinfo
  {author} {\bibfnamefont {C.}~\bibnamefont {Poelking}}, \bibinfo {author}
  {\bibfnamefont {N.}~\bibnamefont {Bernstein}}, \bibinfo {author}
  {\bibfnamefont {J.}~\bibnamefont {Kermode}}, \bibinfo {author} {\bibfnamefont
  {G.}~\bibnamefont {Csanyi}},\ and\ \bibinfo {author} {\bibfnamefont
  {M.}~\bibnamefont {Ceriotti}},\ }\bibfield  {title} {\bibinfo {title}
  {Machine learning unifies the modelling of materials and molecules},\
  }\href@noop {} {\bibfield  {journal} {\bibinfo  {journal} {Science Advances}\
  }\textbf {\bibinfo {volume} {3 (12)}},\ \bibinfo {pages} {e1701816} (\bibinfo
  {year} {2017})}\BibitemShut {NoStop}%
\bibitem [{\citenamefont {Deringer}\ \emph {et~al.}(2020)\citenamefont
  {Deringer}, \citenamefont {Caro},\ and\ \citenamefont
  {Csányi}}]{Deringer2020}%
  \BibitemOpen
  \bibfield  {author} {\bibinfo {author} {\bibfnamefont {V.~L.}\ \bibnamefont
  {Deringer}}, \bibinfo {author} {\bibfnamefont {M.~A.}\ \bibnamefont {Caro}},\
  and\ \bibinfo {author} {\bibfnamefont {G.}~\bibnamefont {Csányi}},\
  }\bibfield  {title} {\bibinfo {title} {A general-purpose machine-learning
  force field for bulk and nanostructured phosphorus},\ }\href@noop {}
  {\bibfield  {journal} {\bibinfo  {journal} {Nature Communications}\ }\textbf
  {\bibinfo {volume} {11 (1)}},\ \bibinfo {pages} {1} (\bibinfo {year}
  {2020})}\BibitemShut {NoStop}%
\bibitem [{\citenamefont {Deringer}\ \emph {et~al.}(2021)\citenamefont
  {Deringer}, \citenamefont {Bernstein}, \citenamefont {Csányi}, \citenamefont
  {Mahmoud}, \citenamefont {Ceriotti}, \citenamefont {Wilson}, \citenamefont
  {Drabold},\ and\ \citenamefont {Elliott}}]{Deringer2021}%
  \BibitemOpen
  \bibfield  {author} {\bibinfo {author} {\bibfnamefont {V.~L.}\ \bibnamefont
  {Deringer}}, \bibinfo {author} {\bibfnamefont {N.}~\bibnamefont {Bernstein}},
  \bibinfo {author} {\bibfnamefont {G.}~\bibnamefont {Csányi}}, \bibinfo
  {author} {\bibfnamefont {C.~B.}\ \bibnamefont {Mahmoud}}, \bibinfo {author}
  {\bibfnamefont {M.}~\bibnamefont {Ceriotti}}, \bibinfo {author}
  {\bibfnamefont {M.}~\bibnamefont {Wilson}}, \bibinfo {author} {\bibfnamefont
  {D.~A.}\ \bibnamefont {Drabold}},\ and\ \bibinfo {author} {\bibfnamefont
  {S.~R.}\ \bibnamefont {Elliott}},\ }\bibfield  {title} {\bibinfo {title}
  {Corigins of structural and electronic transitions in disordered silicon},\
  }\href@noop {} {\bibfield  {journal} {\bibinfo  {journal} {Nature}\ }\textbf
  {\bibinfo {volume} {589}},\ \bibinfo {pages} {59} (\bibinfo {year}
  {2021})}\BibitemShut {NoStop}%
\bibitem [{\citenamefont {Zeni}\ \emph
  {et~al.}(2021{\natexlab{a}})\citenamefont {Zeni}, \citenamefont {Rossi},
  \citenamefont {Glielmo},\ and\ \citenamefont {De~Gironcoli}}]{Zeni2020}%
  \BibitemOpen
  \bibfield  {author} {\bibinfo {author} {\bibfnamefont {C.}~\bibnamefont
  {Zeni}}, \bibinfo {author} {\bibfnamefont {K.}~\bibnamefont {Rossi}},
  \bibinfo {author} {\bibfnamefont {A.}~\bibnamefont {Glielmo}},\ and\ \bibinfo
  {author} {\bibfnamefont {S.}~\bibnamefont {De~Gironcoli}},\ }\bibfield
  {title} {\bibinfo {title} {Compact atomic descriptors enable accurate
  predictions via linear models},\ }\href@noop {} {\bibfield  {journal}
  {\bibinfo  {journal} {Journal of Chemical Physics}\ }\textbf {\bibinfo
  {volume} {154 (22)}},\ \bibinfo {pages} {224112} (\bibinfo {year}
  {2021}{\natexlab{a}})}\BibitemShut {NoStop}%
\bibitem [{\citenamefont {Zeni}\ \emph
  {et~al.}(2021{\natexlab{b}})\citenamefont {Zeni}, \citenamefont {Rossi},
  \citenamefont {Glielmo},\ and\ \citenamefont {De~Gironcoli}}]{Zeni2021}%
  \BibitemOpen
  \bibfield  {author} {\bibinfo {author} {\bibfnamefont {C.}~\bibnamefont
  {Zeni}}, \bibinfo {author} {\bibfnamefont {K.}~\bibnamefont {Rossi}},
  \bibinfo {author} {\bibfnamefont {A.}~\bibnamefont {Glielmo}},\ and\ \bibinfo
  {author} {\bibfnamefont {S.}~\bibnamefont {De~Gironcoli}},\ }\bibfield
  {title} {\bibinfo {title} {Compact atomic descriptors enable accurate
  predictions via linear models},\ }\bibfield  {journal} {\bibinfo  {journal}
  {Journal of Chemical Physics}\ }\textbf {\bibinfo {volume} {154}},\ \href
  {https://doi.org/10.1063/5.0052961} {10.1063/5.0052961} (\bibinfo {year}
  {2021}{\natexlab{b}})\BibitemShut {NoStop}%
\bibitem [{\citenamefont {Lysogorskiy}\ \emph {et~al.}(2021)\citenamefont
  {Lysogorskiy}, \citenamefont {van~der Oord}, \citenamefont {Bochkarev},
  \citenamefont {Menon}, \citenamefont {Rinaldi}, \citenamefont
  {Hammerschmidt}, \citenamefont {Mrovec}, \citenamefont {Thompson},
  \citenamefont {Csányi}, \citenamefont {Ortner},\ and\ \citenamefont
  {Drautz}}]{Lysogorskiy2021}%
  \BibitemOpen
  \bibfield  {author} {\bibinfo {author} {\bibfnamefont {Y.}~\bibnamefont
  {Lysogorskiy}}, \bibinfo {author} {\bibfnamefont {C.}~\bibnamefont {van~der
  Oord}}, \bibinfo {author} {\bibfnamefont {A.}~\bibnamefont {Bochkarev}},
  \bibinfo {author} {\bibfnamefont {S.}~\bibnamefont {Menon}}, \bibinfo
  {author} {\bibfnamefont {M.}~\bibnamefont {Rinaldi}}, \bibinfo {author}
  {\bibfnamefont {T.}~\bibnamefont {Hammerschmidt}}, \bibinfo {author}
  {\bibfnamefont {M.}~\bibnamefont {Mrovec}}, \bibinfo {author} {\bibfnamefont
  {A.}~\bibnamefont {Thompson}}, \bibinfo {author} {\bibfnamefont
  {G.}~\bibnamefont {Csányi}}, \bibinfo {author} {\bibfnamefont
  {C.}~\bibnamefont {Ortner}},\ and\ \bibinfo {author} {\bibfnamefont
  {R.}~\bibnamefont {Drautz}},\ }\bibfield  {title} {\bibinfo {title}
  {Performant implementation of the atomic cluster expansion (pace) and
  application to copper and silicon},\ }\bibfield  {journal} {\bibinfo
  {journal} {npj Computational Materials}\ }\textbf {\bibinfo {volume} {7}},\
  \href {https://doi.org/10.1038/s41524-021-00559-9}
  {10.1038/s41524-021-00559-9} (\bibinfo {year} {2021})\BibitemShut {NoStop}%
\bibitem [{\citenamefont {Kovács}\ \emph {et~al.}(2021)\citenamefont
  {Kovács}, \citenamefont {van~der Oord}, \citenamefont {Kucera},
  \citenamefont {Allen}, \citenamefont {Cole}, \citenamefont {Ortner},\ and\
  \citenamefont {Csányi}}]{kovacs2021}%
  \BibitemOpen
  \bibfield  {author} {\bibinfo {author} {\bibfnamefont {D.~P.}\ \bibnamefont
  {Kovács}}, \bibinfo {author} {\bibfnamefont {C.}~\bibnamefont {van~der
  Oord}}, \bibinfo {author} {\bibfnamefont {J.}~\bibnamefont {Kucera}},
  \bibinfo {author} {\bibfnamefont {A.~E.~A.}\ \bibnamefont {Allen}}, \bibinfo
  {author} {\bibfnamefont {D.~J.}\ \bibnamefont {Cole}}, \bibinfo {author}
  {\bibfnamefont {C.}~\bibnamefont {Ortner}},\ and\ \bibinfo {author}
  {\bibfnamefont {G.}~\bibnamefont {Csányi}},\ }\bibfield  {title} {\bibinfo
  {title} {Linear atomic cluster expansion force fields for organic molecules:
  Beyond rmse},\ }\href@noop {} {\bibfield  {journal} {\bibinfo  {journal} {J.
  Chem. Theory Comput.}\ } (\bibinfo {year} {2021})}\BibitemShut {NoStop}%
\bibitem [{\citenamefont {Rasmussen}\ \emph {et~al.}(2006)\citenamefont
  {Rasmussen}, \citenamefont {Williams}, \citenamefont {Press}, \citenamefont
  {Bach},\ and\ \citenamefont {(Firm)}}]{rasmussen_2006_gpr}%
  \BibitemOpen
  \bibfield  {author} {\bibinfo {author} {\bibfnamefont {C.}~\bibnamefont
  {Rasmussen}}, \bibinfo {author} {\bibfnamefont {C.}~\bibnamefont {Williams}},
  \bibinfo {author} {\bibfnamefont {M.}~\bibnamefont {Press}}, \bibinfo
  {author} {\bibfnamefont {F.}~\bibnamefont {Bach}},\ and\ \bibinfo {author}
  {\bibfnamefont {P.}~\bibnamefont {(Firm)}},\ }\href@noop {} {\emph {\bibinfo
  {title} {Gaussian Processes for Machine Learning}}},\ Adaptive computation
  and machine learning\ (\bibinfo  {publisher} {MIT Press},\ \bibinfo {year}
  {2006})\BibitemShut {NoStop}%
\bibitem [{\citenamefont {Engel}\ \emph {et~al.}(2018)\citenamefont {Engel},
  \citenamefont {Anelli}, \citenamefont {Ceriotti}, \citenamefont {Pickard},\
  and\ \citenamefont {Needs}}]{Engel2018}%
  \BibitemOpen
  \bibfield  {author} {\bibinfo {author} {\bibfnamefont {E.~A.}\ \bibnamefont
  {Engel}}, \bibinfo {author} {\bibfnamefont {A.}~\bibnamefont {Anelli}},
  \bibinfo {author} {\bibfnamefont {M.}~\bibnamefont {Ceriotti}}, \bibinfo
  {author} {\bibfnamefont {C.~J.}\ \bibnamefont {Pickard}},\ and\ \bibinfo
  {author} {\bibfnamefont {R.~J.}\ \bibnamefont {Needs}},\ }\bibfield  {title}
  {\bibinfo {title} {Mapping uncharted territory in ice from zeolite networks
  to ice structures},\ }\bibfield  {journal} {\bibinfo  {journal} {Nature
  Communications}\ }\textbf {\bibinfo {volume} {9}},\ \href
  {https://doi.org/10.1038/s41467-018-04618-6} {10.1038/s41467-018-04618-6}
  (\bibinfo {year} {2018})\BibitemShut {NoStop}%
\bibitem [{\citenamefont {Stone}(1974)}]{Stone1974}%
  \BibitemOpen
  \bibfield  {author} {\bibinfo {author} {\bibfnamefont {M.}~\bibnamefont
  {Stone}},\ }\bibfield  {title} {\bibinfo {title} {Cross-validatory choice and
  assessment of statistical predictions},\ }\href@noop {} {\bibfield  {journal}
  {\bibinfo  {journal} {Journal of the Royal Statistical Society, Series B
  (Methodological)}\ }\textbf {\bibinfo {volume} {36 (2)}},\ \bibinfo {pages}
  {111 } (\bibinfo {year} {1974})}\BibitemShut {NoStop}%
\bibitem [{\citenamefont {Drautz}(2020)}]{Drautz2020}%
  \BibitemOpen
  \bibfield  {author} {\bibinfo {author} {\bibfnamefont {R.}~\bibnamefont
  {Drautz}},\ }\bibfield  {title} {\bibinfo {title} {Atomic cluster expansion
  of scalar, vectorial, and tensorial properties including magnetism and charge
  transfer},\ }\bibfield  {journal} {\bibinfo  {journal} {Physical Review B}\
  }\textbf {\bibinfo {volume} {102}},\ \href
  {https://doi.org/10.1103/PhysRevB.102.024104} {10.1103/PhysRevB.102.024104}
  (\bibinfo {year} {2020})\BibitemShut {NoStop}%
\bibitem [{\citenamefont {Zeni}\ \emph
  {et~al.}(2021{\natexlab{c}})\citenamefont {Zeni}, \citenamefont {Rossi},
  \citenamefont {Pavloudis}, \citenamefont {Kioseoglou}, \citenamefont
  {de~Gironcoli}, \citenamefont {Palmer},\ and\ \citenamefont
  {Baletto}}]{Zeni2021a}%
  \BibitemOpen
  \bibfield  {author} {\bibinfo {author} {\bibfnamefont {C.}~\bibnamefont
  {Zeni}}, \bibinfo {author} {\bibfnamefont {K.}~\bibnamefont {Rossi}},
  \bibinfo {author} {\bibfnamefont {T.}~\bibnamefont {Pavloudis}}, \bibinfo
  {author} {\bibfnamefont {J.}~\bibnamefont {Kioseoglou}}, \bibinfo {author}
  {\bibfnamefont {S.}~\bibnamefont {de~Gironcoli}}, \bibinfo {author}
  {\bibfnamefont {R.~E.}\ \bibnamefont {Palmer}},\ and\ \bibinfo {author}
  {\bibfnamefont {F.}~\bibnamefont {Baletto}},\ }\bibfield  {title} {\bibinfo
  {title} {Data-driven simulation and characterisation of gold nanoparticle
  melting},\ }\href@noop {} {\bibfield  {journal} {\bibinfo  {journal} {Nature
  Communications}\ }\textbf {\bibinfo {volume} {12 (1)}},\ \bibinfo {pages} {1}
  (\bibinfo {year} {2021}{\natexlab{c}})}\BibitemShut {NoStop}%
\bibitem [{\citenamefont {Bonilla}\ \emph {et~al.}(2009)\citenamefont
  {Bonilla}, \citenamefont {Chai},\ and\ \citenamefont
  {Williams}}]{Bonilla2009}%
  \BibitemOpen
  \bibfield  {author} {\bibinfo {author} {\bibfnamefont {E.~V.}\ \bibnamefont
  {Bonilla}}, \bibinfo {author} {\bibfnamefont {K.~M.~A.}\ \bibnamefont
  {Chai}},\ and\ \bibinfo {author} {\bibfnamefont {C.~K.}\ \bibnamefont
  {Williams}},\ }\bibfield  {title} {\bibinfo {title} {Multi-task gaussian
  process prediction}\ }(\bibinfo {year} {2009})\BibitemShut {NoStop}%
\bibitem [{\citenamefont {West}\ \emph {et~al.}(2007)\citenamefont {West},
  \citenamefont {Ventura},\ and\ \citenamefont {Warnick}}]{West2007}%
  \BibitemOpen
  \bibfield  {author} {\bibinfo {author} {\bibfnamefont {J.}~\bibnamefont
  {West}}, \bibinfo {author} {\bibfnamefont {D.}~\bibnamefont {Ventura}},\ and\
  \bibinfo {author} {\bibfnamefont {S.}~\bibnamefont {Warnick}},\ }\bibfield
  {title} {\bibinfo {title} {Spring research presentation: A theoretical
  foundation for inductive transfer},\ }\href@noop {} {\bibfield  {journal}
  {\bibinfo  {journal} {Brigham Young University, College of Physical and
  Mathematical Sciences}\ } (\bibinfo {year} {2007})}\BibitemShut {NoStop}%
\bibitem [{\citenamefont {Jacobsen}\ \emph {et~al.}(2018)\citenamefont
  {Jacobsen}, \citenamefont {Jørgensen},\ and\ \citenamefont
  {Hammer}}]{Jacobsen2018}%
  \BibitemOpen
  \bibfield  {author} {\bibinfo {author} {\bibfnamefont {T.~L.}\ \bibnamefont
  {Jacobsen}}, \bibinfo {author} {\bibfnamefont {M.~S.}\ \bibnamefont
  {Jørgensen}},\ and\ \bibinfo {author} {\bibfnamefont {B.}~\bibnamefont
  {Hammer}},\ }\bibfield  {title} {\bibinfo {title} {On-the-fly machine
  learning of atomic potential in density functional theory structure
  optimization},\ }\bibfield  {journal} {\bibinfo  {journal} {Physical Review
  Letters}\ }\textbf {\bibinfo {volume} {120}},\ \href
  {https://doi.org/10.1103/PhysRevLett.120.026102}
  {10.1103/PhysRevLett.120.026102} (\bibinfo {year} {2018})\BibitemShut
  {NoStop}%
\bibitem [{\citenamefont {Río}\ \emph {et~al.}(2019)\citenamefont {Río},
  \citenamefont {Mortensen},\ and\ \citenamefont {Jacobsen}}]{delrio2019}%
  \BibitemOpen
  \bibfield  {author} {\bibinfo {author} {\bibfnamefont {E.~G.~D.}\
  \bibnamefont {Río}}, \bibinfo {author} {\bibfnamefont {J.~J.}\ \bibnamefont
  {Mortensen}},\ and\ \bibinfo {author} {\bibfnamefont {K.~W.}\ \bibnamefont
  {Jacobsen}},\ }\bibfield  {title} {\bibinfo {title} {Local bayesian optimizer
  for atomic structures},\ }\bibfield  {journal} {\bibinfo  {journal} {Physical
  Review B}\ }\textbf {\bibinfo {volume} {100}},\ \href
  {https://doi.org/10.1103/PhysRevB.100.104103} {10.1103/PhysRevB.100.104103}
  (\bibinfo {year} {2019})\BibitemShut {NoStop}%
\bibitem [{\citenamefont {Yang}\ \emph {et~al.}(2021)\citenamefont {Yang},
  \citenamefont {Jiménez-Negrón},\ and\ \citenamefont {Kitchin}}]{Yang2021}%
  \BibitemOpen
  \bibfield  {author} {\bibinfo {author} {\bibfnamefont {Y.}~\bibnamefont
  {Yang}}, \bibinfo {author} {\bibfnamefont {O.~A.}\ \bibnamefont
  {Jiménez-Negrón}},\ and\ \bibinfo {author} {\bibfnamefont {J.~R.}\
  \bibnamefont {Kitchin}},\ }\bibfield  {title} {\bibinfo {title}
  {Machine-learning accelerated geometry optimization in molecular
  simulation},\ }\bibfield  {journal} {\bibinfo  {journal} {Journal of Chemical
  Physics}\ }\textbf {\bibinfo {volume} {154}},\ \href
  {https://doi.org/10.1063/5.0049665} {10.1063/5.0049665} (\bibinfo {year}
  {2021})\BibitemShut {NoStop}%
\end{thebibliography}%

\end{document}